\documentclass[11pt]{amsart}
\usepackage[dvips]{graphicx} 
\usepackage{amssymb, latexsym, amsmath, amscd, array, epigraph}
\usepackage{epsf}
\usepackage[french]{babel}
\usepackage{url}
\usepackage[pagewise]{lineno}
\usepackage{setspace}
\usepackage{float}
\usepackage{enumitem}
\usepackage[T1]{fontenc}
\usepackage[applemac]{inputenc}
\usepackage{xstring}
\makeatletter
\expandarg
\makeatother
\usepackage{color}

\title{\'Evolution du principe d'exclusion comp\'etitive:\\
\begin{small}
le r\^ole des math\' ematiques. 
\end{small}
}
\author{Claude Lobry}
\address{$^*$Centre de Recherche en Histoire des Idées,  Université de Nice Sophia-Antipolis,
98, Bd Edouard Herriot, BP 3209- 0624 Nice-cedex.}
\date{\today}

\begin{document}
\newcommand{\bsm}{\begin{small}}
\newcommand{\fsm}{\end{small}}
\newcommand{\ANS}{Analyse Non Standard}
\newcommand{\green}{\textcolor{green}}
\newcommand{\red}{\textcolor{red}}
\newcommand{\bleu}{\textcolor{blue}}
\newcommand{\memo}{\footnote{\red{ref}}}
\newcommand{\fin}{\end{document}}
\newcommand{\beq}{\begin{equation}}
\newcommand{\feq}{\end{equation}}
\newcommand{\dcom}{\begin{quote}\begin{small}}
\newcommand{\fcom}{\end{small}\end{quote}}
\newcommand{\mq}{\footnote{\red{ref}}}
\newcommand{\bq}{\begin{quote}}
\newcommand{\fq}{\end{quote}}
\newcommand{\cg}{\big \{ }
\newcommand{\cd}{\big \} }
\newcommand{\nb}{\overline{n} }
\newcommand{\om}{\omega }
\newcommand{\Om}{\Omega }
\newcommand{\Rmat}{\mathbb{R} }
\newcommand{\Nmat}{\mathbb{N} }
\newcommand{\Zmat}{\mathbb{Z} }
\newcommand{\Qmat}{\mathbb{Q} }
\newcommand{\bv}{\big \bracevert }
\newcommand{\e}{\mathrm{e}}
\newcommand{\eps}{\varepsilon}
\newcommand{\xgr}{\textbf{x}}
\newcommand{\igr}{\textbf{i}}
\newcommand{\sgr}{\textbf{s}}
\newcommand{\qS}{S^{n-1^+} }
\newcommand{\gralpha}{ \boldsymbol{\alpha}}
\newcommand{\chn}{\Sigma_{chn}}
\newcommand{\grm}{\textbf{m}}
\newcommand{\gri}{\textbf{I}}
\newcommand{\rgr}{\textbf{r}}
\newcommand{\Ical}{\mathcal{I}}
\newcommand{\grd}{\textbf{d}}
\newcommand{\bitbul}{\begin{itemize}[label = \textbullet]}
\newcommand{\bittiret}{\begin{itemize}[label = -]}
\newcommand{\bito}{\begin{itemize}[label =$\circ$]}
\newcommand{\bit}{\begin{itemize}}
\newcommand{\fit}{\end{itemize}}
\newcommand{\ben}{\begin{enumerate}}
\newcommand{\fen}{\end{enumerate}}
\setlength{\fboxrule}{0.3mm}
\setlength{\fboxsep}{1mm}

\maketitle
\subsection*{Abstract}
Tout le monde peut constater que depuis cent cinquante ans l'écologie théorique s'est considérablement mathématisée.
Mais quelle est la nature de ce phénomène ? Les mathématiques s'appliquent-elles, comme par exemple dans l'usage des tests statistiques, où bien s'impliquent elles, comme dans la physique, où les lois ne peuvent s'exprimer sans elles ?
 A travers l'histoire du {\em Principe d'Exclusion Compétitive}
formulé au tout début du XX me siècle par le naturaliste Grinnell à propos de la répartition des mésanges à dos marron, jusqu'à son intégration moderne dans, ce qu'en mathématiques, on appelle la {\em dynamique des populations}, je mets en évidence
l'efficacité de ce  qu'on pourrait appeler  le {\em roman mathématique} dans le travail de clarification de certains concepts de l'écologie théorique.
\tableofcontents

\section{Introduction : Qu'est-ce que le Principe d'Exclusion Compétitive ?}

 Dans un ouvrage à destination des étudiants en écologie  \cite{BAR92}\footnote{
 {\em \'Ecologie des peuplements,} Masson 1992.} R. Barbault écrit au début du court   paragraphe de trois pages intitulé \textbf{Le principe d'exclusion compétitive} (à partir de maintenant désigné par PEC) : 
 \dcom
 Le principe d'exclusion compétitive, en germe dans ''L'origine des espèces'' de Darwin, a été principalement développé par Grinnell (1904) et Gause (1935). Hardin (1960) le résume en ces  termes :\\
 « Deux ou plusieurs espèces présentant des modes d'utilisation des ressources identiques ne peuvent continuer à coexister dans un environnement stable, la plus apte éliminant l'autre .» 
\fcom  
 Le paragraphe est ensuite consacré à une critique assez serrée de ce qui vient d'être énoncé pour conclure, en s'appuyant sur des travaux postérieurs à 1970,  par :
 \dcom
 En d'autres termes, il y a une autre solution possible que l'exclusion compétitive : la coexistence par partage des ressources.
 \fcom
 Comme je ne voudrais pas que l'on pense que cette opinion négative sur le PEC, est celle d'un ''gauchiste-irresponsable''   formé à l'université de Vincennes \footnote{ 
 La carrière académique de R. Barbault montre qu'il est tout, sauf un ''gauchiste-irresponsable'' : il a notamment dirigé le département {\em Écologie et gestion de la biodiversité} au Muséum National d’Histoire Naturelle.}
 immédiatement après mai 1968, voici, par exemple,  une autre opinion qui nous vient cette fois d'outre Atlantique, 
 sous la plume de P. J. den Boer\footnote{
 {\em The Present Status of the Competitive Exclusion Principle}, Trends in Ecology and Evolution, Vol 1, n°1, (1986) \cite{BOE86}}:
 \dcom
 These tests led to the conclusion that congeneric species coexist more frequently than could be expected from a random distribution of species over habitats (or islands), so that the ‘exclusion principle’ could be replaced by the ‘coexistence principle’ :[...] 
  As a future trend we can expect a further depreciation of competition, both intra and interspecific, as being a major force in ecology and evolution. The manifold influences of weather, climate and other physical factors will assume greater importance, and the claim of universality of competition will probably be replaced by a gradual revaluation of the role of predation.
 \fcom
 Cet auteur qui met en avant le rôle de la {\em prédation} plutôt que celui de la {\em coopération}  n'a peut être pas la sensibilité politique de R. Barbault mais n'en rejette pas moins fermement la pertinence du concept {\em d'exclusion compétitive} dans une revue savante de premier plan. On voit que, du point de vue de l'écologie\footnote{
 Dans tout ce texte le mot ''écologie'' désigne l'écologie scientifique, jamais des mouvements sociaux ou politiques.},
 en un siècle,  l'idée du rôle joué par la compétition dans les rapports entre  les espèces vivantes à pour le moins évolué.
 
 L'écologie est issue du croisement du courant ''naturaliste'' (Linnée) de description des espèces vivantes et de la théorie de l'évolution. Si, comme le dit Darwin, les espèces évoluent sous la pression de l'environnement,  il devient important de comprendre leur relations avec ce dernier. Haeckel, à qui l'on attribue la création du néologisme {\em \' Ecologie} en 1866 écrit, en 1875\footnote{
 Cité par P. Acot, {\em Histoire de l'écologie}, Presses Universitaires de France (1988) \cite{ACO88}, p. 45} : 
 \dcom
 Par \oe cologie, on entend le corps du savoir concernant l'économie de la nature - l'étude de toutes les relations de l'animal à son environnement inorganique et organique ; ceci inclut, avant tout, les relations amicales ou hostiles avec ceux des animaux et des plantes avec lesquels il entre directement ou indirectement  en contact - en un mot, l'\oe cologie est l'étude des relations complexes auquelles Darwin se réfère par l'expression de conditions de la lutte pour l'existence.
 \fcom
 Au XXIème siècle  l'écologie microbienne qui traite des organismes unicellulaires est devenue une branche importante sinon essentielle  de l'écologie (voir à ce sujet l'article de Jessup et al {\em Big questions, small worlds: microbial model systems in ecology}, \cite{JES04}). La tradition a imposé d'y utiliser la terminologie générale introduite par Haeckel et ses contemporains mais
 on mesure à quel point ce vocabulaire, ''amical'', ''hostile'', ''lutte pour l'existence'' serait jugé inapproprié s'il était utilisé, pour la première fois de nos jours,  pour parler des relations entre ces organismes  unicellulaires, bactéries, virus, levures et autres micro organismes qui occupent maintenant une place centrale dans notre compréhension de l'organisation du vivant. La description de ce  qui se passe dans l'écosystème intestinal d'un mammifère  relève plus de la chimie que des arts martiaux.

  Mais à la fin du XIXème siècle on pensait essentiellement organismes macroscopiques et il ne semblait pas incongru d'utiliser ce genre de vocabulaire avec le risque  d'introduire de façon inconsciente dans une science naissante les préjugés idéologiques d'une époque\footnote{Des préjugés qui, par exemple, on conduit Haeckel à développer une théorie raciste de l'évolution humaine expliquant une soi disant supériorité de l'homme blanc.}. 
 Mais mon propos n'est pas de traiter de l'influence de l'air du temps de la fin du XIX siècle sur la création du PEC, ni en retour de son apport au darwinisme sociologique, mais de prendre acte des conditions de sa naissance et d'évaluer le rôle des mathématiques dans son évolution jusqu'à nos jours.
 
 Il y a un accord assez large chez les écologues (il est cité dans tous les cours de base) pour faire remonter l'idée d'exclusion compétitive à un article de 1904, de l'ornithologue  Joseph Grinnell,  intitulé  {\em The origin and distribution of the chestnut-backed chickadee} (Origine et distribution de la mésange à dos marron) \cite{GRI04}\footnote{
The Auk, Vol.21,N°3,pp. 364-382.} 
 où il s'intéresse à la répartition le long de la côte pacifique des Etats Unis de trois variétés de mésanges à dos marron se distinguant uniquement par la variation de la couleur de leurs flancs. 
Dans cet article le mot {\em competition} est utilisé trois fois :
\bito
\item {\small  (...) as to escape intra-specific competition (...)}
\item{ \small(...) their will be increased competition within the species itself (...)}
\item{\small  (...) in other words the extremest of intra-competition does not ensue until after further dissemination is impossible.}
\fit
{\em compete} apparait deux fois :
\bito 
\item{\small  (...) and be less likely to compete with their parents or other of their kind.}
\item{\small  No one of these could probably be successfully competed against by a foreigner.}
\fit
et finalement {\em exclusion} ne figure qu'une seule fois dans le passage crucial :
\dcom
 Every animal tends to increase at a geometric ratio, and is checked only by limit of food supply.
It is only by adaptations to different sorts of food, or modes of food getting, that more than one species can occupy the same locality.
{\em Two species of approximately the same food habits are not likely to remain long evenly balanced in numbers in the same region}. One will crowd out the other; the one longest exposed to local conditions, and hence best fitted, though ever so slightly, will survive, to the \textbf{exclusion}\footnote{Souligné par moi} of any less favored would be invader. 
\fcom
dont la phrase que j'ai mise en italique est si souvent citée. 

Comme on peut le voir  le style est littéraire. La longue citation  à venir, que je ne propose pas à la lecture mais simplement pour  confirmer que c'est bien là le style d"écriture de Grinnel, est représentative de tout l'article. En particulier on n'y trouve pas le moindre symbole mathématique.
\dcom
We come now to consider the origin of the races of {\em Parus rufescens}. In a species of recent arrival into a new region (by invasion from a neighboring faunal area), as it adapts itself better and better to its new surroundings, granted the absence of closely related or sharply competing forms, its numbers will rapidly increase. This means that there will be increased competition within the species itself, on account of limited food supply. The alternative results are either starvation for less vigorous individuals during recurring seasons of unusual food scarcity, or dissemination over a larger area. In a way the first might be considered as beneficial in the long run, as doubtless leading to the elimination of the weaker; such a process evidently does take place to a greater or less degree all the time, and is important for the betterment of the race. But as a matter of observation Nature first resorts to all sorts of devices to ensure the spreading of individuals over all inhabitable regions; in other words, the extremest intra-competition does not ensue until after further dissemination
is impossible. In birds we find a trait evidently developed on purpose to bring about scattering of individuals. This is the autumnal 'mad impulse' which occurs just after the complete annual moult, when both birds-of-the-year and adults are in the best physical condition, and just before the stress of winter food shortage. Even in the most sedentary of birds, in which no other trace of a migratory instinct is discernible, this fall season of unrest is plainly in evidence. I may suggest not unreasonably that autumnal migration may have had its origin in such a trait as this, the return movement in the spring becoming a necessary sequence.
\fcom
 Devons nous en conclure que la démarche de Grinnell n'est pas scientifique ? Certainement pas, comme en témoigne, entre autres, une liste de références extrêmement fournie (figure \ref{Grinnell}); Grinnell sait de quoi il parle, cite ses sources et ce n'est pas un hasard si son nom a été retenu.
\begin{figure}[!tb]
   \centering
  \includegraphics[width=0.8\textwidth]{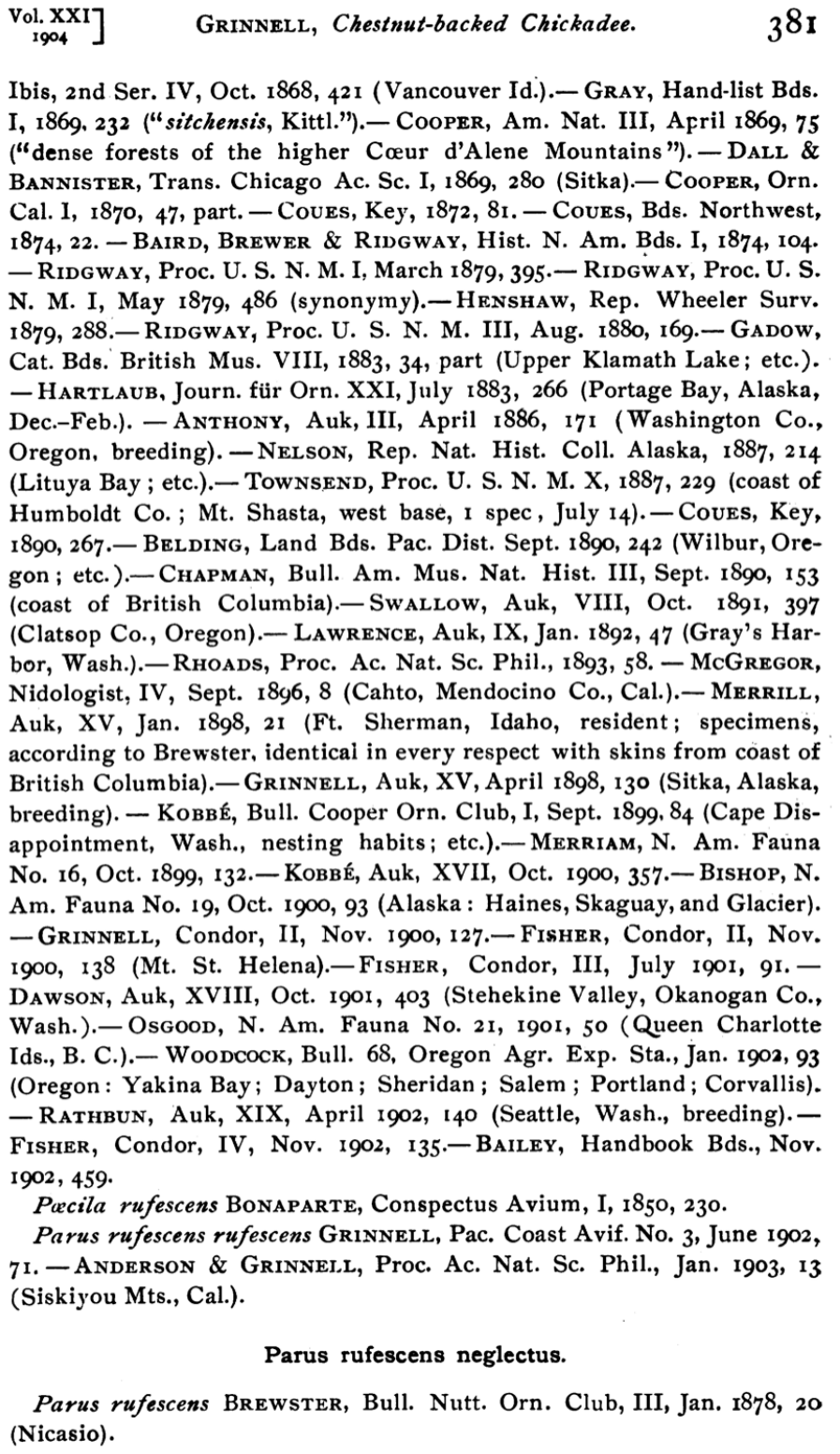} 
  \caption{Quelques références de
  {\em The origin and distribution of the chestnut-backed chickadee} }\label{Grinnell} 										\end{figure}
Mais à la fin du XIXeme siècle l'usage des modèles mathématiques pour analyser la dynamique des populations n'était pas connu chez les naturalistes.

Maintenant que nous ne sommes plus à la fin  du XIXème siècle et que  nous avons vu que le PEC est largement remis en cause voyons ce que Wikipedia, qui est une bonne source pour humer "l'air de notre temps", nous en dit\footnote{
Wikipedia, {\em Competitive exclusion principle}, article consulté le 16/09/2023.}.
 \dcom
 In ecology, the competitive exclusion principle,[1] sometimes referred to as Gause's law,[2] is a proposition that two species which compete for the same limited resource cannot coexist at constant population values.
 \fcom 
 La première référence [1] est : Hardin,  {\em The Competitive Exclusion Principle}, \cite{HAR60}\footnote{Science, vol 131, (1960)}. On reconnait, en version anglaise, la formule prêtée, en traduction française, par R. Barbault à Hardin. L'attribution à ce dernier de la paternité de l'expression "two species which compete for the same limited resource cannot coexist at constant population values" qu'on retrouve {\em ad nauseam} dans la littérature demanderait à elle seule une étude particulière car elle n'est pas de lui ! En effet, dans l'article cité,  Hardin l'attribue à un auteur anonyme qui à son tour l'attribue à Gause ; le texte exact de Hardin est :
 \dcom
In the words of an anonymous reporter\footnote{Anonymous, {\em J. Animal Ecol.} 13, 176 (1944)} "a lively discussion...centred about Gause's contention (1934) that two species with similar ecology cannot live together in the same place.
 \fcom
 et les mots "limited" et "resource" n'y figurent pas.  En revanche il s'agit d'un essai plutôt  brillant \footnote{
 Ce qui ne veut pas forcément dire scientifiquement bien fondé comme nous le verrons plus loin}
 ce qui explique peut être son important succès\footnote{
 Selon Google Scholar, au 16/09/2023 il est cité 4301 fois}. 
 Le style est différent de celui de Grinnel, pas seulement parce qu'il s'agit d'un court  essai et non d'une publication assez technique, mais par l'apparition d'arguments mathématiques  : \og Demonstration of the formal truth of the principle have been given in terms of calculus and set theory.\fg$\,$  Mais Hardin ne veut pas, dans ce texte, utiliser des équations et il enchaine \og Those to whom the mathematics does not appeal may prefer the following intuitive verbal argument\fg$\,$, argument  qu'il développe ensuite. Ainsi, plus  d'un demi-siècle après Grinnel il est amis que des argument mathématiques peuvent être pertinents en dynamique des populations, mais aussi qu'ils peuvent rebuter le lecteur  ou la lectrice naturaliste.
 
 Si maintenant nous voulons savoir  ce qu'est le "style" de l'écologie théorique moderne  nous pouvons regarder le "Aims and scope" de la revue {\em Theoretical Population Biology}, une des revues importantes de l'écologie théorique, dont le premier paragraphe est
 \dcom
 An interdisciplinary journal, Theoretical Population Biology presents articles on \textbf{theoretical} aspects of the biology of populations, particularly in the areas of \textbf{demography, ecology, epidemiology, evolution, and genetics}\footnote{
 Les caractères gras sont de la revue.}.
 Emphasis is on the development of mathematical theory and models that enhance the understanding of biological phenomena.
 \fcom 
 et où l'on trouve couramment des formules du genre de celle de la figure \ref{maths} prise au hasard dans un article.
 \begin{figure}[!tb]
   \centering
  \includegraphics[width=0.8\textwidth]{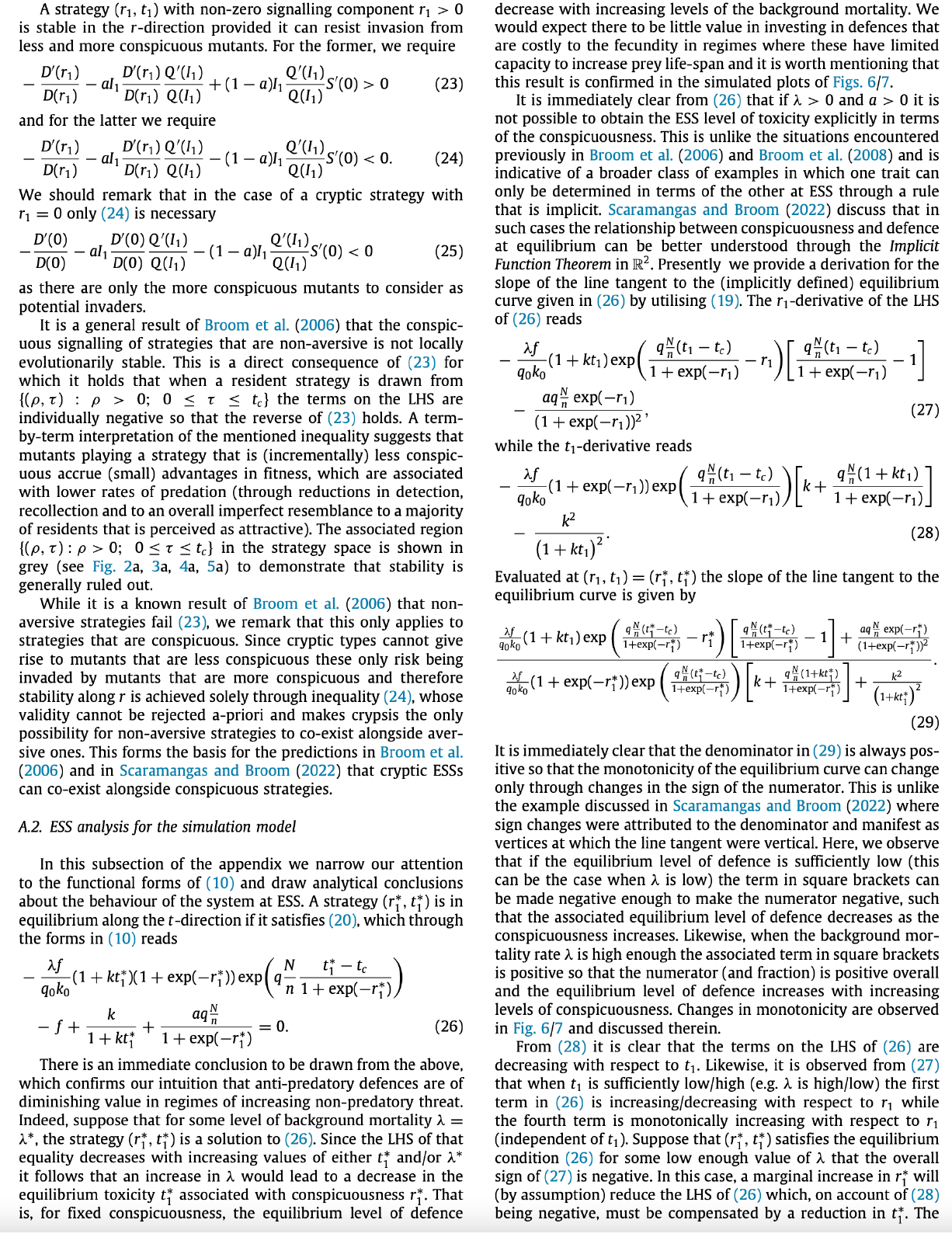} 
  \caption{Une page typique de la revue {\em Theoretical Population Biology} avec de nombreuses formules mathématiques..}\label{maths} 
\end{figure}

Donc nous pouvons conclure que depuis Haekel, en 150 ans, l'écologique s'est incontestablement mathématisée. Mais quel est le sens de cette mathématisation ?
A-t-elle joué un rôle dans l'évolution de la conception du  PEC ? La thèse que je soutiens ici est que la réponse à cette question est {\em oui} et que le style de mathématisation qui est utilisé dans ce domaine mériterait de s'appeler {\em roman mathématique} comme je l'explique en conclusion.
\newpage

\section{1840-1920 : Modèle  ou courbe d'ajustement ? La logistique.}
\textit{Dans cette section il n'est pas encore question de compétition entre deux populations mais des premières tentatives de "mathématisation"  du phénomène de croissance d'une population. Ces mathématisations auront bien entendu une influence sur la façon d'aborder la "compétition".}
\subsection{Modèle et fonction logistique : le point de vue actuel}
Conscient du risque de lire de façon erronée des textes anciens à la lumière de concepts contemporains je commence quand même par exposer ce qu'un étudiant de licence de sciences de la vie pourrait lire dans un cours contemporain sur la  ''logistique''. Je le fais pour attirer l'attention du lecteur ignorant des méthodes de l'écologie théorique actuelle, sur la nature du lien entre mathématiques et réalité qui, en écologie, est très différent de ce qu'il est en physique. 

On appelle {\em modèle logistique} l'équation différentielle
\beq \label{logistic1}
\frac{dx}{dt} = rx\left(1-\frac{x}{K}\right)
\feq 
Le paramètre  $K$ s'appelle la ''capacité de charge'' (en anglais ''carrying capacity''). 
On montre de façon élémentaire (par exemple en remarquant que $y = 1/x$ est régi par une équation linéaire) que la solution de \eqref{logistic1} de condition initiale $x(0) = x_0$ est la fonction :
\beq \label{flogistic}
t \to x(t) =  \frac{x_0 K\e^{rt}}{K-x_0+x_0\e^{rt}}
\feq
dont le graphe est représenté sur la figure \ref{logistique} et qu'on appelle {\em fonction logistique}. La fonction \eqref{flogistic} peut être envisagée comme une simple courbe d'ajustement à des "phénomènes en $S$" dépendant de trois paramètres alors que dans  l'équation différentielle \eqref{logistic1}  on peut voir un "mécanisme" explicatif des propriétés de $t \mapsto x(t)$. 
 \begin{figure}[!tb]
   \centering
  \includegraphics[width=0.8\textwidth]{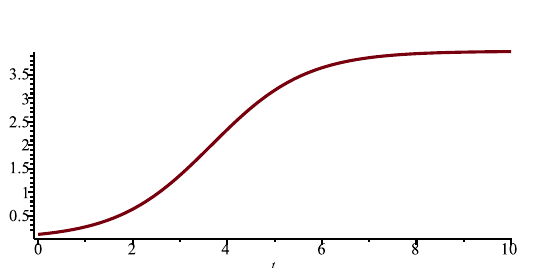} 
  \caption{Graphe de la fonction "logistique" \eqref{flogistic}, $ K = 4, r =1, x_0 = 0.1$}\label{logistique} 
\end{figure}

Un premier mécanisme possible est\\
\textbf{Le modèle  $r-K$ : limitation par disparition de la ressource.} Je considère un  exemple.  Dans une boite de Pétri on cultive un microorganisme (une bactérie, une levure...) qui se nourrit d'un substrat. On désigne par $x(t)$ la quantité de microorganisme présent à l'instant $t$ et par $S(t)$ la quantité de substrat ; par ''quantité'' on entend la concentration. On admet que, pendant une durée $dt$, la quantité de substrat consommé est  proportionnelle à la probabilité de rencontre entre une molécule de substrat et une cellule soit $\mu S(t)x(t)$ ce qui donne 
$
S(t+dt) = S(t) -\mu x(t)S(t)dt
$
ou, en prenant la limite $dt \to 0$
$$\frac{dS}{dt} = -\mu x(t)  S(t)$$
D'autre part on admet qu'une fraction constante du substrat absorbé est transformé en microorganisme donc que
\beq \label{equSx}
\frac{dx}{dt} = + \rho \mu x(t)  S(t)
\feq
La quantité $\rho S(t)+ x(t)$ est donc constante et égale à $\rho S(0)+x(0)$ soit $ S(t) = S(0) + x(0)/\rho - x(t)/ \rho$ ; l'inoculum $x(0)$ est par hypothèse petit et donc négligé devant $S(0)$ ; on reporte $S(t)$ dans \eqref{equSx} et il vient
\beq \label{equx}
\frac{dx}{dt} = +\mu x(t) (\rho S(0) -x(t))
\feq
qui s'écrit aussi
\beq \label{equxbis}
\frac{dx}{dt} = x(t) \mu \rho S(0)  \left(1 -\frac{x(t)}{\rho S(0)}\right)
\feq
qui est l'équation \eqref{logistic1} avec $r = \mu \rho S(0)$ et $K = \rho S(0)$. Les solutions de $\eqref{logistic1}$, pour des conditions initiales petites croissent vers l'équilibre $K = \rho S(0)$ pour lequel tout le substrat initial $S(0)$ a été transformé en la biomasse  $\rho S(0)$. D'où l'expression ''capacité de charge'' ou ''carrying capacity'' : c'est la quantité maximale de population que le milieu (ici notre boite de Petri avec une quantité de substrat donnée) est susceptible de recueillir.

L'hypothèse qui conduit à l'équation \eqref{equSx} est ce que les chimistes appellent la ''loi d'action de masse''. Si $A$ et $B$ désignent deux espèces qui réagissent en produisant une espèce $C$
$$ A + B \longrightarrow C$$
si $[A]$, $[B]$ et [C] désignent les concentrations de $A$, $B$ et $C$ la vitesse de production de $C$ est proportionnelle au produit des concentrations de $A$ et $B$, c'est à dire à la probabilité de rencontre d'une molécule de A avec une molécule de B,  et donc
$$ \frac{d[C]}{dt} = \mu [A][B]$$
Dans le cas de la croissance des microorganismes la ''réaction chimique'' est {\em substrat} + {\em biomasse} $\longrightarrow$ {\em biomasse} ce que les chimistes appellent réaction auto catalytique (i.e. le produit de la réaction est un des constituants de la réaction).

La logique en \oe uvre dans cette interprétation est une logique de croissance de la population jusqu'à épuisement de la ressource. Ici la ressource est la nourriture mais ce pourrait être  un autre type de ressource comme, par exemple, l'espace  dans le cas d'une colonisation d'une ile par des plantes. En effet, si $s(t)$ est la surface colonisée à l'instant $t$ on a 
$$ s(t+dt) = s(t) + \mu s(t) (S-s(t))dt.$$
L'accroissement de la surface colonisée est proportionnel à la quantité de plantes émettrices de graines, donc à la surface actuellement colonisée, et à la surface $S-s(t)$ qu'il reste à coloniser.

Ces exemples et d'autres conduisent à imaginer une ressource abstraite, dont la nature n'est pas précisée, que la croissance de la population épuise progressivement. La ressource disponible initialement (la quantité de substrat, la surface de l'ile) est la ''capacité de charge'' ou  ''carrying capacity" en anglais.\\\\
Une autre interprétation possible est \\
\textbf{Le modèle $r-\alpha$ : limitation par interférence.} Considérons  l'équation
\beq  \label{logistic2}
\frac{dx}{dt} = r x - \alpha x^2
\feq
Mathématiquement parlant c'est la même équation que \eqref{logistic1}(avec $K = \frac{r}{\alpha}$), mais on lui donne une interprétation complètement différente. On imagine une population d'animaux vivant sur un site où la nourriture est en si grande quantité que, en l'absence d'autres facteurs, la croissance serait exponentielle :
\beq\label{lin}
\displaystyle  \frac{dx}{dt} = r x
\feq
Mais ces animaux ont un tempérament agressifs les uns envers les autres si bien que la rencontre de deux individus  se traduit en moyenne par la disparition de $\alpha$ individus. Le résultat est l'équation \eqref{logistic2} dont les solutions positives tendent vers $x^* = \frac{r}{\alpha}$. La quantité $ \frac{r}{\alpha}$ est la ''population à l'équilibre" et la constante $\alpha$ est la ''force de la compétition intraspécifique''. Le  modèle $r-\alpha$ est une autre interprétation possible de l'équation différentielle \eqref{logistic1}

Cette rapide analyse de "la logistique", met bien en évidence la question de l'interprétation. 

\bit
\item Pour le mathématicien ou la mathématicienne, modèle $r-K$ et modèle $r-\alpha$ c'est la même chose puisque, modulo quelques changements de variables on peut toujours se ramener à $\frac{dx}{dt} = x(1-x)$ et donner des formules explicites des solutions à l'aide de fonctions exponentielles.
\item Pour  la dynamique des population, les propriété mathématiques des équations étant établies, ce qui compte, c'est l'interprétation. L'interprétation $r-K$ attribue la diminution du taux de croissance à une {\em modification de l'environnement de la population}- diminution de la ressource disponible, substrat, surface colonisable - alors que l'interprétation $r-\alpha$ l'attribue à une {\em propriété intrinsèque de la population} - conséquences de la rencontre de deux individus -\fit

Dans les manuels d'écologie en usage actuellement c'est encore largement l'interprétation $r-K$ qui prédomine. Toutefois cette dernière est aussi vivement critiquée. Un article de 2012, dont le titre annonce clairement la couleur, 
{\em The struggle for existence: how the notion of carrying capacity, K, obscures the links between demography, Darwinian evolution, and speciation} de J. Mallet \cite{MAL12}\footnote{Evolutionary Ecology Research, 2012, 14: 627–665} fait une synthèse de diverses critiques possibles du modèle $r-K$.
\dcom
Given that both these formulations [la formulation $r-K$ et $r-\alpha$]  represent the same underlying model, it is obviously immaterial which we use. However, as I argue below, the $r-K$ model has nonetheless misled generations of ecologists and evolutionary biologists in a way that the $r-\alpha$ model would not have done.
\fcom 
Un autre article, un peu plus récent, {\em The perfect mixing paradox and the logistic equation: Verhulst vs. Lotka} \cite{ARD16}\footnote{
Arditi, R., Bersier, L. F., \& Rohr, R. P. (2016). {\em Ecosphere}, 7(11), e01599.}
ajoute que certains paradoxes de l'interprétation $r - K$, qui apparaissent lorsqu'elle est utilisée pour des modèles à plusieurs sites, disparaissent dans l'interprétation $r -\alpha$. Il plaide pour l'utilisation prioritaire de cette paramétrisation.
\dcom
We have shown that the logistic equation in its usual r-K parameterization presents paradoxical properties when generalized to a multi-patch situation.\\
(...)\\
With the original Verhulst form (i.e., with the r-a parameterization), Eq. (11) shows that both parameters can simply be averaged independently from one another in order to describe the dynamics of the average population in the two- patch environment.\\
(...)\\
Regrettably, most, if not all, textbooks make uncritical and exclusive use of the r-K form of the logistic model.
\fcom

\subsection{Verhulst le démographe.}
	Revenons à l'histoire de la logistique.
	
Pierre François Verhulst (1804-1849) est un mathématicien Belge, membre de l'Académie Royale, dont l'histoire personnelle et celle  de la logistique sont  racontées dans l'article  bien documenté,  {\em  Les héritiers de Pierre-François Verhulst: une population dynamique},  du mathématicien Belge J. Mawhin \cite{MAW02}\footnote{
 {\em Bulletin de la Classe des sciences}, tome 13, n°7-12, 2002. pp. 349-378} 
 dont je ne garderai que les quelques éléments suivants. Un premier mémoire de 1838 de Verhulst, {\em Notice sur la loi que la population suit dans son accroissement}   \cite{VER38},  est consacré à l'accroissement de la taille, notée $p(t)$, d'une population humaine. Il part de la loi exponentielle qu'il considère être la loi de Malthus (1766-1834) et il écrit :
\dcom
Soit $p$ la population ; représentez par $dp$ l'accroissement infiniment petit qu'elle acquiert pendant le temps infiniment court $dt.$. Si la population croît selon une progression géométrique, nous aurons l'équation $dp / dt = mp$. Cependant, comme le taux de croissance de la population est diminué par l'augmentation même du nombre d'habitants, il faut soustraire une fonction inconnue de $p$ à $mp$, de telle sorte que la formule à intégrer devienne :
$$\frac{dp}{dt} = mp - \varphi(p) $$
\fcom
On appréciera la remarquable modernité de cette proposition où l'évolution d'une quantité discrète, un nombre d'individu, est approchée sans autre forme de procès par l'évolution régie par une équation différentielle d'une quantité continue $p(t)$.   C'est que Verhulst a été formé par le grand savant Belge A. Quetelet (1796-1874), mathématicien, astronome et statisticien. Il poursuit en disant que la fonction la plus simple que l'on puisse prendre pour $\varphi(p) $ et $np^2$. L'équation logistique 
\beq \label{verhulst1}
\frac{dp}{dt} = mp - np^2 
\feq 
 est née et immédiatement intégrée ce qui donne
 \dcom
\beq \label{verhulst2}
p =\frac{mp'\e^{mt}}{np'\e^{mt}- m - np'}
\feq
(...) où l'on note par $p'$ la population qui correspond à $t = 0$
\fcom 
Le mémoire suivant \cite{VER46}, {\em Note sur la loi d’accroissement de la population}, de 1846, qui est beaucoup plus développé reprend l'idée de la diminution du taux de croissance avec des notations un peu différentes et justifie à nouveau le choix de $np^2$ par la simplicité
\dcom
On peu faire une infinité d'hypothèses sur la loi d'affaiblissement du coefficient $\frac{1}{M}$  [$P$ dans les anciennes notations]. La plus simple consiste à regarder cet affaiblissement comme proportionnel à l'accroissement de la population (...)
\fcom
Comme la parabole $p \mapsto mp- np^2$ est symétrique par rapport à $\frac{n}{m}p$ les solution de l'équation logistique ont pour particularité que leur point d'inflexion se trouve être celui où la taille de la population $p(t)$ est la moitié de  la taille limite qu'elle est susceptible d'atteindre. Cette particularité n'échappe pas à Verhulst qui affirme
\dcom 
En général, plus la fonction qui représente la gène éprouvée par la population, sera rapidement croissante, plus l'ordonnée du point d'infexion sera grande par rapport à l'ordonnée limite. Pour le montrer, nous prendrons
$$ \frac{Mdp}{p dt}  = m - f(p)$$
pour l'équation différentielle de la population.
\fcom
et il se livre à une véritable étude qualitative de cette équation, c'est à dire une étude où le second membre n'est pas une fonction spécifique, mais une fonction qui satisfait quelques conditions particulières. C'est un point de vue remarquablement moderne d'étude qualitative d'une classe d'équations différentielles.

Mais l'intérêt de la forme simple \eqref{verhulst1}  est que ses solutions \eqref{verhulst2} peuvent s'exprimer à l'aide de fonctions exponentielles, donc tabulées, et par suite, se prêter à des calculs numériques d'ajustement $-$ nous sommes au  milieu du XIXeme et siècle l'ordinateur n'existe pas ! Une fois les paramètres estimés on pourra extrapoler et prédire la taille de la population à venir. Les calculs que fait Verhulst à partir de statistiques démographiques disponibles pour la Belgique et la France prédisent une taille limite d'un ordre de grandeur tout à fait  raisonnable ( 6,6 M pour la Belgique et   40 M pour la France) mais surtout Verhulst nous avertit sur les limites de ces prédictions
\dcom
Ces résultats numériques nous apprennent que, si les lois et les moeurs de la Belgique n'éprouvaient aucun changement notable, {\em la population de ce royaume, bien que toujours croissante, ne s'élèverait jamais à six millions six cent mille âmes.}
\fcom 
\subsection{Pearl et les croissances de microorganismes}
Cette technique de quantification du phénomène de croissance de population humaine via des équations différentielles publiée par Verhulst en 1835  est restée très longtemps ignorée de la communauté des naturalistes. Ce n'est que dans les années 1920, près d'un siècle plus tard, que l'usage de l'équation logistique y fera son apparition. 

Pour fêter le 50 ême anniversaire de sa création  la revue {\em The Quarterly Review of Biology} re-publiait un article de 1927 de R. Pearl (1879-1940),   {\em The growth of populations} \cite{PEA27} où l'on pouvait trouver l'ajustement de la figure \ref{pearl} de la croissance d'une levure par une logistique.
\begin{figure}[!tb]
   \centering
  \includegraphics[width=0.8\textwidth]{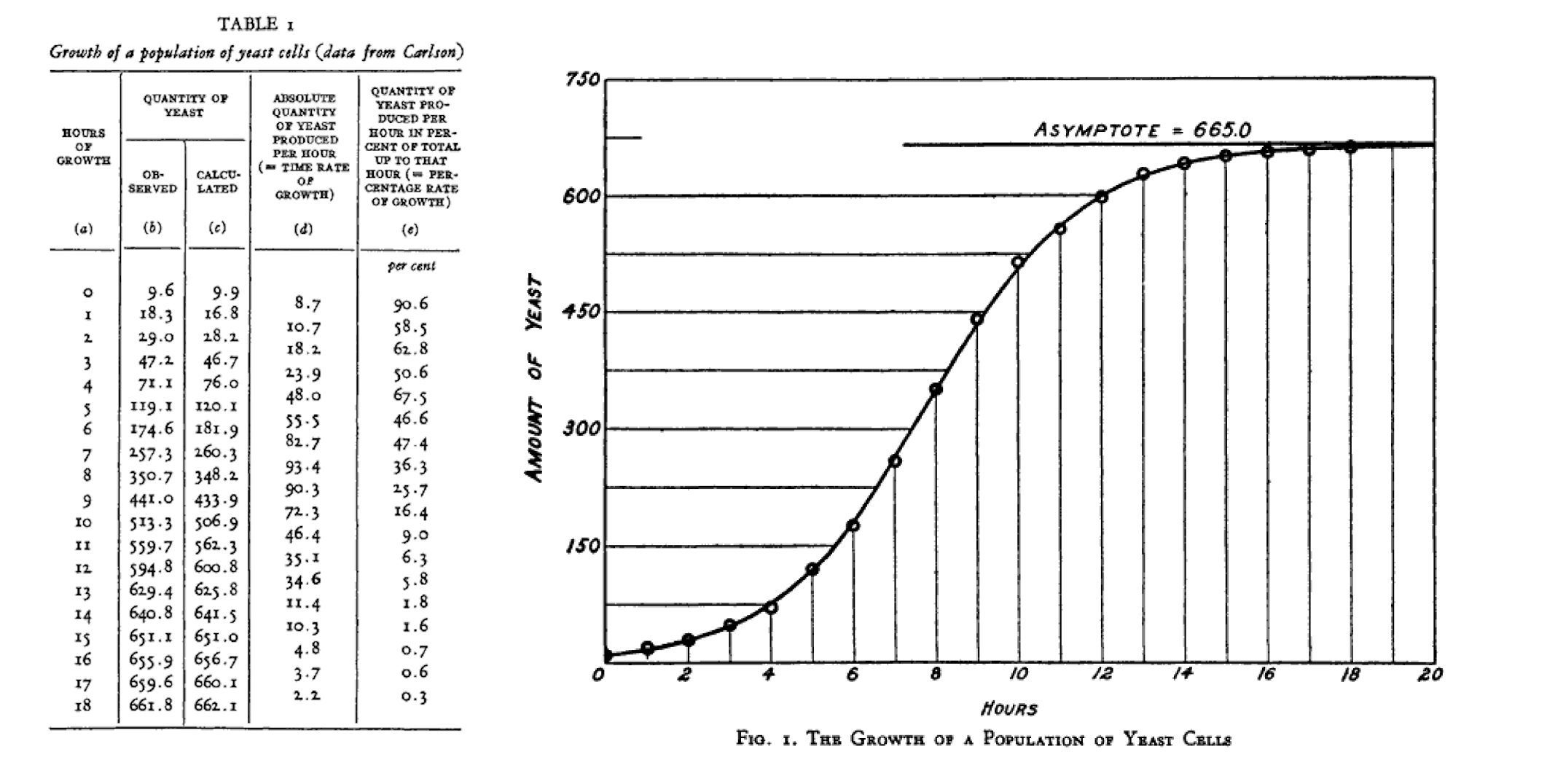} 
  \caption{Ajustement à une logistique de données de croissance d'une population de levures par Pearl \cite{PEA27}. Les chiffres des colonnes (a) (observations) et (b) (ajustement logistique) sont représentés graphiquement à droite.}\label{pearl} 
\end{figure}
Cet ajustement d'une remarquable précision a certainement beaucoup fait pour la popularisation de l'ajustement logistique au delà du raisonnable ce qui a conduit le grand probabiliste  W. Feller (1906-1970) à publier en 1940 \cite{FEL40}\footnote{
On the logistic law of growth and its empirical verifications in biology, {\em  Acta Biotheoretica 5, (1940) pp. 51-66}.} 
un article où il montre ironiquement que la croissance du tournesol peut être tout aussi bien, sinon mieux,  décrite par la fonction $arctg$  que par la logistique. Ici Feller montre bien la différence qu'il y a entre une courbe d'ajustement et un modèle explicatif.

Toutefois le purgatoire de la logistique n'a pas duré jusqu'en 1927. Un peu plus tôt, en 1908, T. B. Robertson (1884-1930), un biochimiste publie  {\em On the Normal Rate of Growth of an Individual, and its Biochemical Significance} et {\em Further remarks on the normal rate of growth of an individual, and its biochemical significance} ( \cite{ROB08a} et \cite{ROB08b}) où l'on peut lire
\dcom
The differential equation which is characteristic of the initial stages of an autocatalytic monomolecular  reaction is as follows:
$$\frac{dx}{dt} = k_1x(a-x)$$
which expresses in mathematical symbols {\em the fact that the velocity of the transformation is, at any instant, proportional to the amount of material which is undergoing change and to the amount of material which has already been transformed}\footnote{L'emphase est de Robertson.}. 
\fcom 
 qui  ne concernent plus seulement la croissance de populations d'humains, mais la croissance d'organismes individuels, humains ou autres.  Il étend hardiment hors du champs de la chimie le concepts de réaction auto-catalytique ce qui est clairement une interprétation $r-K$ de la logistique. Il ne cite pas Verhulst.
 
 Ces deux articles sont remarqués par Pearl dans un article de 1909 \cite{PEA09} consacré à la croissance d'organes (cerveau de rat, moelle épinière ...)
 \dcom
 A new and suggestive view in regard to the ultimate physiology of the growth process has been put forth recently in two papers by Robertson. In brief this view is stated by the author in the following words (first  paper p.612) :
 \ben
 \item(...)
 \item Any particular cycle of growth obeys the formula $ log\frac{ x}{A - x}= K(t-t_1)$   where $x$ is the amount (in weight or volume) of growth
which  has been attained at time $t$ , $A$  is the total amount  of growth attained during, the cycle, $K$ is a constant and $t_1$ 1 is the time at which
growth is half completed.
 \item (...) 
 \fen
  These conclusions, if well founded, are certainly, of very fundamental importance
 \fcom 
Ce n'est qu'en 1920 que, en  collaboration avec J. Reed,  Pearl publie des ajustements de la fonction logistique à des populations humaines, toujours sans référence à Verhulst, en s'appuyant sur les arguments suivants :
\dcom
The following conditions should be fulfilled by any equation which is to describe adequately the growth of population in an area of fixed limits.
\ben 
\item  Asymptotic to a line $y = k$ when $x = + \infty $.
\item Asymptotic to a line $y = 0$ when $x = - \infty$.
\item A point of inflection at some point $x = \alpha$ and $y = \beta$
\item Concave upwards to left of $x = \alpha$ and concave downward to right
of $x = \alpha$
\item No horizontal slope except at $x = ± \infty$.
\item Values of y varying continuously from $0$ to $k$ as $x$ varies from $ -\infty $ to $ +\infty$.
\fen 
In these expressions $y$ denotes population, and $x$ denotes time. An equation which fulfils these requirements is
$$ y = \frac{b\e^{ax}}{1+c \e^{ax}} \quad \quad \quad (ix)$$
\fcom
Il caractérise les asymptotes,  les points d'inflexion de la fonction puis il établit qu'elle est solution d'une équation différentielle
\dcom
Expressing the first derivative of (ix) in terms of $y$, we have
$$\frac{dy}{dx} = \frac{ay(b-cy)}{b}$$
Putting the equation in this form shows at once that it is identical with that describing an autocatalyzed chemical reaction, a point to which we shall return later.
\fcom 
Ce qu'il fait à la fin de l'article
\dcom
The same objections [La symétrie irréaliste du modèle logistique] apply to the use of the equation of an autocatalytic reaction to the representation of organic growth in the individual. This fact has been noted by Robertson who was the first to discover that, in general, growth follows much the same curve as that of autocatalysis. What needs to be done is to generalize (ix) in some such form as will free it from the two restrictive features (location of point of inflection and symmetry) we have mentioned, and will at the same time retain its other essential features. We are working along this line now and hope presently to reach a satisfactory solution.
\fcom
C'est finalement dans un article de 1927 (\cite{PEA27}) que Pearl reconnait l'antériorité de Verhulst dans l'introduction de la logistique en dynamique des populations 
\dcom
The equation to the curve which has been found by experiment and observation to be descriptive of population growth in a wide variety of organisms was first discovered by the Belgian mathematician, Verhulst (\cite{VER38, VER46, VER47}). His pioneer work was forgotten, and consequently overlooked by most subsequent students of the population problem. In 1920 the present writer and his colleague, Lowell J. Reed, without any knowledge of Verhulst's prior work, independently hit upon the same equation.
Verhulst called his curve the "logistic."
This usage we shall follow. 
\fcom
\subsection{Premiers enseignements.}
C'est dans son second mémoire que Verhulst propose l'expression ''logistique'' pour désigner la courbe
\dcom
Nous donnerons le nom de {\em logistique} à la courbe ({\em voyez la figure}) caractérisée par l'équation précédente.
\fcom
De nos jours le mot désigne aussi bien la courbe d'ajustement que le modèle.

Dans mon rappel de ce qu'un étudiant contemporain sait de la logistique j'ai pris soin de distinguer le ''modèle'' \eqref{logistic1} de la ''courbe'' \eqref{flogistic}.
Les citations que j'ai proposées de Verhulst puis de Robertson montrent qu'un mathématicien du milieu du XIX$^e$ siècle ou un biochimiste du début du XX$^e$ maitrisaient parfaitement l'usage des équations différentielles que l'on attend maintenant de tout étudiant en sciences.   Ils utilisent les notations modernes, savent exprimer les solutions à l'aide des fonctions élémentaires et surtout ils maitrisent l'idée selon laquelle des variables continues peuvent rendre compte de quantités discrètes comme le nombre d'individus d'une population pourvu que ce dernier soit grand. En bref ils maitrisent l'art de ce qu'on appelait au milieu du XX$^e$ siècle la ''mise en équation'' et que maintenant on appelle ''modélisation''.

En revanche il est manifeste que pour Pearl l'usage des mathématiques consiste essentiellement à trouver les meilleures courbes d'ajustement à une série de données d'observation ou expérimentales. Son style montre qu'il n'a pas une vision claire de ce qu'est une équation différentielle. Ceci est corroboré par une affaire déplaisante qui est bien documentée par l'historienne S. Kingsland dont j'utilise ici le travail \cite{KIN82}\footnote{ The refractory model: The logistic curve and the history of population ecology. {\em The Quarterly Review of Biology} 57.1 (1982): 29-52}. Kingsland nous dit
\dcom
But his mathematical methods [Celles de Pearl] were equally under attack, and nowhere with greater fervor than from the Harvard physicist turned statistician, Edwin Bidwell Wilson.
\fcom 
Wilson (1874-1964) est reconnu  comme un mathématicien important  par ses contributions et son rôle institutionnel (voir sa biographie  \cite{HUN73}), mais aussi pour sa propension à délivrer des avis définitifs et pas toujours bien fondés comme, en 1903, de critiquer les fondement de la géométrie de Hilbert  dans un article au tire provocateur \cite{WIL03} : ''The so-called foundations of geometry''. Si Hilbert n'a pas souffert des critiques de Wilson il n'en est pas de même pour Pearl.
 Kingsland cite dans   \cite{KIN82} un passage où Wilson s'attaque de façon particulièrement caustique à Pearl et son école  en 1924 dans un séminaire à la Johns Hopkins University (où travaille Pearl) :
\dcom
Their attitude is Shamanistic. They go through with magic propitiatory rites, idolatrous of mathematics, ignorant of one were interested in using the logistic
what it can and can not do for them. And I am not quite sure that the high priests of this pure and undefiled science do not somewhat aid and abet the idolatry.
\fcom 
et réussit à empêcher la nomination de Pearl à Harward où il souhaitait succéder à M. Wheeler.

Même si ses critiques  sont partiellement justifiées (voir \cite{KIN82} pour plus de détails) l'attitude de Wilson est très méprisante et ne rend pas justice aux contributions de Pearl.
En effet ce dernier, par ses travaux et son soutien à Lotka et Gause\footnote{C'est grace à lui que les deux grands livres classiques que sont \cite{LOT25} et \cite{GAU34} ont été publiés.}, a largement contribué à l'introduction des méthodes expérimentales de laboratoire dans le monde des naturalistes à une époque où l'observation directe de la nature prévalait.

\section{1920-1940 : L'âge d'or de l'écologie théorique}\label{agedor}
\textit{ Mon titre est repris  de  l'ouvrage de F. Scudo et J. Ziegler, {\em The Golden Age of Theoretical Ecology}\cite{SCU78} \footnote{ 
 {\em Lecture Notes in Biomathematics} Vol. 22 1978
}de 1978. 
Ce livre est la reproduction de 22 articles importants parus entre 1923 et 1940, ce qui, à une époque où internet n'existait pas, était une entreprise éminemment utile. Toutefois
l'emploi des termes ''écologie théorique'' me semble en partie usurpé puisque, d'une part, presque tous les articles réunis sont l'\oe vre de mathématiciens et, d'autre part,  
n'y figurent pas les travaux de Gause qui, lui, est incontestablement un grand, {\em sinon  le grand}, théoricien  de l'écologie de cette époque. Il semble que {\em \'Ecologie Mathématique} aurait été préférable. En gardant leur titre pour cette section je rends homage au travail de F. Scudo et J. Ziegler mais je ferai aussi une large part aux travaux de Gause pour le rendre plus adéquat.} 
\subsection{Lotka et Volterra}
\subsubsection*{Deux savants hors norme}
Deux mots pour commencer sur les personnages de Lotka (1880-1949) et Volterra (1860-1940) dont les noms sont associés au plus célèbre des modèles de dynamique des population : le {\em modèle proie-prédateur de Lotka-Volterra}\footnote{
G. Israel, historien des mathématiques, a consacré de nombreux travaux à Lotka et à Volterra. On en trouve une expression ''grand public'' dans son ouvrage {\em La mathématisation du réel}, Seuil 1996 \cite{ISR93} qui s'appuie sur plusieurs versions ''savantes'', en particulier \cite{ISR93} qui nous intéresse ici . }

 Alfred Lotka est un chimiste de formation, et un savant à l'esprit curieux qui a abordé une multitude domaines et s'est particulièrement intéressé aux sciences du vivant\footnote{Entre autres l'épidémiologie, la démographie ; il cite (p.7)    de son {\em Elements of Physical Biology.}\cite{LOT25} le {\em Principles of biology} du philosophe le philosophe H. Spencer}. Dans ce dernier domaine il ré-interprète des résultats mathématiques concernant les systèmes d'équations différentielles, qu'il tient de sa formation de chimiste. Ainsi il propose dans son ouvrage de 1925 \cite{LOT25}, {\em Elements of Physical Biology},  une interprétation d'un travail de 1920 sur les oscillations d'une hypothétique réaction chimique\footnote{\label{lotk}
{\em Undamped oscillations derived from the law of mass action}, Journal of the american chemical society, vol 42, n° 8, pp. 1595-1599 (1920)
}
  en terme de relation ''proie-prédateur'' (le modèle (\ref{pred1}) ci-dessous). Il faut également noter que Pearl a remarqué les travaux de Lotka dès 1920, les a soutenu et a finalement invité Lotka pour un long séjours à  Johns Hopkins où il a rédigé son  {\em Elements of physical biology}.

Vito Volterra (1860-1940) est un très grand mathématicien italien. Engagé dans la vie de la citée, il lutte contre la montée du fascisme et refuse de signer le serment d'allégeance au régime. En conséquence  il doit s'exiler à l'étranger à la fin de sa vie. Volterra a de très nombreuses contributions à la physique mathématique, notamment en mécanique. C'est  peu après 1920, à l'âge de la retraite (il a plus de soixante ans), alors qu'il est couvert d'honneurs, qu'il s'intéresse à une question d'halieutique qui lui a été posée par l'halieute Umberto d'Ancona (son beau fils) : comment expliquer que durant la guerre 1914-1918 la proportion de poissons ''prédateurs'' dans les pêches de la mer Adriatique semble augmenter ? Il propose, de façon tout à fait indépendante, le même modèle (\ref{pred1}) que Lotka pour en donner une explication. Il développera ses idée dans un long mémoire de 1927
{\em Variazioni e fluttuazioni del numero d'individui in specie animali conviventi}\footnote{ publié comme {\em Memorie del R. Comitato talassographico italiano}, Mem. CXXX1 (1927) qui reprend en grande partie des travaux publiés en 1926 (sous le même titre dans les {\em Memorie della R. Academia dei Lincei}, s. VI, vol II, pp. 31-113 ) où il signale les travaux de Lotka.  }
où il propose une théorie assez générale de l'interaction des espèces vivantes. De 1926 à la fin de sa vie (1940) le catalogue de ses \oe uvres complètes mentionne 70 publications dont environ la moitié sont consacrées à ce sujet, ce qui montre bien l'importance qu'il donne à ce travail\footnote{Il a une grande activité de communication. A côté de travaux techniques il multiplie les interventions devant des larges  publics scientifiques ou inversement fait des cours développés \cite{VOL31} devant des mathématiciens.}.

Lotka et Volterra ont tous deux considéré l'interaction de deux espèces dans les cas ou l'une est consommée par l'autre (proie-prédateur) et le cas où elle sont en compétition pour une même ressource.
\subsubsection*{Lotka, Volterra et la prédation : la loi d'action de masse.}

La {\em prédation} est un concept relativement clair : une espèce se développe au détriment d'une autre. Le cas extrême est celui ou l'espèce prédatrice se nourrit de l'espèce exploitée mais il peut s'agir aussi de parasitisme ou encore de consommation d'une ressource (par exemple de l'herbe par une vache, ou  du sucre par une bactérie) cette dernière étant la "proie". Sa représentation mathématique est plus claire que celle de la {\em compétition} et dans une certaine mesure lui est préalable puisque une forme de la compétition est la {\em compétition pour une ressource}. C'est pourquoi je commence par 
le  {\em Modèle Proie-Prédateur} de Lotka-Volterra. C'est  le système de deux équations différentielles\footnote{
Pour faciliter la lecture et la comparaison des modèles les notations ne sont pas les notations originales mais choisies pour cet exposé.}  :\\\\
 \fbox{
 \begin{minipage}{0.97\textwidth}
 \begin{center}
 \textbf{Le modèle de Lotka-Volterra} 
 \end{center}
\beq \label{pred1}
\begin{array}{rcl} 
\displaystyle  \frac{dx(t)}{dt}& =&\displaystyle ax(t)-bx(t)y(t) \\[8pt]
\displaystyle  \frac{dy(t)}{dt}& =&\displaystyle cx(t)y(t) - dx(t)  
  \end{array}
\feq
 où $x(t) $ représente la concentration à l'instant $t$ (quantité par unité de volume) des proies, $y(t)$ la concentration des prédateurs et les trois constantes $a,b,c,d$ sont strictement positives.  En général le temps est implicite dans le modèle et on préfère écrire :
 \beq \label{pred1bis}
\begin{array}{rcl} 
\displaystyle  \frac{dx}{dt}& =&\displaystyle ax-bxy \\[8pt]
\displaystyle  \frac{dy}{dt}& =&\displaystyle cxy - dx  
  \end{array}
\feq
 \end{minipage}}\\\\
 Dans ce modèle, en l'absence de prédateurs la population des proies obéit à $\frac{dx(t)}{dt} = ax(t)$, c'est à dire une croissance exponentielle indéfinie,  et en l'absence des proies celle des prédateurs  obéit à $\frac{dy(t)}{dt} = -dy(t)(t)$, c'est à dire une décroissance exponentielle conduisant à l'extinction.  Le terme $b x(t) y(t)$ doit retenir notre attention.
 Il est très directement inspiré de la loi d'action de masse des chimistes.

Pour Lotka, chimiste de formation, les populations sont des ''espèces biologiques'' analogues à des ''espèces chimiques''. Dans son article de 1920, {\em Undamped oscillations derived from the law of mass action} \cite{LOT20}  il démontre qu'une certaine réaction chimique {\em hypothétique} (voir figure \ref{articleLotka}) possède des solutions oscillantes non amorties.
Plus tard, dans son livre {\em Elements of physical biology} \cite{LOT25} il n'hésite pas à appliquer la même méthodologie pour retrouver ces mêmes équations comme modèle d'une  relation hôte-parasite (p. 88).

Pour Volterra la justification des équations est identique. Il a en tête une question où des poissons prédateurs se nourrissent d'autres poissons ; les poissons sont assimilés à des particules qui se rencontrent au hasard, comme il l'explique dans {\em Variations and Fluctuations of the Number of Individuals in Animal Species living together}\cite{VOL28}\footnote{
ICES Journal of Marine Science, 1928 (troisième version de \cite{VOL26} )} p. 22 :
\dcom
Let us suppose we have two species living together, and let $N_1$  and $N_2$ be respectively the numbers of individuals in each. The number of encounters of individuals of the first species with those of the second species, which occur in a unit of time, will be proportional to $N_1N_2$ and can therefore be assumed equal to $\alpha N_1N_2$, $\alpha$ being a constant. 

\fcom
Selon Volterra, dans la ''lutte pour la vie'' sauvage,  la probabilité de rencontre d'une ''proie'' de A avec son ''prédateur'' B est donc proportionnelle au produit des concentrations. C'est bien la loi d'action de masse des chimistes même s'il ne s'y réfère pas explicitement.
\begin{figure}[!tb]
   \centering
  \includegraphics[width=0.8\textwidth]{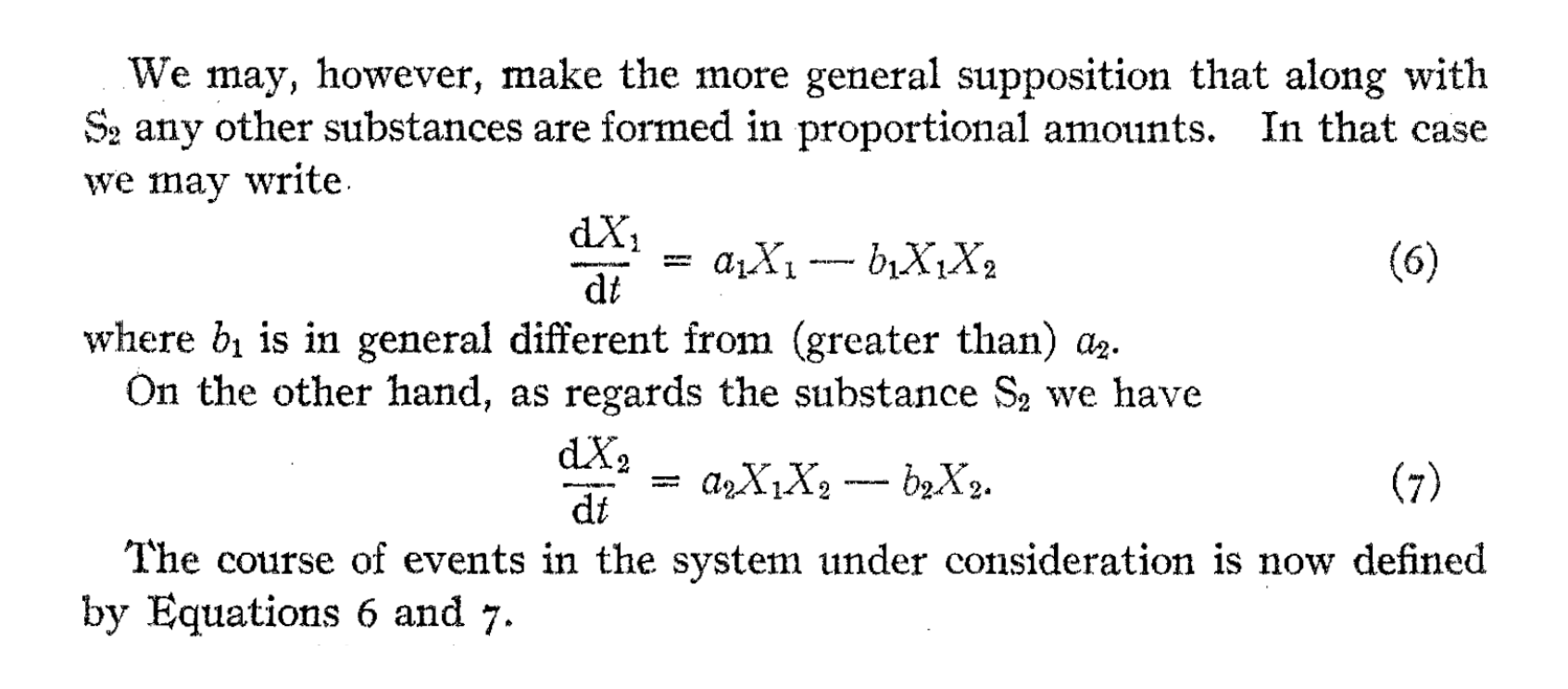} 
  \caption{La réaction chimique hypothétique de Lotka. Extrait de \cite{LOT20}}\label{articleLotka} 
\end{figure}

 Cette assimilation de populations d'organismes vivants à des substances chimiques ou à des molécules identiques et sans intention pose naturellement problème. Les deux auteurs en sont bien conscient et multiplient les remarques de prudence. Ainsi Lotka dans son  {\em Elements of physical biology} \cite{LOT25} p. 93 à la fin du traitement du système (\ref{pred1}) :
\dcom
(...) that the species $S_2$ cannot exterminate $S_1$ under the conditions here considered.

A word of caution, however, is perhaps in order. Although $S_2$ cannot exterminate $S_1$, it may so reduce the latter in numbers as to render it very vulnerable, and liable to extinction from those other influences which have deliberately been ignored in the development of the equations.
\fcom
De son côté, Volterra, dans l'introduction de son  {\em Variations and Fluctuations of the Number of Individuals in Animal Species living together} \cite{VOL28} déjà mentionné ci-dessus insiste sur les hypothèses qui sous-tendent sa théorie. 
\dcom
 On the basis of the ideas expressed above, in order to simplify the treatment, we shall assume that the species increase or decrease in a continuous way, that is to say we shall assume that the number which measures the quantity of individuals of a species is not an integer, but any real positive number whatever which varies continuously. In general the hatchings take place in definite periods separated from each other
by an interval of time; we shall neglect these particulars assuming that births may take place with continuity every moment and that, on a parity with all the other conditions, they may be verified proportionally to the number of living individuals of the species. Let the same assumption be made on death and, according as births may prevail over deaths, or vice versa, an increase or diminution of individuals will occur. Thus we shall assume the homogeneity of the individuals of each species neglecting the variations of age and size.
\fcom

Mais pour Volterra, la meilleure justification de son modèle qu'il puisse donner est qu'il fournit une explication possible au phénomène observé par d'Anconna. En effet, soumettons les deux espèce du modèle (\ref{pred1bis}) à un effort de pêche $\mathcal{E}$ et supposons que le prélèvement, dans chaque population, soit proportionnel à la concentration. Le modèle devient :
 \beq \label{pred2}
\begin{array}{rcl} 
\displaystyle  \frac{dx}{dt}& =&\displaystyle ax-bxy - \mathcal{E} x\\[8pt]
\displaystyle  \frac{dy}{dt}& =&\displaystyle cy - d\,x+\mathcal{E} y
  \end{array}
\feq
Il a par ailleurs démontré (comme l'avait fait Lotka) que les solutions de ces équations sont périodiques et que la moyenne de $x$ ou $y$ sur une période est toujours la même, donc  il suffit de s'intéresser à la valeur de $x$ et $y$ à l'équilibre, soit :
$$x_e = \frac{d+ \mathcal{E}}{c} \quad \quad y_e = \frac{a- \mathcal{E}}{b}$$
et par suite, dans ce modèle, la quantité de proie augmente sous l'action de la pêche alors que celle des prédateurs diminue. L'augmentation de la présence de poissons prédateurs pendant la guerre 1914-1918, où l'effort de pêche était moindre, est ainsi expliquée\footnote{
En fait cette confirmation empirique est moins solide qu'il semble car les statistiques effectuées par d'Ancona ont été contestées.}. Volterra insiste beaucoup, ce qui est très novateur, que plus qu'un bon ajustement à des données empiriques, ce qui fait la pertinence de son modèle, ce sont les conséquences indirectes qu'il est possible d'en retirer à travers l'analyse mathématique.

\subsubsection*{Lotka, Volterra et la compétition}
\paragraph{\textbf{La compétition : vision moderne}}\label{competemoderne}$\,$\\
Comme je l'ai fait pour la logistique je donne ici une vision "moderne" de la {\em compétition} telle qu'on la trouve par exemple dans le grand classique  {\em Population Ecology : A Unified Study of Animals and Plants} \cite{BEG02}\footnote{M. Begon, M. Mortimer and D. J. Thompson Blackwell Science 2002},
 dont voici le plan
\ben
\item Part 1 : Single-Species populations.\\
1) Describing populations, 2) Intraspecific competition, 3) Models of single-species populations.
\item Part 2 : Interspecific Interactions.\\
4) Interspecific competition, 5) Predation.
\item Part 3 : Synthesis.\\
6) Population regulation. 7) Beyond population ecology.
\fen
qui met en évidence l'importance de la distinction entre  compétition {\em intra} et {\em inter} spécifique. Je reprends ce point de vue à partir de la logistique, sous sa forme $r-\alpha$. Nous avons vu comment on peut interpréter le $\alpha$ de l'équation 
\beq \label{ralfa1}
\displaystyle \frac{dx}{dt} = rx - \alpha x^2
\feq
comme une diminution de la croissance de la population causée par la rencontre de deux  de ses individus. Si deux populations, 1 et 2,  sont mélangées la rencontre  entre des individus d'espèce différentes peut également avoir un effet négatif  ce qui conduit à écrire pour la première espèce
\beq \label{ralfa21}
\displaystyle \frac{dx_1}{dt} = r_1x_1 - \alpha_1 x_1^2 - \alpha_{12} x_2x_1
\feq
et pour la seconde 
\beq \label{ralfa12}
\displaystyle \frac{dx_2}{dt} = r_2x_2 - \alpha_2 x_2^2 - \alpha_{21} x_1x_2
\feq
ce qui donne le modèle global 
\beq \label{ralfa1221}
\begin{array}{rclcl}
\displaystyle \frac{dx_1}{dt} &=& r_1x_1 - \alpha_1 x_1^2 - \alpha_{12} x_2x_1&=& x_1\big( r_1- \alpha_1 x_1- \alpha_{12}x_2\big)\\[8pt]
\displaystyle \frac{dx_2}{dt} &=& r_2x_2 - \alpha_2 x_2^2 - \alpha_{21} x_1x_2&=& x_2\big( r_2- \alpha_{21}x_1 - \alpha_2 x_2\big)
\end{array}
\feq
où tous les paramètres sont positifs ou nuls.
L'analyse par isoclines, maintenant banale dans les ouvrages d'enseignement, fait ressortir 4 cas de figure en fonction des 6 paramètres, deux cas d'exclusion forte (une des espèces exclut  l'autre quelle que soit la condition initiale), un cas d'exclusion "faible" (une des espèces exclut l'autre, mais savoir laquelle dépend des conditions initiales, et enfin, c'est le point le plus important pour mon étude, les deux espèces coexistent en un équilibre stable (voir la figure \ref{BEG} extraite de \cite{BEG02}).
\begin{figure}[!tb]
   \centering
  \includegraphics[width=0.8\textwidth]{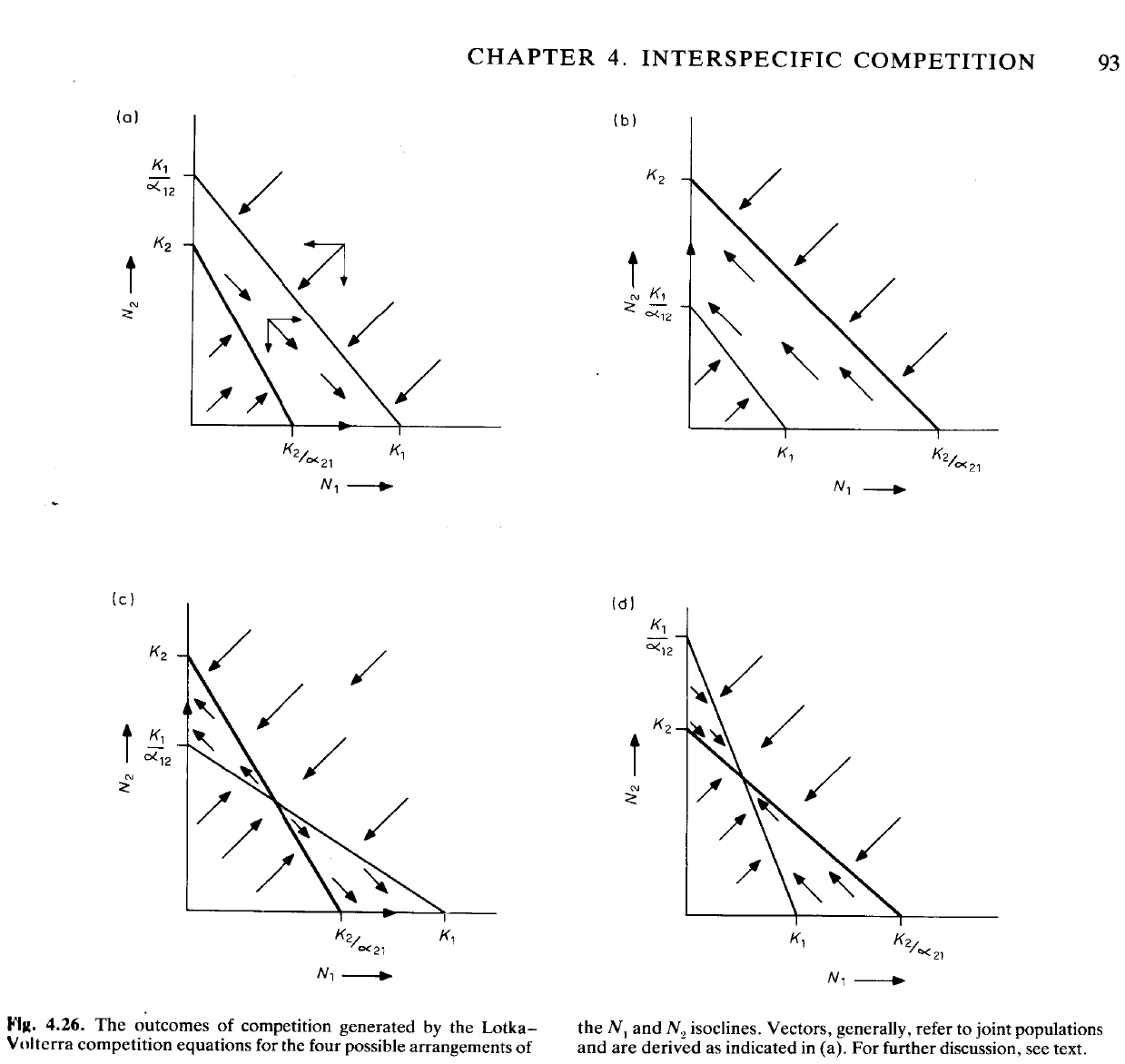} 
  \caption{Analyse par isocline du modèle \eqref{ralfa1221} dans un ouvrage moderne tel que \cite{BEG02}.}\label{BEG}
\end{figure} 
L'équilibre de coexistence à lieu lorsque $\alpha_1\alpha_2 > \alpha_{12}\alpha_{21}$ ce que l'on interprète en disant que la "pression de compétition intra-spécifique" est plus forte que la "pression de compétition inter-spécifique".

\paragraph{\textbf{La compétition chez Volterra}}$\;$\\
Dans son ouvrage de 1931 {\em Le\c cons sur la Théorie Mathématique de la lutte pour la vie} \cite{VOL31}\footnote{
Il s'agit du compte rendu de conférences données à l'Institut Henry Poincaré en 1927-28 à l'invitation de E. Borel  écrit par le jeune mathématicien  M. Brelot (qui deviendra le spécialiste reconnu de théorie du potentiel).}
 qui donne une vison mathématiquement assez achevée de Volterra sur cette question, il aborde la compétition de la façon suivante. 
 
 La chapitre I s'intitule {\em Coexistence de deux espèces} et se décline en trois sections : 1) {\em Deux espèces se disputant la même nourriture}, 2) Deux espèces dont l'une se nourrit de l'autre, 3) Deux espèces dans les divers cas d'action mutuelle. Ce chapitre I est à la fois une introduction au cas plus complexe de plus de deux espèces. La section 2) est consacré au modèle proie prédateur dont nous venons de parler. Je cite très largement la section 1).
 \dcom
 Supposons que, avec une nourriture en quantité suffisante pour satisfaire complètement la voracité de ces êtres, il y ait des coefficients d'accroissement positifs et constant $\eps_1, \eps_2$
 \fcom
 Le "coefficient d'accroissement" $\eps$ est le taux de croissance donc $\frac{1}{N} \frac{dN}{dt} = \eps$ 
\dcom
Si nous nous plaçons maintenant dans le cas réel d'espèces vivant dans un milieu délimité, la nourriture diminuera quand les nombre $N_1$ et $N_2$ des individus des deux espèces 
augmenteront et cela fera baisser la valeur des coefficients d'accroissement. Si l'on représente la nourriture dévorée par unité de temps par $F(N_1,N_2)$ fonction nulle avec $N_1$ et $N_2$ ensemble, tendant vers l'infini avec chacune des variables et fonction croissante de chacune d'elles, il sera assez naturel de prendre comme coefficients d'accroissement 
$$ \eps_1 - \gamma_1 F(N_1,N_2)\quad \quad \eps_2 - \gamma_2 F(N_1,N_2)$$
$\gamma_2$ et $\gamma_2$ étant des constantes positives correspondant aux deux espèces et à leurs besoins respectifs de nourriture.

D'ou le système différentiel traduisant le développement des espèces 
\beq \label{volterracompete}
\begin{array}{lcl}
\displaystyle \frac{dN_1}{dt} &=& [\eps_1-\gamma_1F(N_1,N_2)]N_1\\[8pt]
\displaystyle \frac{dN_2}{dt} &=& [\eps_2 - \gamma_2F(N_1,N_2)]N_2
\end{array}
\feq
\fcom
L'argumentation est claire. Il s'agit de deux espèces {\em se disputant} (on ne parle pas de compétition) une même nourriture. Il est incontestable que la quantité consommée pendant un temps $dt$ est $dt$ que multiplie une certaine fonction $F(N_1,N_2)$ ce qui aura pour effet de diminuer les taux de croissance de $N_1$ et $N_2$. A quelqu'un qui aurait critiqué la phrase  un peu vague  \og...étant des constantes positives correspondant aux deux espèces et à leurs besoins respectifs de nourriture\fg$\,$ 
Volterra aurait certainement pu répondre en explicitant l'équation de la "quantité de nourriture" en écrivant par exemple le système
\beq \label{volterracompetesubstrat}
\begin{array}{lcl}
\displaystyle \frac{ds}{dt} &=& - \rho_1a_AsN_1-\rho_2a_2sN_2\\[8pt]
\displaystyle \frac{dN_1}{dt} &=& a_1 s N_1\\[8pt]
\displaystyle \frac{dN_2}{dt} &=& a_2 S N_2
\end{array}
\feq
où la consommation de la nourriture est explicite, puis éliminer $s$  après avoir constaté que $\frac{d}{dt}(s+\rho_1N_1+\rho_2N_2)= 0$. Ce parti pris de ne considérer que le partage de la nourriture comme possibilité de compétition a pour conséquence que dans le modèle \eqref{volterracompete} les isoclines (autres que les axes) \textbf{sont des droites parallèles} ce qui exclu la possibilité de l'équilibre de coexistence que présente le modèle \eqref{ralfa1221} qui, lui, considère l'interférence entre les espèces.

Le reste de l'ouvrage ne revient pas sur cette réduction de la compétition au partage de la nourriture et généralise l'exclusion au cas de $n$ espèces. Une seule des espèces en présence subsiste.

\paragraph{\textbf{La compétition chez Lotka}}$\;$\\
Dans  le chapitre III {\em  Fundamental equations of Kinetics (Continues) - Special case of:  Two and Three variables} de son  {\em Elements of physical biology}, Lotka précise ce qu'il entend par cas de deux variables
\dcom
The case of two variables [entendre "espèces"], $X_1$, $X_2$ which we now approach, is of interest as the simplest example exhibiting the relations between interdependant species. This relation can take a variety of forms.
\fcom  
S'il  distingue dans un premier groupe, où une espèce $S_1$  sert de nourriture  à une seconde, $S_2$ pas moins de 4 types d'interaction qui sont
\bito 
\item a) \og The organism $S_2$ kills $S_1$ (...)\fg
\item b) \og $S_2$ lives on $S_1$ without killing it outright (...)\fg$\,$ 
\item c) \og Ici $S_2$ "profite" de $S_1$ sans lui porter préjudice en se nourrissant de ses cadavres ou déchets.
\item d) La symbiose où \og (...) $S_1$ and $S_2$ live in partnership which, as a rule, is in some degree mutually beneficial.\fg
\fit
il est très bref en revanche  dans la description de la competition
\dcom
In addition to these types (1a) to (1d), another large group of case (2) are those in which two or more species compete for a common food supply.
\fcom
qu'il réduit strictement à la compétition pour une nourriture.

Dans un article postérieur de 1932 {\em The growth of mixed populations : Two species competing for a commun food supply} \cite{LOT32}\footnote{{\em  Journal of the Washington Academy of Sciences}, Vol 22, N°16 p. 461-469}, remarqué par Scudo et Ziegler dans leur compilation \cite{SCU78} il revient sur cette question à partir de l'analyse de Volterra dont il a maintenant connaissance. Il écrit les équations comme Volterra le préconise mais avec une petite réticence cependant (soulignée par moi dans la citation)
\dcom
\beq \label{Lotka}
\begin{array}{rclcl}
\displaystyle \frac{dN_1}{dt} &=& r_1N_1\big(1 - p_1(hN_1+kN_2)\big)\\[8pt]
\displaystyle \frac{dN_2}{dt} &=&r_2N_2\big(1 - p_2(hN_1+kN_2)\big)
\end{array}
\feq
a system of equation that may be regarded as an almost self-evident extension of the equation [La logistique], \underline{except that one may question why} \underline{ the same constants $h,k$ appear in the two equations}\footnote{C'est moi qui souligne.}. We shall take up this question later. For the present we shall accept Volterra's original setting.
\fcom
Ensuite il se livre à une étude du système de Volterra et montre que les méthodes qu'il avait développées dès avant 1925 fonctionnent également sur ce modèle. Il y a certainement là le dépit de ne pas voir son travail suffisamment reconnu mais, pour l'histoire du PEC  le plus intéressant sont ces quelques  lignes que je reproduit in extenso :

\dcom Let us now briefly consider the implication of Volterra's restriction that $h_1 = h_2$, and $k_1 = k_2$. The physical significance of this restriction is, essentially, that the two species consume one and the same food material, or, if they consume a mixed diet, that the proportion of each ingredient of the diet which they consume is the same for both species.

Now this is a rather narrow an unrealistic restriction. Moreover if we adopt the general method of treating the subject , it is unnecessary. The solution applies as well if $h_1 \not = h_2$ and $k_1 \not = k_2$ . Certain significant differences, however, appears in the result.
\fcom
Effectivement, maintenant les isoclines de \eqref{Lotka} ne sont plus nécessairement parallèles, elles peuvent se couper, et donc :
\dcom
Instead of three  equilibria in the finite field, there are now four, and one of these may be such that \underline{not only one species survives, but both}.
\fcom 
Il peut y avoir coexistence à l'équilibre de deux espèces. Malheureusement Lotka ne développe pas plus ce point. Il remarque simplement :
\dcom
This is  more in keeping with the facts of nature, since it is a matter  of the most commun knowledge that a graet varieties of species of organims sharing certain sources of food do live together in essentially stable equilibrium.
\fcom

Dans l'introduction de la section consacrée à la compétition de \cite{SCU78} Scudo et Ziegler écrivent
\dcom These works [Cet article et le livre de Gause \cite{GAU34}] where seminal to the controvertial concepts of "competitive exclusion", "ecological displacement", and "nich width."
\fcom

\subsection{Gause : mathématiques et expériences.}
Georgy Gause (1910-1986) est un biologiste Russe qui a été très tôt influencé par les travaux de Pearl sur la croissance des population (c.f. note \ref{pearl}). Convaincu que les études sur le terrain des interaction de populations sont trop complexes pour être analysées, il a développé l'écologie de laboratoire en interaction avec des modèles mathématiques. Plus encore que Lotka ou Volterra, on peut considérer qu'il a introduit les méthodes de la physique  : {\em  implication des mathématiques + méthode expérimentale}, en dynamique des populations. Voici le traitement théorique qu'il fait de la compétition dans son ouvrage de 1934\cite{GAU34}\footnote{On remarquera l'extrême jeunesse de l'auteur de cet ouvrage reconnu comme fondateur.},  {\em The struggle for existence}\footnote{Qu'il publie aux états-unis grâce à Pearl. Voici ce qu'il dit dans la préface : «I wish to express my sincere thanks to Professor W. W. Alpatov for interest in the experimental investigations and for valuable suggestions. To Professor Raymond Pearl I am deeply indebted for great assistance in the publication of this book, without which it could never have appeared before the American reader. I am also grateful to the Editors of The Journal of Experimental Biology and Archiv fur Protistenkunde for permission to use material previously published in these periodicals.»} , traitement suivant d'assez près ceux  de Volterra et Lotka  auquel il renvoie.
\dcom
We are now sufficiently prepared for the acquaintance with the mathematical equations of the struggle for existence, and for a critical consideration of the premises implied in them. Let us consider first of all the case of competition between two species for the possession of a common place in the microcosm. This case was considered theoretically for the first time by Vito Volterra in 1926. An experimental investigation of this case was made by Gause [référence à ses travaux], and at the same time Lotka  [il s'agit de \cite{LOT32}]  submitted it to a further analysis along theoretical lines. If there is competition between two species for a common place in a limited microcosm, we can quite naturally extend the premises implied in the logistic equation.
\fcom 

Il part de la logistique pour une seule espèce $x_1$ qu'il écrit :
 \beq \label{gause1}
\begin{array}{rcl} 
\displaystyle  \frac{dx_1}{dt}& =&\displaystyle b_1 x_1\left(\frac{K_1- x_1}{K_1}\right) \\[8pt]
  \end{array}
\feq
interprétant le terme $b_1 x_1$ comme ''potentiel biotique'' (aujourd'hui le terme $b_1$ est le ''taux de croissance intrinsèque''). Le terme $K_1$ est la "taille maximale possible de la population"   et le terme $\frac{K_1-x_1}{K_1}$ est interprété comme le ''nombre relatif de places encore vacantes'' étant entendu qu'une ''place'' n'est pas  un espace physique mais quelque chose d'assez abstrait :
\dcom
The difference between the maximally possible and the already accumulated population $(K_1-x_1)$, taken in
a relative form, i.e., divided by the maximal population $\left( \frac{K_1-x_1}{K_1}\right)$ , shows the relative number of the "still vacant places" for definite species in a given microcosm at a definite moment of time.
\fcom
plus proche du concept de {\em niche écologique} qui commence à se répandre chez les biologistes. Ensuite il introduit une nouvelle espèce $x_2$ qui à son tour va diminuer le ''nombre de places restantes'' de façon à ce que la croissance de $x_1$ soit maintenant seulement :
$$\displaystyle  \frac{dx_1}{dt} =\displaystyle b_1 x_1\left(\frac{K_1- (x_1+\alpha x_2)}{K_1}\right)$$
La même équation s'appliquant à la seconde espèce il vient le système :
 \beq \label{gause2}
\begin{array}{rcl} 
\displaystyle  \frac{dx_1}{dt}& =&\displaystyle b_1 x_1\left(\frac{K_1-( x_1+\alpha x_2)}{K_1}\right) \\[8pt]
\displaystyle  \frac{dx_2}{dt}& =&\displaystyle b_2 x_2\left(\frac{K_2-( x_2+\beta x_1)}{K_2}\right)
  \end{array}
\feq
Gause appelle les constantes $\alpha$ et $\beta$ les ''coefficients de lutte pour l'existence''. On voit que lorsque $\alpha = \beta$ il n'est pas possible d'annuler simultanément les deux parenthèses de (\ref{gause2}), donc d'avoir un équilibre où $x_1$ et $x_2$ sont strictement positifs simultanément ; donc, si deux espèces ''exploitent'' de façon identique  le milieu ($\alpha = \beta =1$), l'une élimine l'autre.

Gause insiste sur ce que l'écriture mathématique apporte à la description purement verbale comme on peut le voir dans le passage reproduit sur la figure \ref{gausephilo}.
\begin{figure}[!tb]
   \centering
  \includegraphics[width=0.8\textwidth]{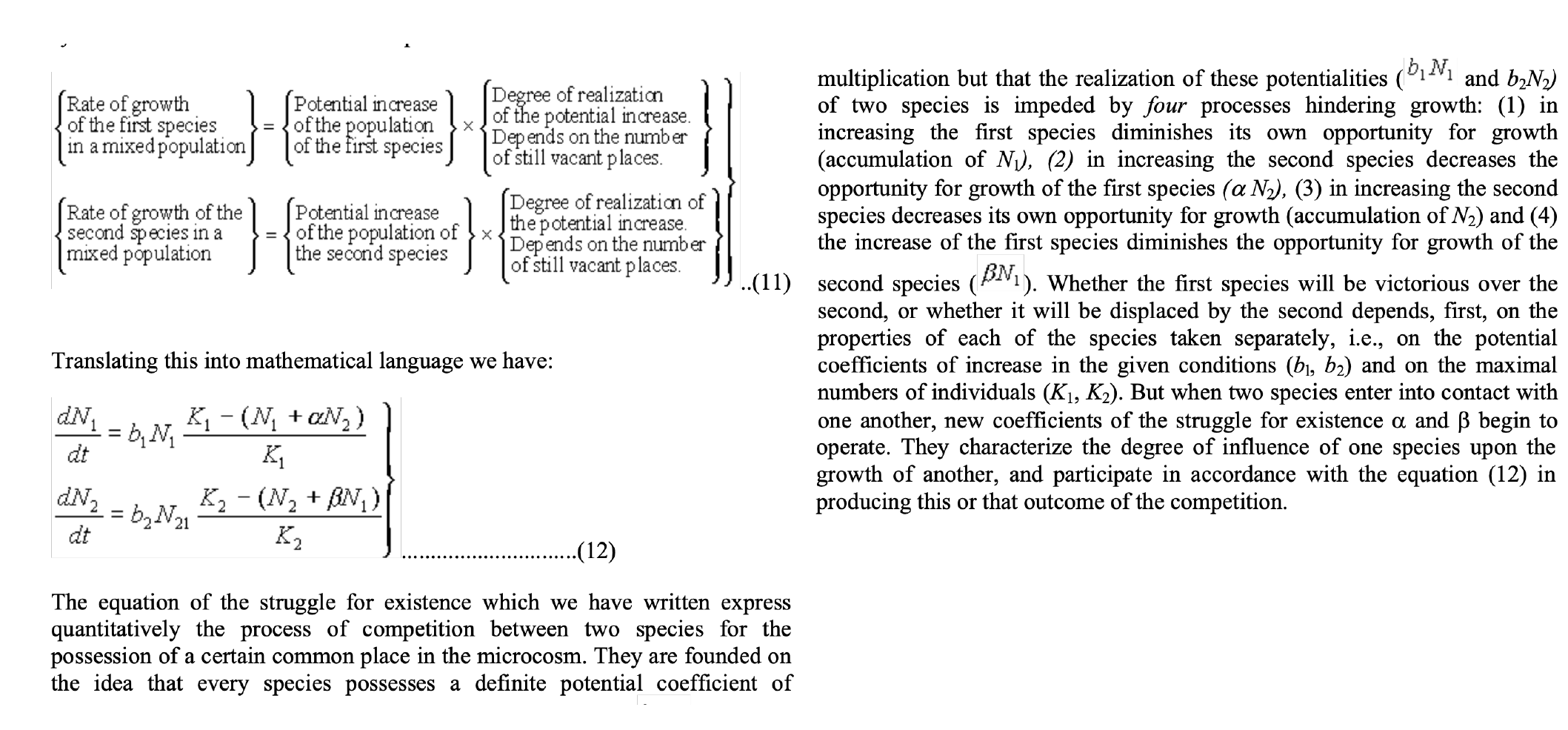} 
  \caption{Le passage du ''discours'' aux équations chez Gause.}\label{gausephilo} 
\end{figure}
Gause confronte ces équations à des expériences de laboratoire soigneusement conduites, mets en évidence quelques prédictions correctes, mais insiste aussi sur les échecs et leurs causes possibles. A ce sujet voici la conclusion de son chapitre 4 consacré à une expérience  compétition entre des levures :
\dcom
 Such theoretical calculations agree completely with the experimental data only under aerobic conditions, where the limitation of growth in both species depends almost completely on the ethyl alcohol. In the case of anaerobic conditions the situation becomes more complicated as a result of the influence of certain other waste products. This shows that extreme care is necessary in the investigation of biological systems, because various and often unexpected factors may participate in the process of interaction between two species.
\fcom
Le chapitre suivant (IV) est consacré à une expérience de compétition entre levures et le suivant (V) 
\begin{figure}[!tb]
   \centering
  \includegraphics[width=0.8\textwidth]{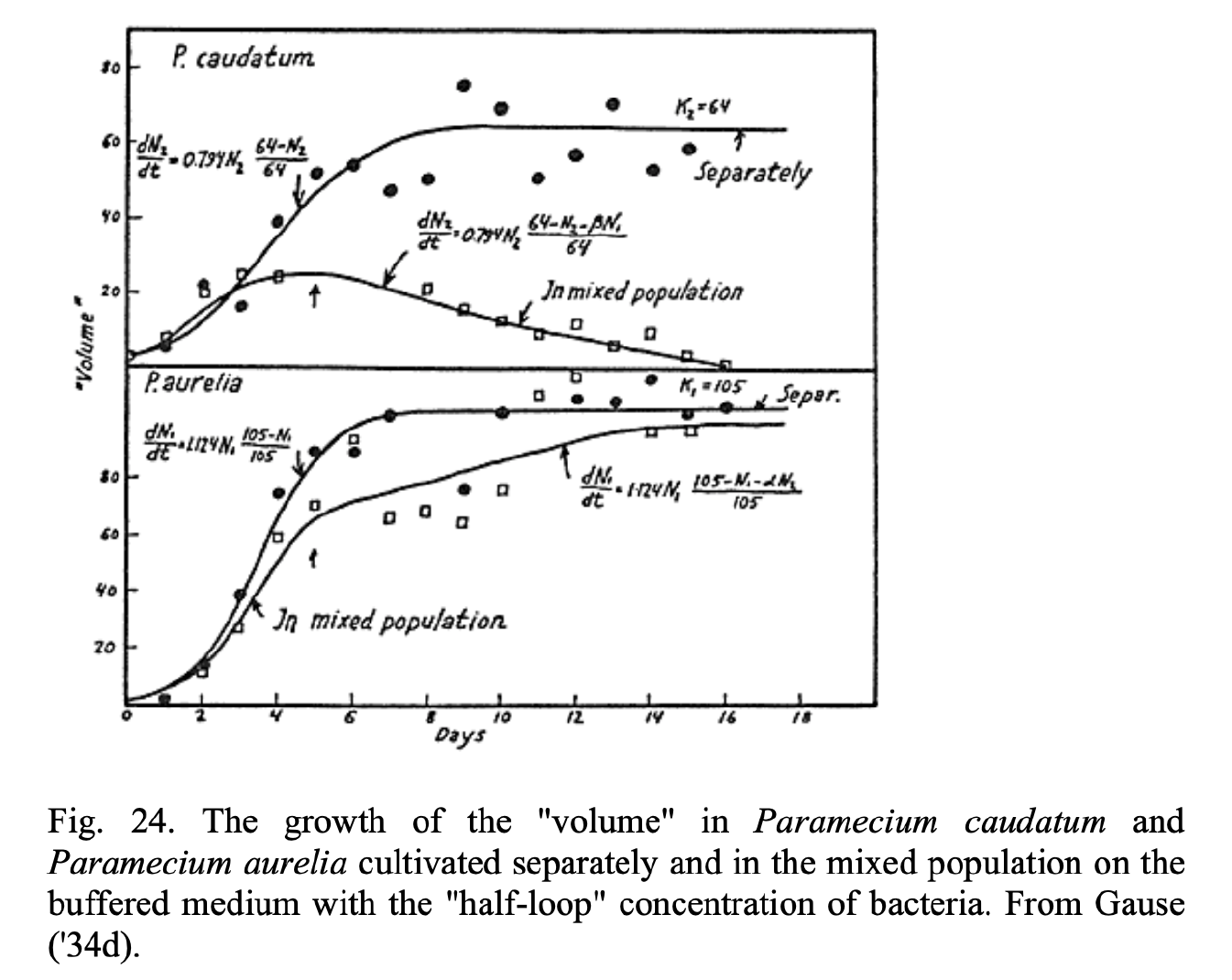} 
  \caption{Une expérience de Gause.}\label{gauseexp} 
\end{figure}
On y trouve quelques résultats  d'expériences comme celui présenté figure \ref{gauseexp} qui expliquent probablement que l'on ait appelé ''principe de Gause'' le principe d'exclusion compétitive. 

Nous n'en avons pas fini avec Gause qui reparaitra dans la section \ref{GRMA}

Pour conclure sur la séquence Lotka-Volterra-Gause je dirai que,  sauf pour Lotka qui a failli y échapper dans \cite{LOT32},  l'interprétation $r-K$ de la logistique est restée tout le temps prégnante au point d'occulter la compétition intra-spécifique qui est restée largement ignorée jusqu'à ce que s'impose la {\em ratio-dépendance} dans le modèle proie-prédateur (voir section \ref{ratiodep})

\subsection{1936 Un O.V.N.I. piloté par un extraterrestre.}
Cette courte section est consacré à un très court article de mathématiques de A. Kolmogorov.

Andreï  Kolmogorov (1903-1987) est une sorte de mathématicien extraterrestre. On lui reconnait des contributions fondamentales \footnote{On pourra lire à ce sujet l'ouvrage collectif {\em L'héritage de Kolmogorov en Mathématiques}, Belin, 2004} dans presque tous les domaines des mathématiques :
\bito
\item La théorie des nombres.
\item La théorie des Séries de Fourier.
\item La logique intuitionniste et le constructivisme,  
\item Fondement axiomatique de la théorie  des probabilités.
\item La définition, indépendamment de G. Chaikin, de la complexité algorithmique.
\item Le célèbre test statistique de Kolmogorov-Smirnov.
\item La célèbre théorie K.A.M. (Kolmogorov, Arnold, Moser) en mécanique céleste.
\item La théorie des systèmes dynamiques.
\item ...
\fit

En 1936 il lança un O.V.N.I. qui, comme il se doit pour un tel objet, fut peu remarqué. Il s'agit d'une courte note (6 pages) {\em Sulla teoria di Volterra della lotta per l’esistenza} \cite{KOL36}\footnote{Giornale d'el  Instituto Italiano degli Attuari 7, 74-80, 1936. Comme on le voit, Kolmogorov n'avait pas un gros souci de communication.} qui comme son titre l'indique reprend les équations de Volterra de la relation proie-prédateur qu'il écrit :

 \beq \label{Kolmo1}
\begin{array}{rcl} 
\displaystyle  \frac{dN_1}{dt}& =&\displaystyle \left(\varepsilon_1- \gamma_1 N_2\right) N_1 \\[8pt]
\displaystyle  \frac{dN_2}{dt}& =&\displaystyle \left(\varepsilon_2 +  \gamma_2 N_1\right) N_2 
\end{array}
\feq
et qu'il remplace immédiatement par :
 \beq \label{Kolmo2}
\begin{array}{rcl} 
\displaystyle  \frac{dN_1}{dt}& =&K_1(N_1, N_2) N_1 \\[8pt]
\displaystyle  \frac{dN_2}{dt}& =&K_2(N_1, N_2) N_2
\end{array}
\feq
mettant en évidence les taux de croissance des espèces 1 et 2 comme des fonctions de $N_1,N_2)$ qui {\em ne sont pas précisées}   et sur lesquelles on se contente de faire des hypothèses {\em purement qualitatives} telles que :
\ben 
\item $\frac{dK_1}{dN_1} < 0$ : Pour une quantité $N_1$ donnée, le taux de croissance de $N_1$ diminue quand $N_2$ augmente.
\item $K_1(0,0) = 0$ : Lorsque les proies et les prédateurs sont en nombre suffisamment faible le taux de croissance des proies est positif.
\item $\frac{dK_2}{dN_2} < 0$ : Le taux de croissance de l'espèce 2 est une fonction décroissante de $N_2$ 
\item etc...
\fen
Il y a en tout 9 hypothèses de ce type qu'il traduit par la figure \ref{figKolm1} dans l' espace des phases :
\begin{figure}[!tb]
   \centering
  \includegraphics[width=0.8\textwidth]{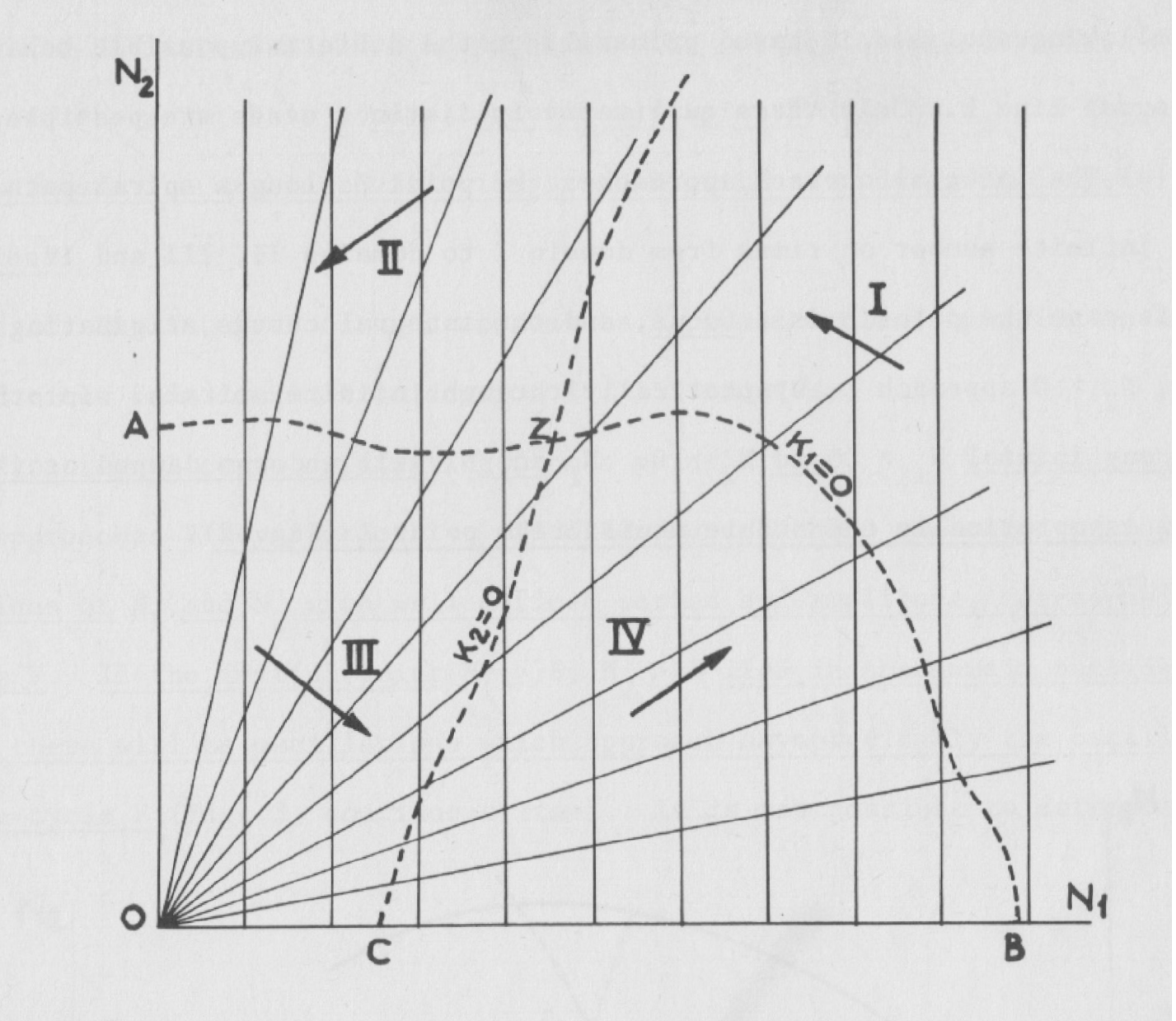} 
  \caption{Méthode d'analyse de Kolmogorov.}\label{figKolm1} 
\end{figure}
On remarque sur cette figure l'importance des droites passant par l'origine, c'est à dire de l'ensemble des points où le {\em ratio} $N_1/N_2 = S$ est constant. Une des hypothèses porte sur les variations des $K_2$ relativement à $S$ :  $\frac{dK_2}{dS} > 0$ qu'il interprète : «Pour tout {\em ratio} du nombre de proies par prédateur, l'augmentation des proies et des prédateurs favorise l'espèce prédatrice». C'est la première fois qu'un modèle mathématique met en évidence l'importance du {\em ratio}  entre proies et prédateurs pour expliquer la dynamique de l'interaction entre deux population.
\begin{figure}[!tb]
   \centering
  \includegraphics[width=0.8\textwidth]{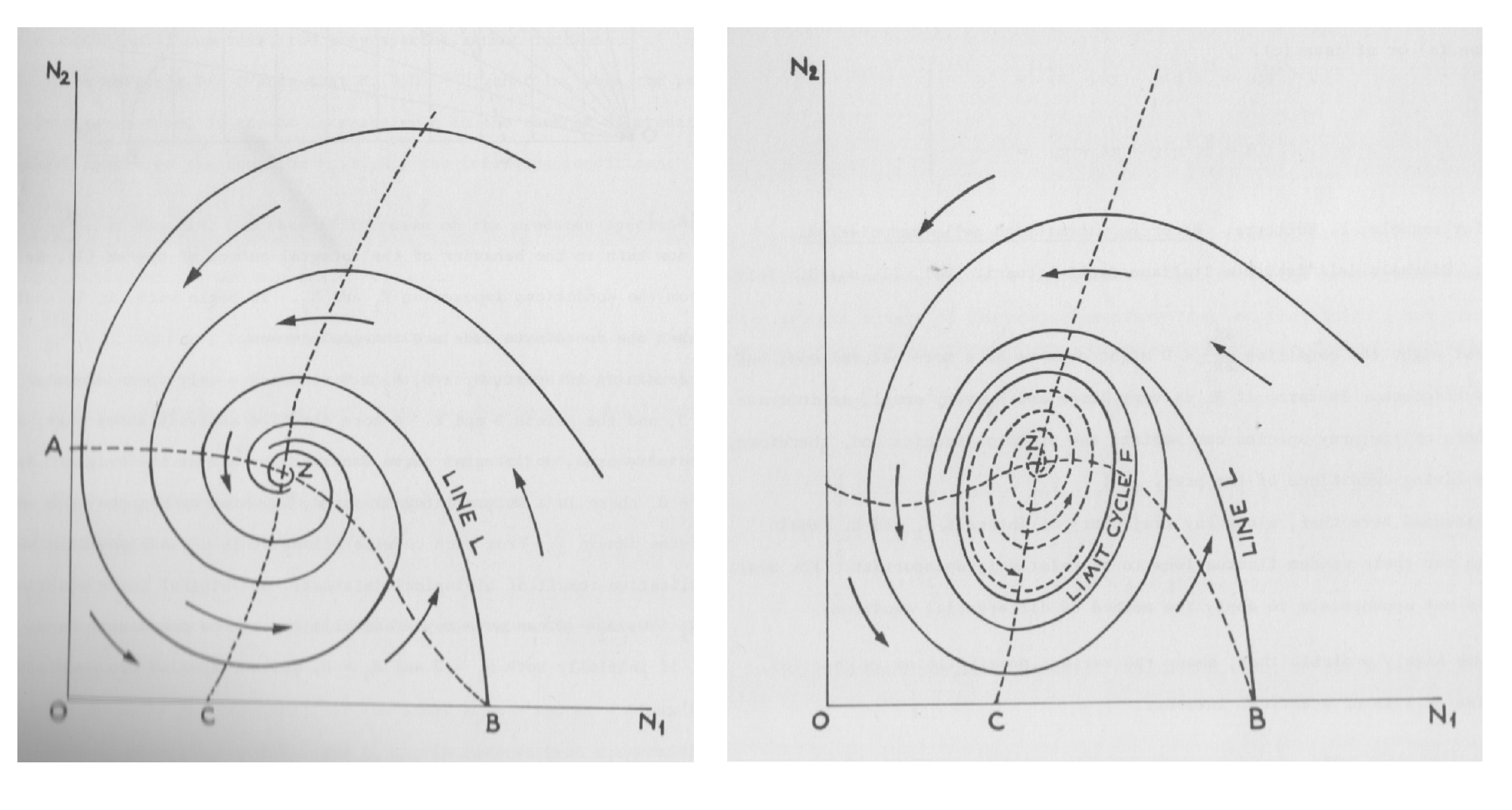} 
  \caption{Situation générale sous les hypothèses qualitatives du modèle (\ref{Kolmo2}).}\label{figKolm2} 
\end{figure}
De ses hypothèses il déduit qu'essentiellement deux cas peuvent se produire : un équilibre (figure \ref{figKolm2} gauche), ou un cycle limite (figure \ref{figKolm2} droite) ou, enfin, mais comme {\em cas limite} le cas du centre (famille de solutions périodiques) de Lotka-Volterra.

On a là, en germe, plusieurs idées de la {\em théorie qualitative} naissante des systèmes dynamiques. 
\subsection{Derrière l'O.V.N.I. des mathématiques en devenir.}
L'objet de cet article n'est pas d'écrire une histoire de la théorie mathématique des systèmes dynamiques mais d'élucider la façon dont elle est  intervenue dans la clarification d'une question d'écologie. Mais pour ce faire il faut évidemment prendre en considération d'état de la théorie mathématique en question  au moment de son intervention. C'est pourquoi je me permets ici un petit rappel sur la naissance du concept de ''stabilité structurelle" essentiel en dynamique des populations moderne. 
On trouvera dans l'article {\em Writing the History of Dynamical Systems and Chaos: Longue Durée and Revolution, Disciplines and Cultures} de D.David Aubin et Amy Dahan Dalmedico \cite{AUDAL}, sur lequel je vais m'appuyer en partie, une description très bien documentée sur cette question. 

A la fin du XIXème siècle, le grand ancêtre  de la théorie des systèmes dynamiques, H. Poincaré, a en tête la mécanique céleste. Les équations de la mécanique ne sont pas des "modèles" mais des "lois" de la nature. J'entends par là que, dans le domaine qui est le leur, celui  des vitesses faibles par rapport à celle de la lumière, les lois de Newton exprimées à travers l'équation différentielle 
$$\frac{d\overrightarrow{x}}{dt} = m \overrightarrow{a}$$
ont une force prédictive  qui défie toute expérimentation. Cela conduit à une certaine façon d'envisager les mathématiques des équations différentielles. Les équations sont l'expression mathématique d'une loi qu'il n'est pas question de "bricoler". Les équation du problème des trois corps sont ce qu'elles sont et c'est leur mystère qu'il faut élucider. Mais, au tournant du XXème siècle nombre d'autres usages des équations différentielles voient le jour.  Aubin et  Dahan Dalmedico notent \cite{AUDAL} p. 286
\dcom
Contrary to celestial mechanics (where Newton’s law was supposed to be exactly true), in considering “real physical systems [which] we are always forced to simplify and idealize” [Andronov et al. 1966, xv ]\footnote{\label{AVC}
Il s'agit du célèbre traité {\em Theory of oscillattors} de Andronov A.A., Vitt, A.A. et Khaikin S. E. de 1937 traduit et re-édité.}. In a lengthy introduction they explained that : It is evident that since small random perturbations are inevitable in all physical systems, processes which are possible only in the absence of any random deviations or perturbations whatsoever cannot actually occur in them.
\fcom 
A partir du moment où le système d'équations différentielles n'est plus une  représentation (quasi) exacte du phénomène mais un "modèle" qui en possède (on l'espère) un certain nombre de propriétés, il convient que ce modèle soit "robuste". Quel sens cela  aurait-il de pratiquer des mathématiques sophistiquées pour décrire le comportement du système différentiel $\frac{dx}{dt} = f(x)$ si une petite perturbation de ce système $\frac{dx}{dt} = f(x)+\eps h(x)$ a un comportement totalement différent. On reconnait le concept de "stabilité structurelle" qui est (aux détails techniques près).
\ben
\item Deux systèmes définis sur $\Rmat^n$,  $\frac{dx}{dt} = f(x)$ et $\frac{dx}{dt} = g(x)$ sont équivalents si et seulement si il existe un homéomorphisme de
$\Rmat^n$ qui échange leurs trajectoires.
\item Le système $\frac{dx}{dt} = f(x)$ est {\em structurellement stable} si pour toute perturbation $\frac{dx}{dt} = f(x)+\eps h(x)$, pour $\eps$ suffisamment petit, le système perturbé est équivalent au système de départ.
\fen
La condition 1) définit mathématiquement un comportement "qualitativement" identique, la condition 2) la robustesse. En fait cette terminologie moderne ne s'est pas imposée immédiatement. Toujours dans\cite{AUDAL}  on lit peu après la citation précédente :
\dcom
From such concerns emerged the notion of “coarse systems.” First introduced in the literature as “systèmes grossiers” by Andronov and Pontryagin [\cite{ANPO} 1937], this term has been diversely translated in English, as “coarse” or “rough systems.” In his 1949 translation [de \cite{ANPO}, Lefschetz called these systems “structurally stable systems.” As Arnol’d [\cite{ARN}1994] has emphasized, this notion appeared in Andronov’s work as both a mathematically rigorous definition and a general idea about the type of systems useful for mathematical modeling in physics and engineering.
\fcom 
Ces questions abordées dès les années 1930 par l'école Russe  ne seront considérées que plus tardivement en occident\footnote{
C'est un raccourci trop rapide, il faudrait aussi  parler des travaux des années trente de Van der Pol, Levinson, et autres... Voir \cite{AUDAL} pour un compte rendu plus complet.} sous la houlette énergique de S. Lefchetz pour aboutir au tournant des années 1960 aux célèbres travaux de Smale et de son école.\footnote{Synthétisés par Smale en 1967 dans \cite{SMA67}. On trouvera également dans un essai de M. Hirsch  \cite{HIR84}, un autre très grand contributeur de la théorie des systèmes dynamiques, des informations sur des applications des systèmes dynamique à la dynamique des populations.}
A partir de cette date la théorie qualitative des systèmes dynamiques devient une branche de plus en active de la recherche mathématique et ses concepts et résultats de base s'installent dans la culture générale de tous les mathématiciens. 

Avant de clore cette section une remarque en deux points :
\bittiret 
\item Le travail sur la dynamique des populations de  1936 de Kolmogorov a échappé aux investigations de Aubin et Dalmedico \cite{AUDAL}. A-t-il eu une influence sur les idées d'Andronov et Pontryagin ?
\item Le travail de pionnier de Gause et ses collaborateurs {\em Further studies of interaction between predators and prey} \cite{GAU36} de 1936 est co-signé par A.A. Witt. De son côté le {\em Theory of oscillators} est co-signé par A.A. Vitt
avec un ''V'' mais il s'agit bien de la même personne
\footnote{Witt a été victime de purges staliniennes, son nom a disparu un temps du  {\em Theory of oscillators} puis y est revenu. On peut consulter sa fiche Wikipedia \url{https://de.wikipedia.org/wiki/Alexander_Adolfowitsch_Witt} dont je donne une traduction en annexe \ref{witt}}. Son intérêt pour la dynamique des populations a-t-il influencé Andronov et Pontryagin  dans lzur réflexion sur la stabilité structurelle (les systèmes grossiers) ?
\fit
Ces deux questions sur l'éventuelle influence de questions écologiques sur le développement des mathématiques ne rentrent pas dans le cadre de mon étude mais elles méritent d'être posées.

\textit{ Que retenir de l'âge d'or ?
Une première lecture des travaux des années 1920-1940 donne l'impression que la collaboration de mathématiciens (Lotka, Volterra) et de biologistes (Pearl, Gause)  fournit des bases solides à la formulation de Grinnell du PEC. 
 Il s'agit  des premiers travaux  
d'une branche  en train de se construire de l'Ecologie Théorique : La dynamique des populations.}

\section{ 1940-1970 : Un peu plus de biologie}
{\em Entre le livre de Gause (1936) et l'article considéré à son tour comme fondateur de Rosenzweig et MacAthur (1963) il s'écoule plus d'un quart de siècle où l'on ne note pas grand chose dans le domaine de la mathématisation de l'écologie, sauf, peut être, par le biais de la microbiologie et du développement des cultures en continu.  }
\subsection{L'apport de la microbiologie : Monod-Novick et Szilard} \label{th-chemostat}
\begin{figure}[!tb]
   \centering
  \includegraphics[width=0.95\textwidth]{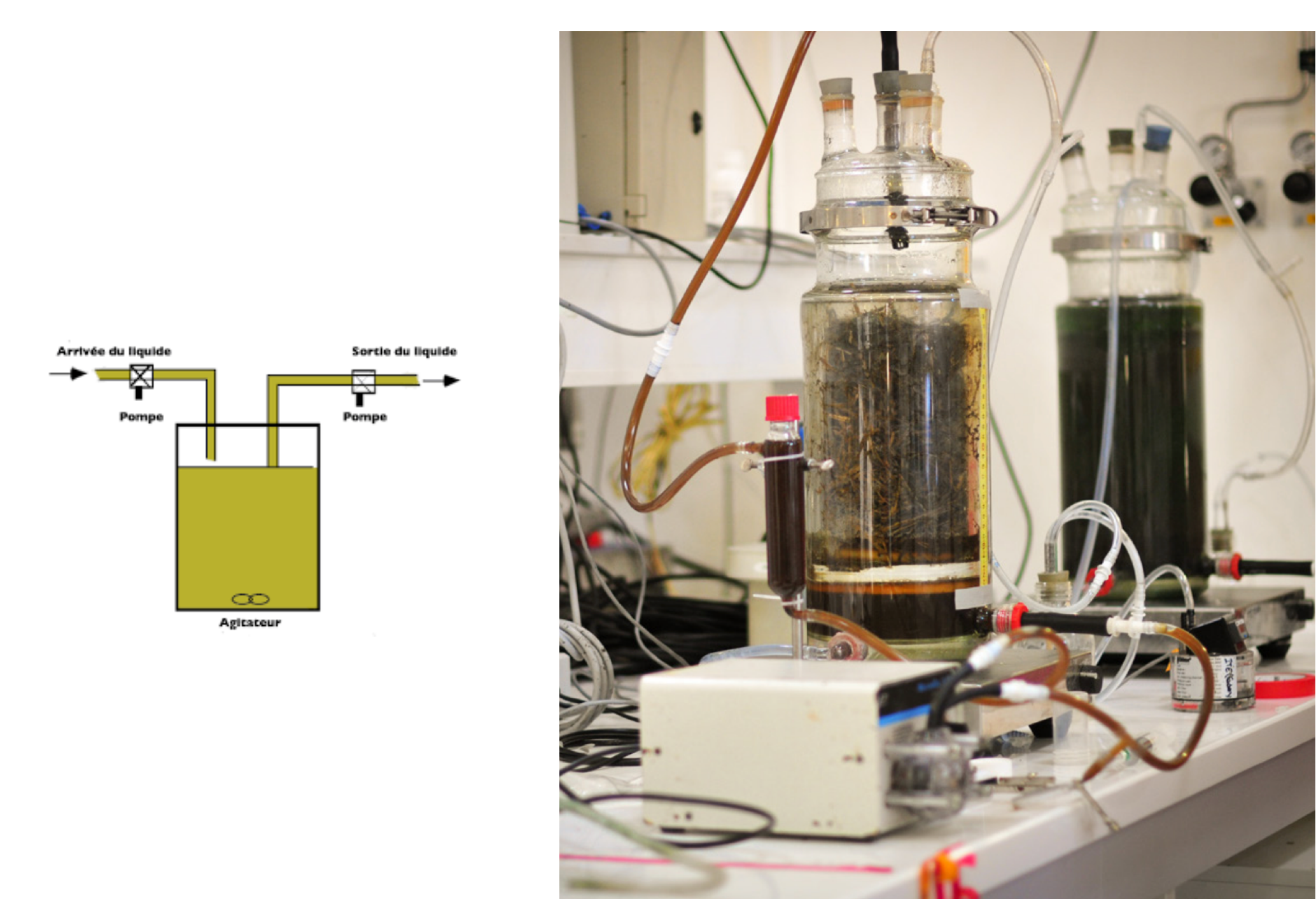} 
  \caption{Schéma de principe et chémostat de laboratoire.}\label{figchem} 
\end{figure}
Le chémostat est un dispositif de laboratoire destiné à cultiver des microorganismes. On attribue son invention à J. Monod \cite{MON50}\footnote{
J. Monod (1950), "La technique de culture continue; théorie et applications," Annales de L'Institut Pasteur 79: 390-401 (1950)}(en France) et indépendamment à A. Novick and L. Szilard \cite{NOV50}\footnote{A. Novick and L. Szilard , "Experiments with the chemostat on spontaneous mutations of bacteria," Proceedings of the National Academy of Science 36: 708-19 (1950).} en 1950. Il s'agit d'un réacteur alimenté en {\em substrat} nutritif de façon continue et destiné à la culture de microorganismes unicellulaires (bactéries, levures, cellules planctoniques) dont la théorie est clairement détaillée par les inventeurs. Les équations (avec des notations modernes) en sont les suivantes :\\\\
 \fbox{
 \begin{minipage}[c]{0.97\textwidth}
 \begin{center}
 \textbf{Les équations du chémostat}
 \end{center} 
\beq \label{chem}
\begin{array}{rcl} 
\displaystyle \frac{ds}{dt}& =&\displaystyle d(S_{in}-s) - \frac{\mu(s)}{Y}x\\[8pt]
\displaystyle \frac{dx}{dt} &=&\displaystyle ( \mu(s) - d)x
  \end{array}
\feq
où  :
\bittiret 
\item $d$ est le débit du flux qui traverse le réacteur (voir figure \ref{figchem})
\item $S_{in} $ est la concentration du substrat en amont du réacteur, $s$ la concentration dans le réacteur.
\item $x$ est la concentration de la biomasse dans le réacteur.
\item $\frac{\mu(s) }{Y} y $ est la quantité de substrat consommé par unité de temps transformée en biomasse $ \mu(s)x$ ; le coefficient $Y$ est un terme de rendement (plus petit que 1).
\fit
\end{minipage}}\\\\

Le terme :
$$\frac{\mu(s)}{Y}x$$
qui représente la quantité de substrat consommé par unité de temps et la quantité $\mu(s)$ qui est le taux de croissance de la biomasse, méritent un peu d'attention.
Supposons que $\mu (s) = \alpha s$ soit linéaire. Alors on retrouve le terme $ \frac{\alpha}{Y} s\times y$ de la loi d'action de masse utilisé par Lotka et Volterra. Mais ici la fonction $s \mapsto \mu$  n'est pas linéaire. 
A l'équilibre on a :
$$ \mu(s_e) = d$$
Le taux de disparition de la biomasse entrainée par le flux est $d$ est exactement compensé par le taux de croissance $\mu(s)$. Comme la concentration  $s_e$ de substrat à l'équilibre  peut être mesurée nous avons la possibilité, en variant le débit $d$, de {\em mesurer} $\mu(s)$ en fonction de $s$. Monod nous dit :
\dcom
On sait que le taux de croissance varie avec la concentration de l'aliment carboné suivant une loi assez bien exprimée par la relation hyperbolique\footnote{
Cette partie de la fonction homographique qui représente une croissance monotone et bornée est ce que la tradition a décidé d'appeler {\em modèle de Monod} dans les milieux  de la microbiologie, {\em de Michaelis-Menten} chez les chimistes et que la tradition de l'écologie appelle modèle {\em de Holling} de type II.}:
$$\mu = \mu_0\frac{s}{S_K+s} $$
\fcom 
Par rapport à la loi d'action de masse $-$ proportionnalité au produit des concentrations $s \times x$ $-$ le modèle de consommation d'un substrat par un microorganisme introduit une dissymétrie essentielle qui traduit un fait biologique observable : 
{\em Le taux de consommation du substrat $\frac{\mu(s)}{Y}$ } est borné. Aussi grande que soit la concentration de substrat il arrive un moment ou le microorganisme ne peut pas en absorber plus par unité de temps.
 En revanche, pour les faibles concentrations en substrat on peut prendre l'approximation linéaire et l'on  retrouve la loi d'action de masse.
Les théories que font du chémostat,  Monod d'une part, Novick et Szilard de l'autre, sont des théories parfaitement satisfaisantes du point de vue d'un physicien et elles sont précisées mathématiquement en 1955 par C.C. Spicer , {\em  The Theory of Bacterial Constant Growth Apparatus}, dans la revue Biometrix \cite{SPI55}.
Ce n'est que plus tard, après 1970,  (voir section \ref{sectioncompeteRMA} ) que les mathématiciens s'empareront de ces équations pour les décortiquer, en particulier du point de vue de la question de l'exclusion compétitive.
En attendant, il semble que le premier article de microbiologie qui s'empare de la question de la compétition soit celui de Powell {\em Criteria for the growth of contaminants and mutants in continuous culture} \cite{POW58}\footnote{Journal of General Microbiology 18: 259-68. 5 (1958)} de 1958 où la théorie de la compétition dans un chémostat est clairement exposée.  Voici  ce qu'il dit en substance, exposé dans une langue un peu plus moderne.

Si l'on admet que (\ref{chem}) décrit correctement la dynamique de croissance d'une espèce il est naturel de proposer pour deux espèces en compétition pour le même substrat le modèle :
\beq \label{compchem}
\begin{array}{lcl} 
\displaystyle \frac{ds}{dt}& =&\displaystyle d(S_{in}-s) - \frac{\mu_1(s) x_1}{Y_1} -  \frac{\mu_2(s) x_2}{Y_2}\\[8pt]
\displaystyle \frac{dx_1}{dt} &=&\displaystyle ( \mu_1(s) - d)x_1\\[8pt]
\displaystyle \frac{dx_2}{dt} &=&\displaystyle ( \mu_2(s) - d)x_2
  \end{array}
\feq
Le dessin de la figure \ref{2mu}  
\begin{figure}[!tb]
   \centering
  \includegraphics[width=0.8\textwidth]{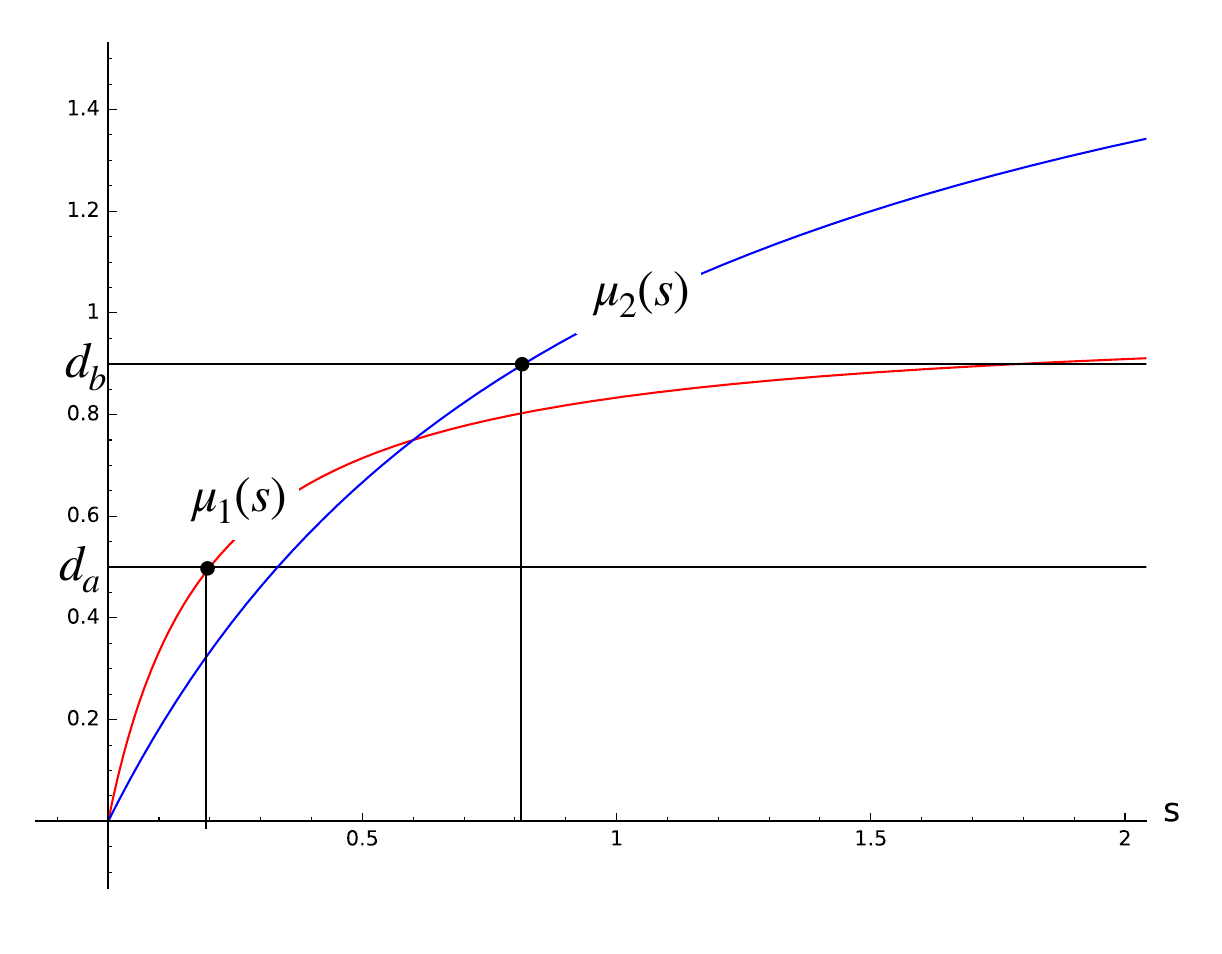} 
  \caption{Le comment de l'exclusion.}\label{2mu} 
\end{figure}
explique clairement  pourquoi les deux espèces ne peuvent coexister à l'équilibre. En effet, à l'équilibre, par définition, on doit avoir simultanément toutes les vitesses nulles, donc, les concentrations $s_e,x_{1e}, x_{2e}$ qui réalisent un équilibre doivent satisfaire les trois équations :

\beq \label{compchemeq}
\begin{array}{rcl} 
0& =&\displaystyle d(S_{in}-s_e) - \frac{\mu_1(s_e) x_{1e}}{Y_1} -  \frac{\mu_2(s_e) x_{2e}}{Y_2}\\[8pt]
0 &=&\displaystyle ( \mu_1(s_e) - d)x_{1e}\\[8pt]
0 &=&\displaystyle ( \mu_2(s_e) - d)x_{2e}
  \end{array}
\feq
Concentrons nous sur les deux dernières équations.  Sur la figure \ref{2mu} j'ai représenté deux graphes possibles de $\mu_1(s) $ et $\mu_2(s)$ ; ils se croisent mais l'argument serait identique s'ils ne se croisaient pas. Nous voyons qu'il y a une seule valeur du débit $d$ pour lequel il existe une valeur de $s$ pour laquelle on a simultanément  $\mu_1(s) = \mu_2(s) = d$ pour les autres valeurs de $d$ il y n'est pas possible d'avoir simultanément $\mu_1(s) = d$ et $\mu_2(s) = d$ ; il faut choisir. Une fois ce choix fait, pour annuler l'autre équation, il faut annuler la valeur de $x_i$. Pour fixer les idées, soit $d_a = 0.5$ comme sur la figure ; nous avons pour $s = 0.2$, $\mu_1(0.2) = 0.5$ et $\mu_2(0.2) = 0.333... < 0.5$. Donc, pour $s_e = 0.2$ l'espèce n° 2 ne peut se maintenir puisque son taux de croissance est inférieur au taux de disparition. Pour les mêmes raisons, pour $d_b = 0.9$ c'est l'espèce $1$ qui disparait et l'espèce 2 qui se maintient pour une concentration $s_e = 0.81...$

C'est en substance, sinon avec le langage mathématique moderne,  ce que Powell écrit dans \cite{POW58}.

Vingt ans plus tard ce thème sera repris, j'y reviendrai dans la section \ref{tempsmathématiciens}.

\subsection{Holling et la {\em réponse fonctionnelle}.} 
Quittons la microbiologie pour des entités visibles à l'\oe il nu : des larves et des petits mammifères. 
Dans un article fondateur  de 1959, {\em Some Characteristics of Simple Types of Predation and Parasitism},\cite{HOL59b}\footnote{The Canadian Entomologist, Vol 91(7),  1959, p.385-398} l'entomologiste C.S. Holling décrit l'expérience de laboratoire suivante :

\texttt{Des  petits disques de papier de verre (de quatre cm de diamètre), les {\em proies}, sont disposés sur une table de un mètre carré ; \\un expérimentateur (le {\em prédateur}) aux yeux bandés}\\
 \begin{minipage}[c]{0.47\textwidth}
\texttt{
 cherche à l'aveugle, pendant une durée donnée, les disques en frappant la table avec le doigt ; chaque disque détecté est retiré de la table. L'expérience consiste à mesurer le nombre de disques retirés (que Holling appelle la {\em réponse fonctionnelle})  en fonction de la densité initiale de disques.}
\end{minipage}
 \begin{minipage}[c]{0.57\textwidth}

  \includegraphics[width=1\textwidth]{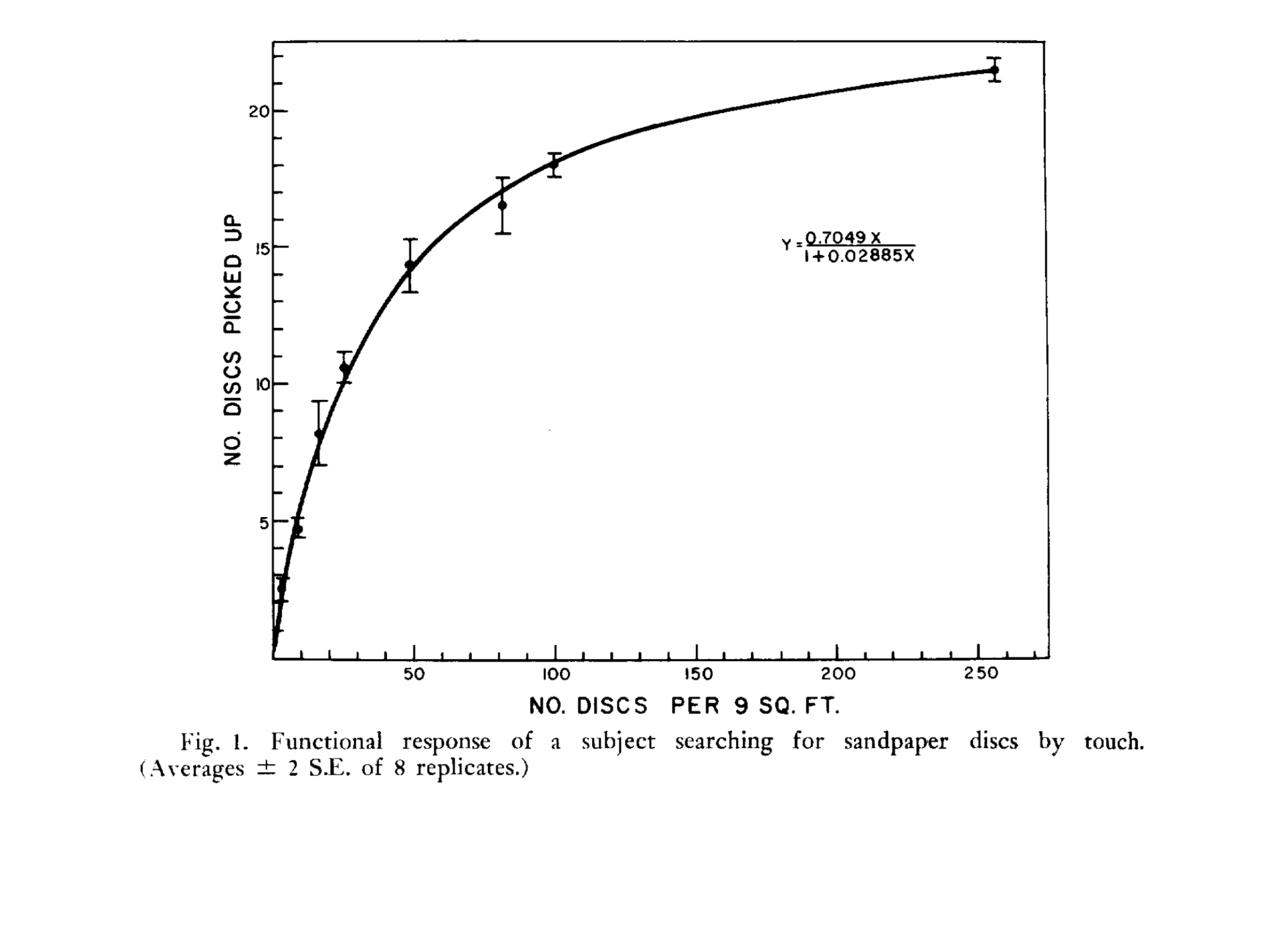} 

\end{minipage}
Le graphe de la figure est celui de la fonction $x \to \frac{0.7049\,}{1 + 0.02885\,x}$. Cette expérience, à première vue puerile, n'est pas gratuite. Elle se veut une idéalisation d'une situation que Holling connait bien, celle de "(...) small mammals preying upon sawfly cocoons" qu'il a largement étudié et décrite dans un article précédent 
{\em The Components of Predation as Revealed by a Study of Small Mammal Predation of the European Pine Sawfly} \cite{HOL59a}\footnote{The Canadian Entomologist, 1959, Vol 91, p. 293-320. }. Voir la figure \ref{Holling2} extraite de  \cite{HOL59a}, qui met bien en évidence le phénomène de saturation, cette fois en situation naturelle.\\
\begin{figure}[!tb]
   \centering
  \includegraphics[width=0.8\textwidth]{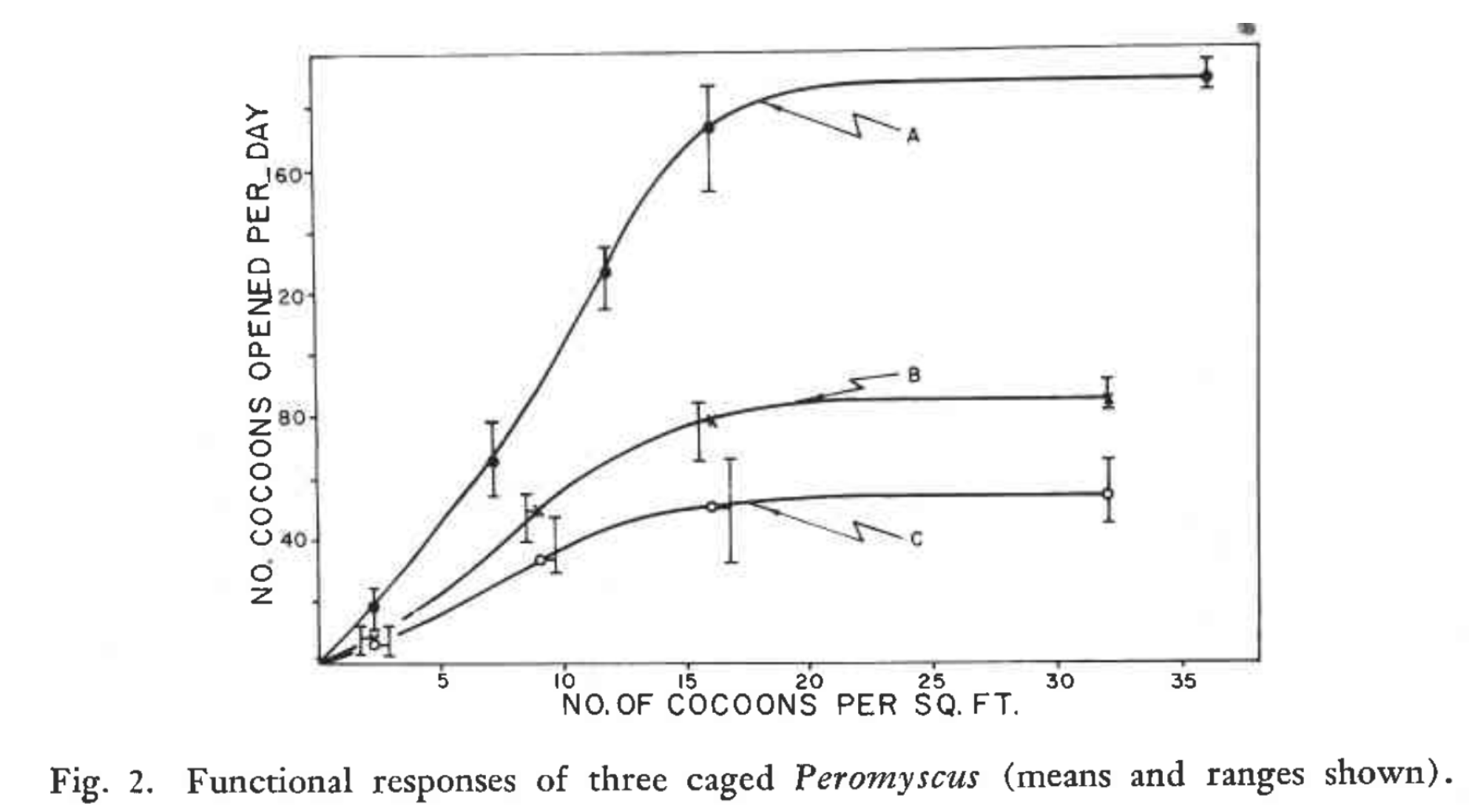} 
  \caption{Mise en évidence de la saturation par Holling.}\label{Holling2} 
\end{figure}

Holling théorise cette saturation de la manière suivante :
\bito 
\item
On admet que pendant une durée $T$ le nombre $y$  de disques détectés est proportionnel à la densité $x$ et la durée soit :
$$ y = aTx$$
où $a$ mesure l'efficacité de la recherche. 
\item Mais pendant l'expérience, toute la durée n'est pas consacrée à la recherche de proies. Une fois une proie trouvée il faut un certain temps pour la "traiter"  (la retirer du pool de disques dans le laboratoire ou la "dépecer" dans la vraie vie), qui ne sera pas consacré à la recherche. Le temps effectif de recherche est donc le temps total $T$ moins   celui, $b\,y $, nécessaire au traitement des $y$ proies. 
\fit
Donc
$$ y =  a (T- by) x$$
et en résolvant par rapport à $y$, 
\beq \label{foncHolling}
 y = \frac{Tax}{1+ab x}
\feq 
Ainsi Holling retrouve la fonction de Monod dont il ignorait l'existence mais surtout donne une interprétation en terme de dynamique des populations d'organismes non microscopiques. Dans leur livre de 2012 {\em How species interact: altering the standard view on trophic ecology}\cite{ARD12}\footnote{Oxford University Press, 2012.}Arditi et Ginzburg la décrivent ainsi
\dcom
As mentioned earlier, Holling’s model is, in fact, identical to the Monod and the Michaelis-Menten models. With a predator-prey interpretation, it can be written as
\beq
g(N)= \frac{aN}{ 1+ahN}
\feq 
The handling time, $h$, is the time spent by the predator handling each prey encountered, and during which it stops searching. The parameter $a$, called searching efficiency (also called attack rate by some authors), can be interpreted as the proportion of prey killed per predator per unit of searching time.
\fcom 
Ces deux articles de Holling ont eu une grande influence (cités plus de 4000 et 5000 fois) et en dynamique des populations il est plus courant de faire référence à Holling qu'à Monod pour désigner une fonction homographique.

\subsection{Le modèle de prédation : (Gause)-Rosenzweig-MacArthur.}\label{GRMA}

A propos du modèle proie-prédateur de Lotka-Volterra deux critiques peuvent être immédiatement formulées.
\ben
\item En l'absence de prédateur la croissance de la proie est exponentielle, illimitée.
\item Aucun phénomène de satiété vient freiner l'appétit des prédateurs : Si pour une quantité $x$ de proies les prédateurs consomment $bxy$ pour $100$ fois plus de proies ils consommeront  $ 100 \,b\, xy$.     
\fen
Le modèle, dit de Rosenzweig-MacArthur, répond à ces deux critiques de la façon suivante :
\ben
\item La croissance de la proie est du type logistique : $\frac{dx}{dt} = rx(1-\frac{x}{K})$
\item La prédation est modélisée par :
$$ - \left(\frac{\mu_{\max}x}{e+x}\right)y$$
\fen
ce qui donne :\\\\
 \fbox{
 \begin{minipage}[c]{0.95\textwidth}
 \textbf{Modèle de Rosenzweig-MacArthur}
 \beq \label{RMA}
\begin{array}{rcl} 
\displaystyle  \frac{dx}{dt}& =&\displaystyle rx\left(1-\frac{x}{K} \right)- \left(\frac{\mu_{\max}x}{e+x}\right)y \\[8pt]
\displaystyle  \frac{dy}{dt}& =&\displaystyle c \left(\frac{\mu_{\max}x}{e+x}\right)y - my  \end{array}
\feq
et sous une forme un plus moderne, où l'on sépare la structure du modèle, exposée dans (\ref{RMAbis}), de l'expression des fonctions $f$ et $\mu$ qui est plus contingente :
\beq \label{RMAbis}
\begin{array}{rcl} 
\displaystyle  \frac{dx}{dt}& =&f(x) - \mu(x) y \\[8pt]
\displaystyle  \frac{dy}{dt}& =& (c\mu(x) - m)y
  \end{array}
\feq
où $x \mapsto f(x)$ est nulle en $0$, croissante puis décroissante et nulle pour $x =K$ et $x \mapsto(\mu(x)$ est nulle en $0$ , croissante et bornée.
\end{minipage}
}\\

On remarquera que si l'on prend pour fonction $f$ la fonction linéaire : $m(X_{in} -x)$ à la place de la logistique on retrouve exactement le modèle du chémostat. Le modèle du chémostat décrit une croissance en milieu ouvert, celui de Rosenzweig-MacArthur une croissance en milieu fermé.

L'article de Rosenzweig-MacArthur {\em Graphical representation and stability conditions of predator-prey interactions} \cite{ROS63}\footnote{The American Naturalist, XCVII, n° 895 (1963)} s'appuie sur l'idée suivante. Si $x(t)$ et $y(t)$ sont respectivement les quantités de proies et de prédateurs, dans certaines conditions leurs vitesses de croissance sont des fonctions de $x$ et de $y$ seulement. Il existe un ensemble de couples $(x,y)$ pour lesquels  $\frac{dx}{dt} =0$, ensemble qu'on appellera {\em isocline des proies}, dont il est admis que ce sera une courbe, et de même une {\em isocline des prédateurs},  $\frac{dy}{dt} =0$. Ensuite, de considérations générales on tentera de déduire la forme des isoclines :
\dcom
We shall start with mental construction of the prey isocline to determine its general shape.
\fcom
De considérations assez subtiles à suivre ils déduisent que l'isocline des proies a la forme de la figure \ref{RMA1}; pour l'isocline des prédateurs ils argumentent simplement que c'est une droite verticale :
\begin{figure}[!tb]
   \centering
  \includegraphics[width=0.8\textwidth]{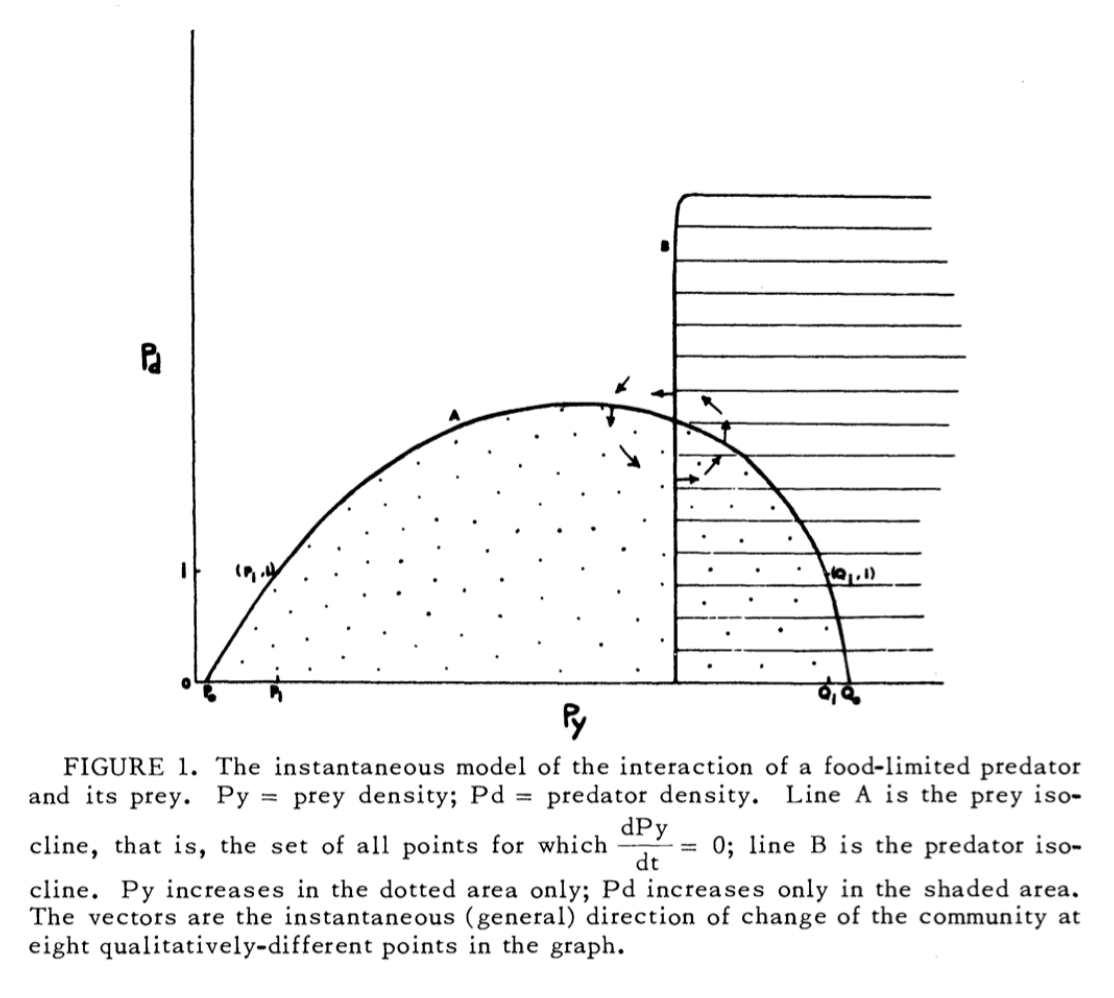} 
  \caption{L'isocline des proies d'après Rosenzweig-MacArthur}\label{RMA1} 
\end{figure}
\dcom
The predator isocline, $\frac{dy}{dt} = 0$, is simpler to deduce. The predator is depending on the prey for its ability to increase. When the prey fall below a certain level, the predator decrease ; when they are greater than this level, the predator increase.
\fcom
Nous verrons comment, bien plus tard (1989) Arditi-Ginzburg contesteront, avec pertinence, cet argument. Pour le moment contentons nous de constater que, en dépit de ce que la tradition a décidé, nulle part dans l'article ne figurent des équations différentielles qui pourraient faire penser de près ou de loin aux équations (\ref{RMA}).\\

En revanche, on peut les reconnaitre 28 ans plus tot chez Gause !\\

 En effet ans l'article {\em Further studies of interaction between predators and prey} \cite{GAU36}\footnote{ G. F. Gause, N. P. Smaragdovaand  et A. A. Witt, The Journal of Animal Ecology 5, 1–18 (1936)} il aborde, avec ses collègues, la question de façon tout à fait originale :
Ils représentent l'observation de  l'évolution de la taille de deux populations en interaction, non pas sous la forme traditionnelle de deux chroniques temporelles mais comme une succession de points dans l'espace des couples $(x,y)$ comme on peut le voir sur la figure \ref{gausePP1}.
\begin{figure}[!tb]
   \centering
  \includegraphics[width=0.8\textwidth]{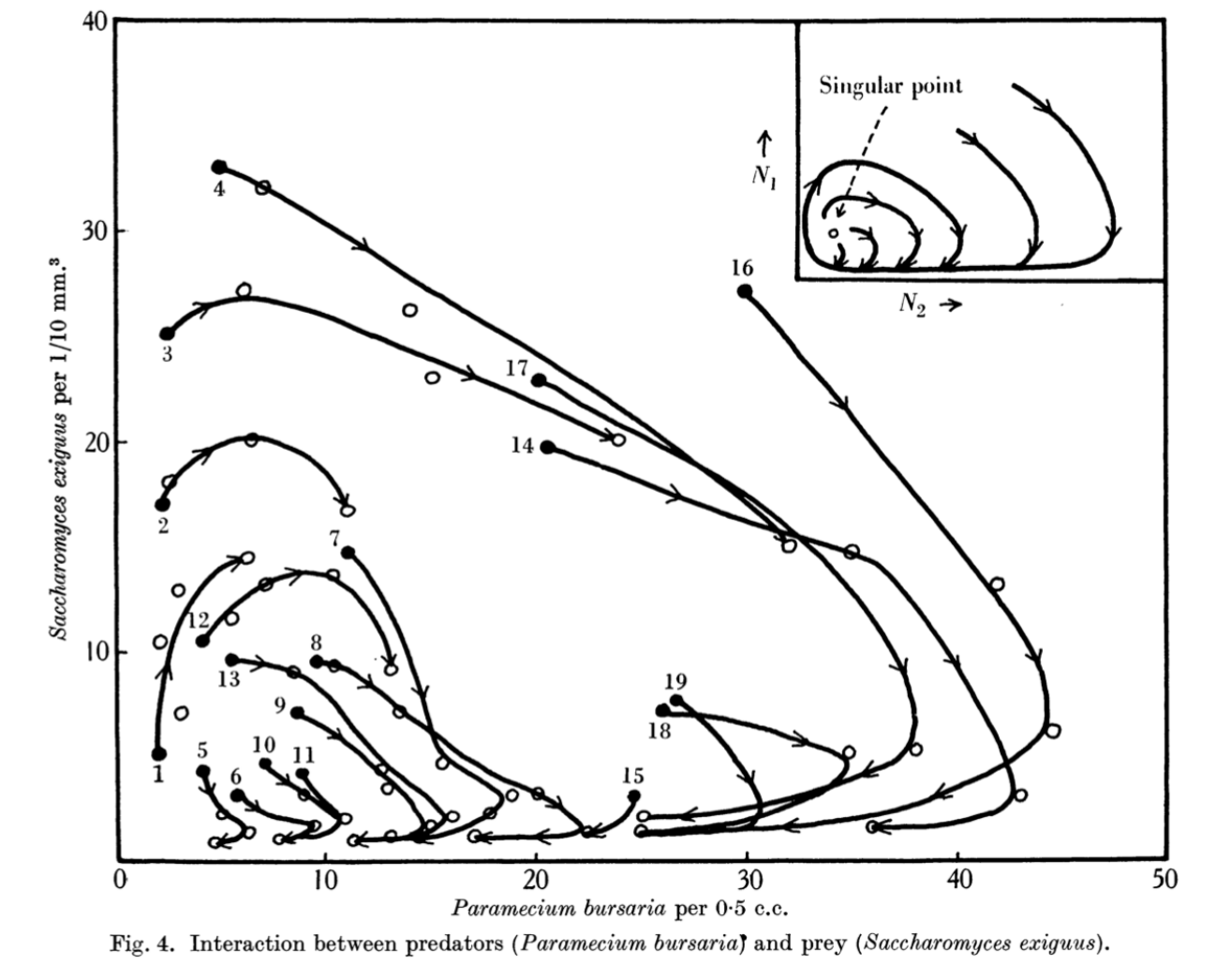} 
  \caption{Portrait de phase empirique de l'interaction {\em Paramecium bursaria} avec {\em Saccharomices exiguus} (\texttt{Attention, contrairement à l'habitude c'est la proie qui est représentée en ordonnée}). }\label{gausePP1} 
\end{figure}
Sur cette représentation on voit clairement se dessiner le portrait de phase d'un système différentiel du plan (notion largement popularisée de nos jours, mais certainement pas à l'époque). De cette image et d'autres considérations ils infèrent :
\dcom
It can be seen that a certain threshold concentration of yeast cells sedimenting on the bottom and elsewhere cannot be destroyed by predators; and predators, even under artificial rarification, do not seriously decrease in concentration until the destruction of the prey down to this threshold (which is partially connected with the decrease in the size of the predators). The threshold has in no way a firmly fixed value, but depends upon the concentration of yeast; when concentration is high, more cells sediment firmly on the bottom and escape destruction and the threshold is more pronounced. The escapes due to sedimentation and other causes, and consequently the threshold, intrinsically accompany the interaction between Paramecia and yeast cells.
\fcom
En d'autres termes l'interaction entre les deux espèces n'est pas de même nature selon que la quantité de proies est au dessous ou au dessus d'un certain seuil : en dessous du seuil les proies ne sont plus ''atteintes'' par le prédateur. Ensuite dans un paragraphe intitulé : {\em Equations of interaction} on trouve les deux équations (cette fois avec les notations des auteurs) :
\beq \label{G1}
\frac{dN_1}{dt} = bN_1\frac{K-N_1}{K} - \frac{N_2}{n_2} f(n_2,N_1) 
\feq
\beq\label{G2}
\frac{dN_2}{dt} = \frac{N_2}{n_2}k f(n_2,N_1) - f_1(n_2) N_2
\feq
où $N_1$ et $N_2$ sont les quantités de proies et de prédateurs et $n_2$ est une quantité qui représente la biomasse d'un prédateur quantité qu'ils déclarent régie par troisième équation ; oublions cette dernière subtilité en supposant que $n_2$ est constant, les deux équations (\ref{G1}) et (\ref{G2}) se réécrivent :
\beq \label{G3}
\begin{array}{rcl} 
\displaystyle \frac{dN_1}{dt}& =&\displaystyle bN_1\frac{K-N_1}{K} - N_2 f(N_1)   \\[8pt]
\displaystyle \frac{dN_2}{dt} &=&\displaystyle N_2k f(N_1)  - f_1 N_2
  \end{array}
\feq
ce qui est, aux notations près (\ref{RMA}) ou (\ref{RMAbis}).

Il semble que cet article, contrairement aux travaux de Gause  sur la compétition soit passé relativement inaperçu\footnote{ L'article de Rosenzweig-MacArthur ne le mentionne pas, ce qui ne s'explique pas par l'état des relations entre le monde occidental et l'Union Soviétique pendant la guerre froide puisque l'article est paru dans une revue américaine.}. Il faut noter aussi que si, formellement, le modèle (\ref{G3}) est le même que (\ref{RMA}) il en diffère en ce que la fonction $f$ est une \textbf{fonction discontinue} or les mathématiques de 1936
ne savent pas encore bien traiter les équations différentielles dont le second membre est discontinu, ce n'est qu'en 1960 que les premiers traitements de tels systèmes on lieu\footnote{Voir Krivan {\em On the Gause predator-prey model with a refuge: a fresh look at the history}, Journal of Theoretical Biology, 274, pp.67-73 (2011) pour un traitement mathématiquement moderne de l'article de Gause et al.}.  L'article était peut être trop en avance sur son époque pour être remarqué mais maintenant que nous avons un recul suffisant il serait équitable, au minimum, de rebaptiser le système d'équation (\ref{RMA}) comme le {\em modèle de Gause-Rosenzweig-McArthur}. L'article 
{\em  On the Gause predator–prey model with a refuge: A fresh look at the history}\cite{KRI11}\footnote{Journal of theoretical biology, 2011, vol. 274, no 1, p. 67-73} de V. Krivan est une tentative dans ce sens.

La citation suivante de {\em The impact of Robert MacArthur on ecology} \cite{FRE75}\footnote{ Annual Review of Ecology and Systematics, 6(1), 1-13.} du biologiste S.D. Fretwell  
\dcom
Rosenzweig's predator-prey work, done with MacArthur, also  made an important advance in population regulation theory by providing a needed analytical tool.It is interesting to note that Kolmogoroff (reviewed in Rescigno and Richardson (...)) had early accomplished a similar developments. Kolmogoroff's work is rather more elegant, but was obscurely published. Still, there is a simplicity about the Rosenzweig-McArthur formulation that makes it one of the most cited textbook graphs, one of the most tested predator-prey theories (...)
\fcom
de 1975 témoigne de la lente progression du langage et des concepts de la théorie des systèmes dynamiques dans la communauté de l'écologie théorique.

\section{1960 :  Hardin. Une opération idéologique ? }
\begin{figure}[!tb]
   \centering
  \includegraphics[width=0.6\textwidth]{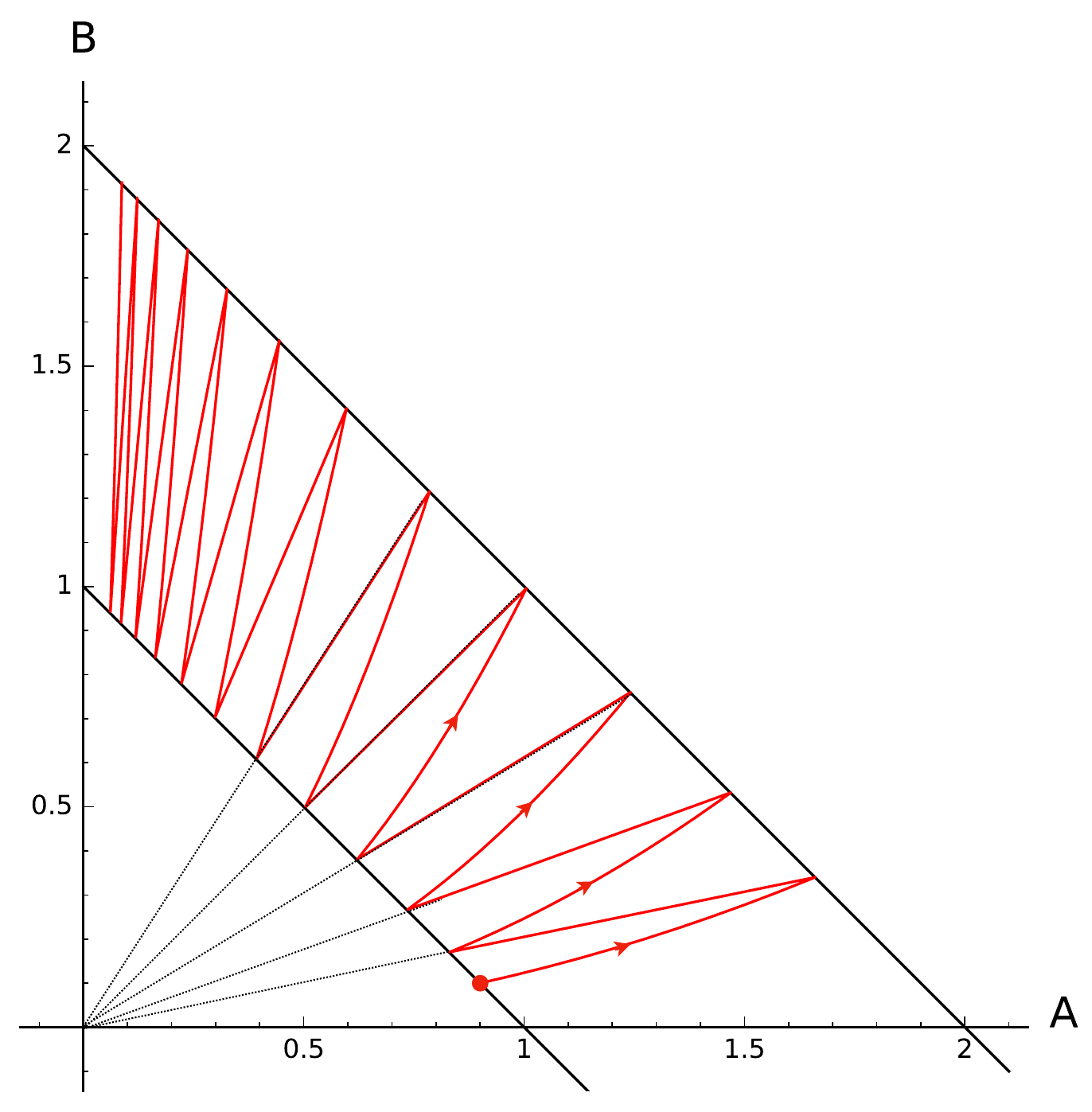} 
  \caption{L'exclusion par la compétition selon Hardin.}\label{Hardin} 
\end{figure}
Garrett Hardin (1915-2003) est un biologiste de formation qui s'est illustré chez les économistes à travers l'article, {\em The tragedy of the communs} \cite{HAR68} un court essai paru dans Science en 1968. {\em The Garrett Harding Society} maintient un site dédié à sa mémoire. Sur la page de présentation du site on lit :
\dcom
 Humankind is embedded in a finite biological setting. Garrett Hardin's writings enable us to responsibly assess our surroundings to optimize the quality of life for present and future generations. By perpetuating Dr. Hardin's writings, the Garrett Hardin Society will provide a meaningful framework to discuss ecological, economic, demographic, and ethical issues.
\fcom
On le voit, l'ambition est grande ! De fait Hardin est une personnalité controversée. Les idées de {\em The tragedy of the communs}, sont loin de faire l'unanimité chez les économistes et il n'est pas besoin d'être économiste pour être surpris par certaines affirmations comme  
 \dcom
 If we love the truth we must we must openly deny the validity of the Universal Declaration of Human Rights, event though it is promoted by the United Nations.
 \fcom
 On s'attend plus à trouver  ce genre de provocation dans des publications engagées politiquement que dans une revue qui s'affiche comme scientifique !
 
 Effectivement, ses productions intellectuelles  sélectionnées par la {\em Garrett Harding Society}, le font apparaitre plutôt comme un militant  politique qu'un scientifique de laboratoire et il ne semble pas que ses premiers travaux en microbiologie aient eu une grande postérité. Toutefois ce n'est pas le personnage de Hardin\footnote{Sur qui on peut lire l'article bien documenté  de F. Locher {\em Les Pâturages De La Guerre Froide : Garrett Hardin et la « Tragédie Des Communs »}, Revue d'histoire moderne et contemporaine, n° 60 p-7-36 (2013) ; on y trouve \dcom
 Il soutient sa thèse en 1941, juste avant Pearl Harbor, mais il ne part pas au front car les séquelles d’une polio le rendent inapte au service. Il passe la période de la guerre au sein du laboratoire de biologie végétale que la Carnegie Institution entretient sur le campus de Stanford, à travailler sur les algues de culture. Mais sa vocation pour la recherche est limitée : en 1946, il s’en détourne en acceptant un poste d’enseignant au Santa Barbara College, en passe d’être intégré à l’Université d’État de Californie.  Il passera plus de quarante années à l’UCSB. Il ne travaille plus en laboratoire mais se consacre à des travaux théoriques, à l’enseignement, à l’écriture et à des activités militantes sur les thématiques sociales et environnementales.
 \fcom
 }
 qui m'intéresse ici mais un article particulier : {\em The Competitive Exclusion Principle. An idea that took a century to be born has implications in ecology, economics, and genetics} \cite{HAR60}\footnote{SCIENCE, vol 131 (1960)} qui, s'il n'atteint pas les 50000 citations du {\em tragedy of the communs},  dépasse les 3000 ce qui est encore un score très respectable. Comme on ne prête qu'aux riches, au vu de ce titre on s'attend à une opération à caractère idéologique. Essayons de voir ce qu'il en est.

L'article parait peu de temps avant celui de Rosenzweig-MacArthur  \cite{ROS63} et largement après celui de Powell \cite{POW58}. C'est une sorte de synthèse historique (modérément bien documentée, le microbiologiste qu'il est aurait pu s'intéresser aux travaux de Powell) de l'idée de compétition et d'exclusion associée ; il en recherche les origines chez Ricardo, Darwin, Grinnell, cite Lotka, Volterra et MacArthur. Il définit le principe ainsi :
\dcom
What does the exclusion principle mean? Roughly this: that (i) if two non interbreeding populations "do the same thing"  $-$ that is, occupy precisely the same ecological niche in Elton's sense (4) $-$ and (ii) if they are "sympatric" $-$ that is, if they occupy the same geographic territory $-$ and (iii) if population A multiplies even the least bit faster than population B, then ultimately A will completely displace B, which will become extinct. This is the"weak form "of the principle. Always in practice a stronger form is used, based on the removal of the hypothetical character of condition(iii).
\fcom 
Il considère ce ''principe'' comme une sorte d'affirmation logique et considère qu'elle a été démontrée :
\dcom
Demonstrations of the formal truth of the principle have been given in 
terms of the calculus (5, 7) and set  theory(8).
\fcom 
Les références (5) et (7) sont celles de Lotka et de Volterra que nous connaissons déjà. 

La référence (8) mérite que l'on s'y arrête. En 1957 se tenait le 
{\em Cold Spring Harbor  Symposium on  Quantitative Biology} et  E. Hutchinson\footnote{G. Evelyn Hutchinson (1903-1991) est un grand nom de l'Ecologie. } en donnait la conclusion. Un paragraphe de cette dernière est intitulé {\em The formalisation of the niche and the Volterra-Gause principle} \cite{HUT57}, paragraphe dans lequel Hutchinson essaye de formaliser le concept de niche écologique.
La niche est un hyper volume $N$ dans un espace abstrait  à $n$ dimensions ou chaque composante représente la valeur d'une ''variable environnementale''. $N$ est la ''niche'' d'une espèce $S$ si l'espèce ne peut survivre que lorsque les valeurs des paramètres environnementaux appartiennent à $N$. Il distingue  bien cette ''niche'' abstraite d'une partie $B$ de l'espace ordinaire physique $\Rmat^3$. A chaque point de $\Rmat^3$ correspondent les valeurs des paramètres environnementaux en ce point et, donc, un volume (ordinaire) de $\Rmat^3$ où les paramètres environnementaux appartiennent à $N$, donc sont acceptables pour l'espèce $S$. Aussitôt ces points développés Hutchinson critique le formalisme qu'il vient de proposer. Je le cite :
\dcom 
{\em Limitations of the set-theoretic mode of expression.} 
\ben
\item It is supposed that all points in the fundamental niche impliy equal probability of persistance of the species (...)
 \item It is assumed that all environmental variables   can be linearly ordered. In the present state of knowledge it is obviously impossible.(...) 
 \item The model refers to a single instant of time.
\item Only a few species are to be considered at once, so that abstraction of these makes little difference to the whole community.
\fen
\fcom
En dépit de ces critiques il développe le formalisme introduit puis, en regard des travaux de Volterra et Gause il conclut le paragraphe par :
\dcom
{\em Validity of the Gause-Volterra Principle.} The set-theoretic approach outlined above permits certain refinements which, however obvious they may seem, apparently require to be stated formally in an unambiguous way to prevent further confusion. This approach however \underline{tells us nothing about the } \underline{validity of the principle}\footnote{Souligné par moi}, but merely where we should look for its verification or falsification.
\fcom
et un peu peu plus loin :
\dcom
The only conclusion that one can draw at present from the observations is that although animal  communities appear qualitatively to be constructed as if competition were regulating their structure, even in the best studied cases there are nearly always difficulties and unexplored possibilities. These difficulties suggest that if competition is determinative it either acts intermittently, as in
abnormally dry seasons for {\em Plethodon}, or it is a more subtle process than has been supposed.
\fcom
Nous sommes vraiment très loin du lapidaire : « Demonstrations of the formal truth of the principle have been given in 
terms (...) of set  theory » de Hardin. 

Ensuite Hardin s'adresse à ceux que les mathématiques rebutent :
\dcom
Those to whom mathematics does not appeal may prefer the following intuitive verbal argument (...) which is based on an economic analogy that is very strange economics but quite normal biology.
\fcom
Il s'agit d'une banque qui a seulement deux déposants, $A$ et $B$. Le déposant $A$ touche un intérêt au taux $r$ et $B$ au taux  $r+\varepsilon$ ; dès que la somme des deux comptes atteint 2 unités, on divise par deux chaque compte ; ce processus est répété et il est facile de se convaincre que la fortune de $A$ va inexorablement tendre vers $0$ (voir figure \ref{Hardin}). S'il n'y a aucun doute, dans cette analogie, que le déposant qui a le moins bon taux de croissance disparait, en revanche, ce qui est moins clair, c'est en quoi $A$ et $B$ sont en compétition ? Ils subissent tous deux la même ''punition'' lorsque la fortune totale atteint $2$ unités : une division par deux. En réalité $B$ ayant un taux de croissance plus grand que celui de $A$, dans une croissance illimitée,   la fortune $a(t)$ de $A$ tend vers l'infini plus vite que celle $b(t)$ de $B$. Le rapport $a(t)/b(t)$ tend vers $0$ et c'est la renormalisation constante à $a(t)+b(t)= 1$ qui fait que $A$ disparait. Où est la compétition ?

La conclusion de l'article de Hardin est un appel à reconnaitre le {\em Principe de l'exclusion compétitive} comme un principe organisateur  :
\dcom
To assert the truth of the competitive exclusion principle is not to say that nature is  and always  must be, everywhere, ''red in tooth and claw.'' Rather, it is to point out that every instance of apparent coexistence must be accounted for.
\fcom
Hutchinson aurait peut être du s'offusquer de l'usage fait par Hardin de son travail de clarification du concept de niche écologique mais, s'il  n'en a rien fait, cela lui a peut être inspiré en partie un célèbre article, postérieur de peu à \cite{HAR60}. Dans {\em The paradox of the plancton} \cite{HUT61}\footnote{
The American Naturalist, Vol. 95, No. 882. 1961.}Hutchinson constate que, alors qu'ils partagent un petit nombre de ressources, de l'ordre d'une dizaine, des centaines d'espèces différentes de phytoplankton coexistent dans les lacs de montagne. Le PEC\footnote{sous une forme plus générale {\em de $n$ espèces en compétition pour $p$ ressources, au plus $p$ peuvent subsister.}} est donc infirmé. Il conclut :
\dcom
Apart from providing a few thoughts on what is to me a fascinating, if somewhat specialized subject, my main purpose has been to show how a certain theory, namely, that of competitive exclusion, can be used to examine a situation where its main conclusions seem to be empirically false. Just because the theory is analytically true and in a certain sense tautological, we can trust it in the work of trying to find out what has happened to cause its empirical falsification. It is, of course, possible that some people with greater insight might have seen further into the problem of the plankton without the theory that I have with it, but for the moment I am content that its use has demonstrated possible ways of looking at the problem and, I hope, of presenting that problem to you.
\fcom 
Le principe n'est donc pas une loi idéale dont la réalité s'approcherait plus ou moins, mais un questionnement : pour quelles raisons observe-t-on la coexistence de ces espèces alors que le PEC dit que cela ne devrait pas être ?

 Pour conclure je dirais que les articles de Hardin et Hutchinson, s'ils sont si souvent cités (surtout Hardin) par les écologues, ce n'est pas directement pour leur contenu réel (et souvent ignoré dans le cas de Hardin) mais parce qu'ils  apparaissent comme  les marqueurs d'une thématique. Par exemple, dans l'article de Sommer {\em The paradox of the plankton: Fluctuations of phosphorus availability maintain diversity of phytoplankton
in flow-through cultures} \cite{SOM84}\footnote{Limnol. Oceanogr., 29(3), 1984, 633-636 (1984)} on peut lire :
\dcom
The diversity of natural phytoplankton seems to contradict the competitive exclusion principle (Gause 1932; Hardin 1960). Although most algae compete for the same inorganic nutrients, often more than 30 species coexist even in small parcels of water (the “paradox of the plankton” sensu Hutchinson 1961) (...)

Although interesting from a theoretical point of view, limitation of different species by different nutrients explains only a small proportion of the diversity and species richness of natural phytoplankton, even if allowance is made for the usually neglected trace elements. Therefore I attempted here to increase the number of coexisting species by modifying the classical chemostat experiments by the pulsed addition of phosphate. 
\fcom
On le constate, la charge émotionnelle de l'expression ''exclusion compétitive'' s'estompe au profit de questions techniques.

Pour ce qui m'intéresse, le rôle des mathématiques dans l'histoire du PEC, le texte de Hutchinson que je viens d'analyser est important. En effet, il utilise les mêmes arguments que Volterra pour justifier l'introduction d'un formalisme mathématique. Pourquoi l'appelle-t-il "set-theoretic mode of expression" plutôt que formalisme mathématique? Pour impressionner la galerie, en mettre plein la vue comme on le devine chez Hardin ? Certainement pas, son analyse critique immédiate de ce qu'il vient de proposer le prouve. 
 Je pense qu'il faut plutôt y voire la maladresse de quelqu'un qui n'a pas une formation de base en mathématiques et n'est pas très bien informé sur ses développements. 
 
 \textbf{Mais, et  c'est là le point important ce ne sont plus les mathématiciens comme Volterra qui prônent les mérites du formalisme, ce sont des biologistes "purs et durs" comme Hutchinson qui en ressentent l'utilité, sans en être dupes.
}

\section{1970 - à nos jours : Vers le temps  des modèles }\label{tempsmathématiciens}

{\em Les modèles de Verhulst, Lotka et Volterra sont des équations différentielles ''non linéaires''. Ce n'est pas nouveau en mathématiques ! Très tôt, pour ainsi dire dès que le formalisme fut inventé,  par exemple pour la description du mouvement de deux corps sous l'effet de leur attraction réciproque, les mathématiques ont considéré des équations différentielles non linéaires. A côté de la mécanique céleste, des problèmes 
posés par les développements scientifiques ont conduit les mathématiciens à s'intéresser à d'autres  types d'équations différentielles non linéaires, par exemple dans  le problème de la stabilité de marche des machines à vapeur (Maxwell, Vichnégradskii $\approx 1870)$ ou de la production de courant alternatif (B. van der Pol {\em 
A theory of the amplitude of free and forced triode vibrations} \cite{VAN20}\footnote{
Radio Review, 1, 701-710, 754-762, 1920.}).
On ne peut donc pas dire que c'est la dynamique des populations qui a engagé les mathématiques sur le "non linéaire". En revanche, à partir de 1970 les concepts et outils abstraits de la théorie des systèmes dynamiques 
 qui se sont développés à partir de 1940 font maintenant partie, à un degré plus ou moins grand de sophistication technique, du bagage de tout jeune mathématicien qui se lance dans la recherche. A ces outils d'exploration strictement mathématiques des systèmes dynamiques, se sont ajoutées les simulations sur ordinateur qui depuis les années 1980 sont accessibles à un coût matériel (un simple ordinateur personnel suffit) et intellectuel (les logiciels plus ou moins conviviaux de simulation abondent) à peu près nul. }
{\em C'est pourquoi, à partir de 1970 la littérature mathématique en dynamique des populations explose. Voici quelques thèmes qui sont loin de constituer une revue complète du sujet :
\bit 
\item Théorie mathématique complète du chémostat et toutes ses variantes.
\item Analyse des chaines trophiques ressource abiotique - consommateur primaire- prédateur.
\item Modèle épidémiologiques de type SIR (Susceptible, Infecté, Retirés). 
\item etc....
\fit
C'est dans ce contexte de développement des mathématiques de la dynamique des populations que se produit, en 1976, un évènement important de l'histoire du PEC où 
 les mathématiques vont apporter une contribution décisive.}
\subsection{La compétition dans le chémostat}
Nous avons vu que dès 1958, Powel décrivait dans un journal de microbiologie comment  le PEC intervient  dans le chémostat \cite{POW58}. Il faut attendre près de vingt ans pour qu'en 1977,  Hsu, S. Hubbell et Waltman publient dans une revue mathématique {\em A mathematical theory for single nutrient competition in continuous cultures of micro-organisms} \cite{HSU77}\footnote{
 SIAM Journal on Applied Mathematics 32: 366-83 (1977).}. Le premier et le dernier auteur sont des mathématiciens, S. Hubbell est un biologiste
 \footnote{S. Hubbel ( 1942...) est célèbre pour son {\em The Unified Neutral Theory of Biodiversity and Biogeography}, Princeton University Press 2001.}.
 Après un rappel sur le chémostat ils insistent sur l'intérêt théorique du dispositif 
 \dcom
 The chemostat is perhaps the best laboratory idealization of nature for population studies (Williams [..]). All natural systems are open systems for energy and material substances. The input and removal of nutrients to and from the chemostat represent the continuous turnover of nutrients in nature. The outflow of organisms is formally equivalent to nonspecific death, predation, or emigration, which always occur in nature.
 \fcom 
 et, comme il se doit dans une revue de mathématiques, ils insistent sur la "rigueur" de leur traitement
 \dcom
 This paper uses the general deterministic model for one substrate and n competing species or strains, and presents a rigorous mathematical analysis of the asymptotic behavior of this system (...)
 \fcom 
et sur l'aspect général et synthétique de leur travail
\dcom
Although some partial results exist in the literature on this problem, we believe that this paper represents the most complete treatment of the system yet available. In particular, it generalizes the work of Powell [...], makes his conclusions mathematically rigorous, and gives a mathematical explanation to some observations of Taylor and Williams [...] in their numerical experiments.
\fcom

Le long délai entre l'explication convaincante par Powel du mécanisme du PEC dans le chémostat et des démonstrations mathématiquement rigoureuses n'est pas due à la difficulté du sujet. Les mathématiques de 1960 pouvaient certainement formaliser le PEC dans le chemostat, mais elles ignoraient qu'elles avaient à le faire !

Un peu plus tard (1980) S. Hubbell publiera dans Science
avec son collègue S. Hansen du département de Zoologie de l'Université de Iowa, {\em Single-Nutrient Microbial Competition: Qualitative Agreement between Experimental and Theoretically Forecast Outcomes} \cite{HAN80}\footnote{
Science, New Series, Vol. 207, No. 4438. (Mar. 28, 1980), pp. 1491-1493.}. Partant de la théorie mathématique de l'article précédent ils associent à chaque espèce $i$ présente dans le chémostat un indice $J_i = K_{s_i}\left(\frac{D}{r_i}\right)$ qui s'exprime (peu importe ici exactement comment) à partir des deux paramètres  de la fonction de Monod. Le critère est que l'espèce qui a le plus petit indice $J_i$ est celle qui gagne la compétition. Les auteurs insistent sur le fait qu'il n'est pas intuitif, que ce sont les mathématiques qui le montent, et qu'il mérite une vérification expérimentale.
\dcom
The J criterion for competitive ability is nonobvious and requires experimental verification. It could not have been predicted from classical theories of competition (8)\footnote{
La note (8) est une assez longue description verbale des travaux de Lotka et Volterra
}. A priori it might have been expected that the winner would always be the species with the highest affinity (lowest $K_s$) for the nutrient, or perhaps the organism with the highest intrinsic rate of increase; in fact there are conflicting opinions on this question (9)\footnote{
la note (9) donne quelques références sur cette controverse}. However, the extended theory of Monod and of Novick and Szilard asserts that it is actually a weighted $K_s$ value which is critical to competitive success-weighted by the ratio of the death rate to intrinsic rate of increase. Thus, a species with a higher affinity for the resource may nevertheless lose if it also has a lower intrinsic rate or higher death rate. The theory also asserts that winning will be independent of the growth efficiency (yield) of the species grown on the limiting resource.
\fcom 
Aussi des expériences seront menées
\dcom
 In the first experiment, {\em Esscherichia coli} strain C-8 was opposed by {\em Pseudonzonas aeruginosa} strain PA0283.
\fcom
sur des espèces de laboratoire et les prédictions mathématiques  du modèle seront vérifiées. Il s'agit donc d'un raffinement du  travail de Powell de 1958 d'ailleurs cité dans \cite{HAN80}.

On retrouve exactement la démarche de Gause un demi siècle plus tôt, y compris dans la collaboration avec des mathématiciens. Cependant les choses ont évolué. Les mathématiques sont plus élaborées $-$ la théorie des systèmes dynamiques a eu le temps de se développer $-$ et les dispositifs expérimentaux sont beaucoup plus sophistiqués avec le chémostat et les techniques d'observation des microorganismes.

\subsection{Le modèle de compétition issu de Gause-Rosenzweig-MacArthur}\label{sectioncompeteRMA}

Comme l'avaient fait Lotka et Volterra  en combinant deux logistiques  pour obtenir un modèle de compétition, on peut combiner deux modèles de Gause-Rosenzweig-MacArthur pour exprimer la compétition de deux espèces pour une ressource. 
Nous partons donc du modèle (\ref{RMAbis}) que je rappelle 

$$
\begin{array}{rcl} 
\displaystyle  \frac{dx}{dt}& =&f(x) - \mu(x) y \\[8pt]
\displaystyle  \frac{dy}{dt}& =& (c\mu(x) - m)y
  \end{array}
$$
où le consommateur $y$ consomme la ressource biotique $x$. Si nous appliquons la même logique à deux consommateurs $y_1$ et $y_2$  qui se partagent la même ressource nous sommes conduits à écrire :
\beq \label{competeRMA}
\begin{array}{lcl} 
\displaystyle  \frac{dx}{dt}& =&f(x) - \mu_1(x) y_1 - \mu_2(x)y_2\\[8pt]
\displaystyle  \frac{dy_1}{dt}& =& (c_1\mu_1(x) - m_1)y_1 \\[8pt]
\displaystyle  \frac{dy_2}{dt}& =& (c_2\mu_2(x) - m_2)y_2 \\[8pt]
  \end{array}
\feq
Le même argument que celui que nous avons exposé à propos du chémostat s'applique ici. Si nous désignons par $(x_e, y_{1e}, y_{2e})$ un équilibre possible, comme il n'est pas possible en général de satisfaire simultanément $  (c_1\mu_1(x_e) - m_1) =0$ et  $  (c_2\mu_2(x_e) - m_2) =0$ on a nécessairement un des deux $y_{1e}$, $y_{2e}$ qui est nul. Ce qui veut dire la disparition d'une des espèces.\\

Toutefois il y a une différence importante avec le cas du chémostat dont je rappele les équations pour la compétition :

\beq \label{compchembis}
\begin{array}{rcl} 
\displaystyle \frac{ds}{dt}& =&\displaystyle d(S_{in}-s) - \frac{\mu_1(s) x_1}{Y_1} -  \frac{\mu_2(s) x_2}{Y_2}\\[8pt]
\displaystyle \frac{dx_1}{dt} &=&\displaystyle ( \mu_1(s) - d)x_1\\[8pt]
\displaystyle \frac{dx_2}{dt} &=&\displaystyle ( \mu_2(s) - d)x_2
  \end{array}
\feq
Dans le modèle (\ref{competeRMA}) la dynamique de la ressource (biotique) est donnée par la fonction $x \mapsto f(x)$ qui est non linéaire (en général croissante puis décroissante) alors que dans le cas de la ressource (abiotique) du chémostat la dynamique est  $s \mapsto d(S_{in}-s)$ qui est linéaire ce qui a une conséquence importante :  on montre que {\em dans une compétition pour une ressource abiotique, dans le modèle du chémostat (\ref{compchembis}) l'équilibre d'exclusion est globalement asymptotiquement stable}, ce qui veut dire que, quelle que soit la condition initiale, après un transitoire plus ou moins long, l'espèce la plus performante aura éliminé l'autre. Nous venons de voir que ce résultat, considéré comme acquis depuis Powell, n'a été établi rigoureusement en tant que théorème mathématique qu'en 1977 par Hsu, Smith et Waltman.
 
Mais,  {\em en l'absence de l'hypothèse de linéarité sur la fonction $f$, il n'est pas possible de montrer que l'équilibre d'exclusion est globalement stable.} Alors, si l'on ne tend pas forcément vers un équilibre d'exclusion, que peut-il se passer ? Dans les années 1970 on sait depuis longtemps qu'un système dynamique non linéaire peut avoir une solution périodique stable (souvenons nous de l'article de 1936 de Kolmogorov) et, pour les systèmes de {\em deux équations}, les méthodes ne manquent pas pour prouver l'existence de tels cycles. Malheureusement le système 
 (\ref{competeRMA}) est un système de \underline{trois équations} et, ce qui est relativement aisé pour \underline{deux équations} devient redoutable pour les dimensions supérieures.
\begin{figure}[!tb]
   \centering
  \includegraphics[width=0.8\textwidth]{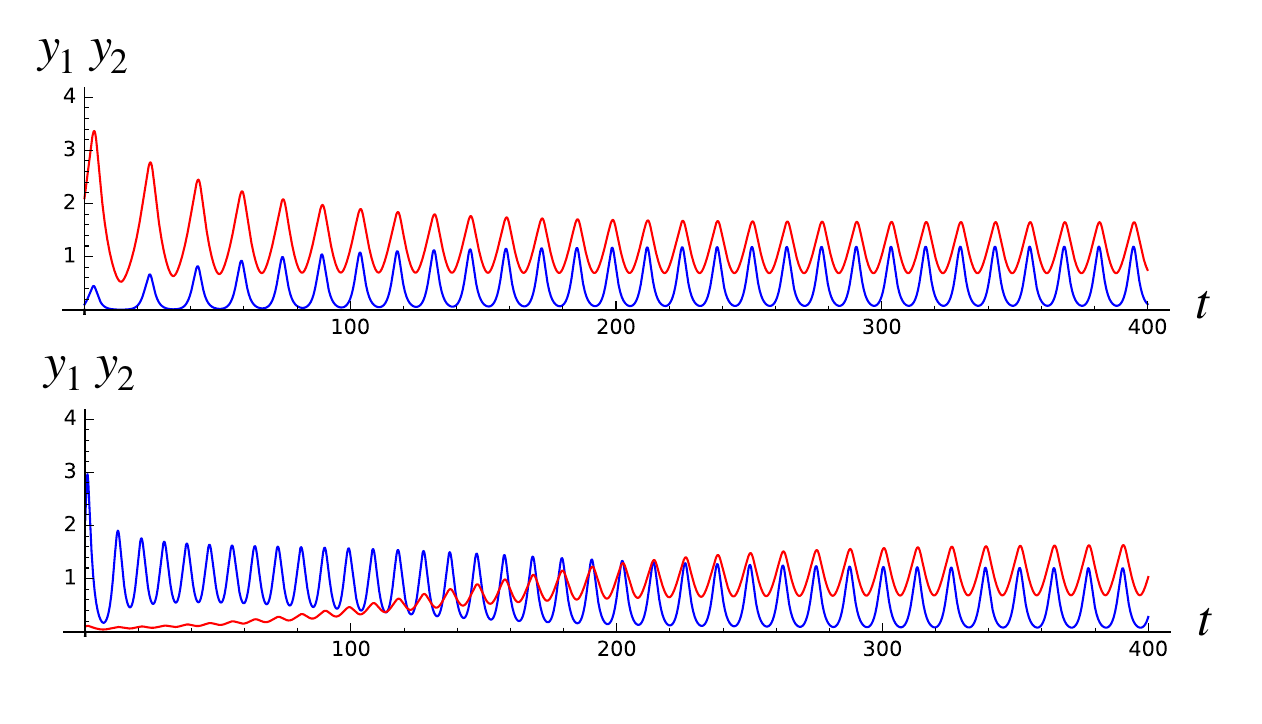} 
  \caption{Simulation, en fonction du temps, des solutions de (\ref{compRMA}). }\label{compRMA1} 
\end{figure}\begin{figure}[!tb]
   \centering
  \includegraphics[width=0.8\textwidth]{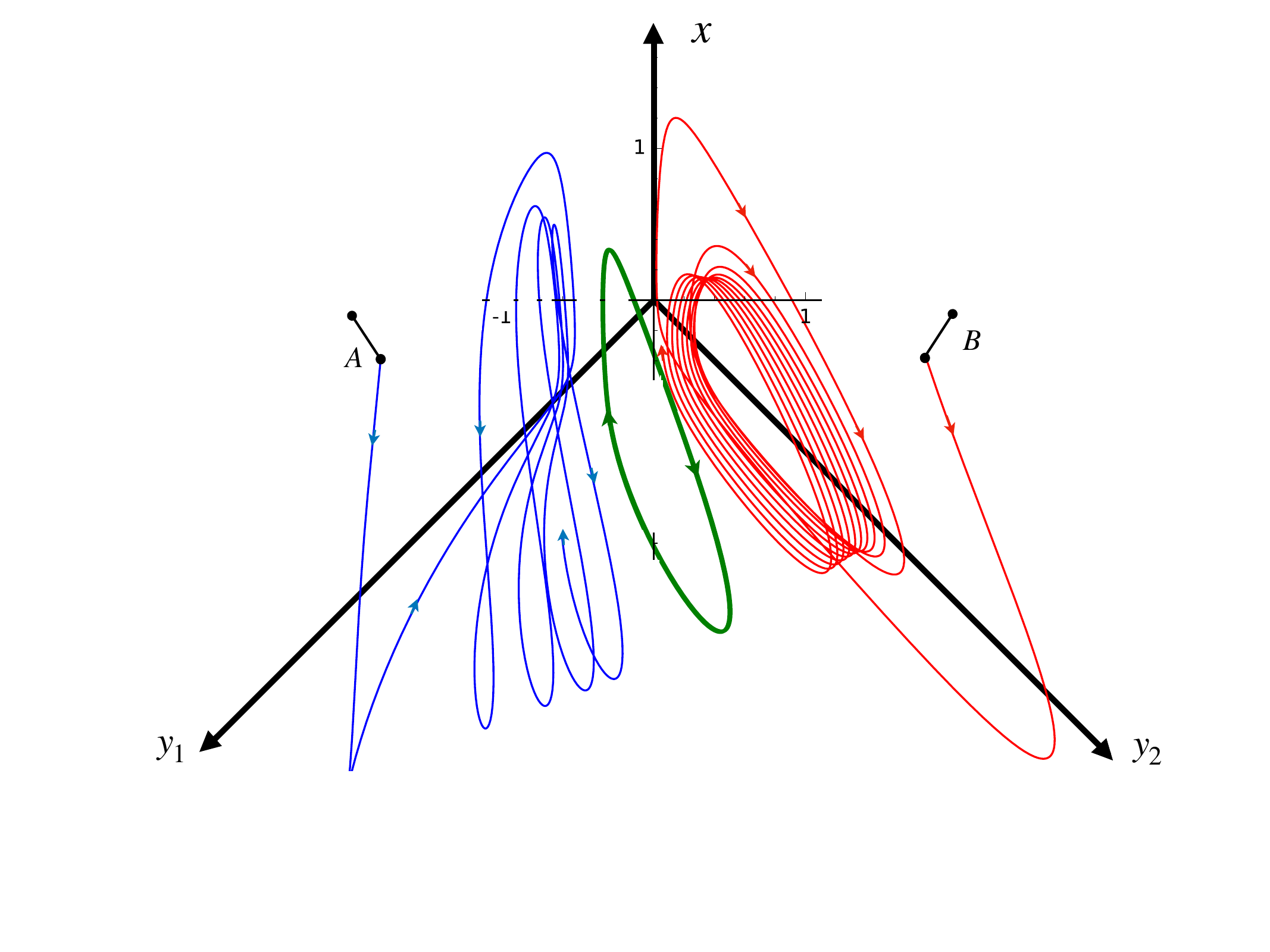} 
  \caption{Simulation, dans l'espace des phases, des solutions de (\ref{compRMA}). }\label{compRMA2} 
\end{figure}

 \subsection{L'apport d'Armstrong-MacGehee} \label{Armstrong-MacGehee}
 C'est dans ce contexte que, un biologiste, Armstrong, et un mathématicien, MacGehee, vont complètement renouveler la perspective dans une série de quatre articles débutant en 1976 et dont le plus cité est le dernier : {\em Competitive exclusion} \cite{ARM80}\footnote{{\em The American Naturalist} , Vol 115 n0°2 (1980)}. Ils montrent, par preuve mathématique et par simulations numériques que, pour certaines valeurs des paramètres, le modèle de compétition construit sur Gause-Rosenzweig-McArthur possède un cycle limite globalement asymptotiquement stable.
 Ainsi, Armstron et MacGehee ont accompli un  exploit technique (pour l'époque) et  ce n'est pas un hasard si MacGehee s'est aussi illustré à propos du problème mathématique des trois corps ! 
 
 Sur les figures \ref{compRMA1} et \ref{compRMA2} on peut observer le résultat de simulations modernes (ici faites à l'aide du logiciel libre  SageMath) du modèle :
 \beq \label{compRMA}
\begin{array}{rcl} 
\displaystyle \frac{dx}{dt}& =&\displaystyle 2 x(1-0.5 )  - \mu_1(x) y_1-  \mu_2(x) y_2\\[8pt]
\displaystyle \frac{dy_1}{dt} &=&\displaystyle ( \mu_1(x) - 0.58)y_1\\[8pt]
\displaystyle \frac{dy_2}{dt} &=&\displaystyle ( \mu_2(x) - 0.45)y_2
  \end{array}
\feq
avec 
$$\mu_1(x) = \frac{2x}{1+x} \quad \quad \quad \mu_2(x) = \frac{x}{0.4+x}$$
 On voit (figure \ref{compRMA2}) comment les trajectoires issues de A (petite présence de l'espèce 2) et celles issues de B (petite présence de l'espèce 1) tendent toutes vers la solution périodique verte. Le long de cette solution les deux espèces oscillent sans jamais disparaitre (figure \ref{compRMA1}).
  \begin{figure}[!tb]
 \centering
  \includegraphics[width=0.8\textwidth]{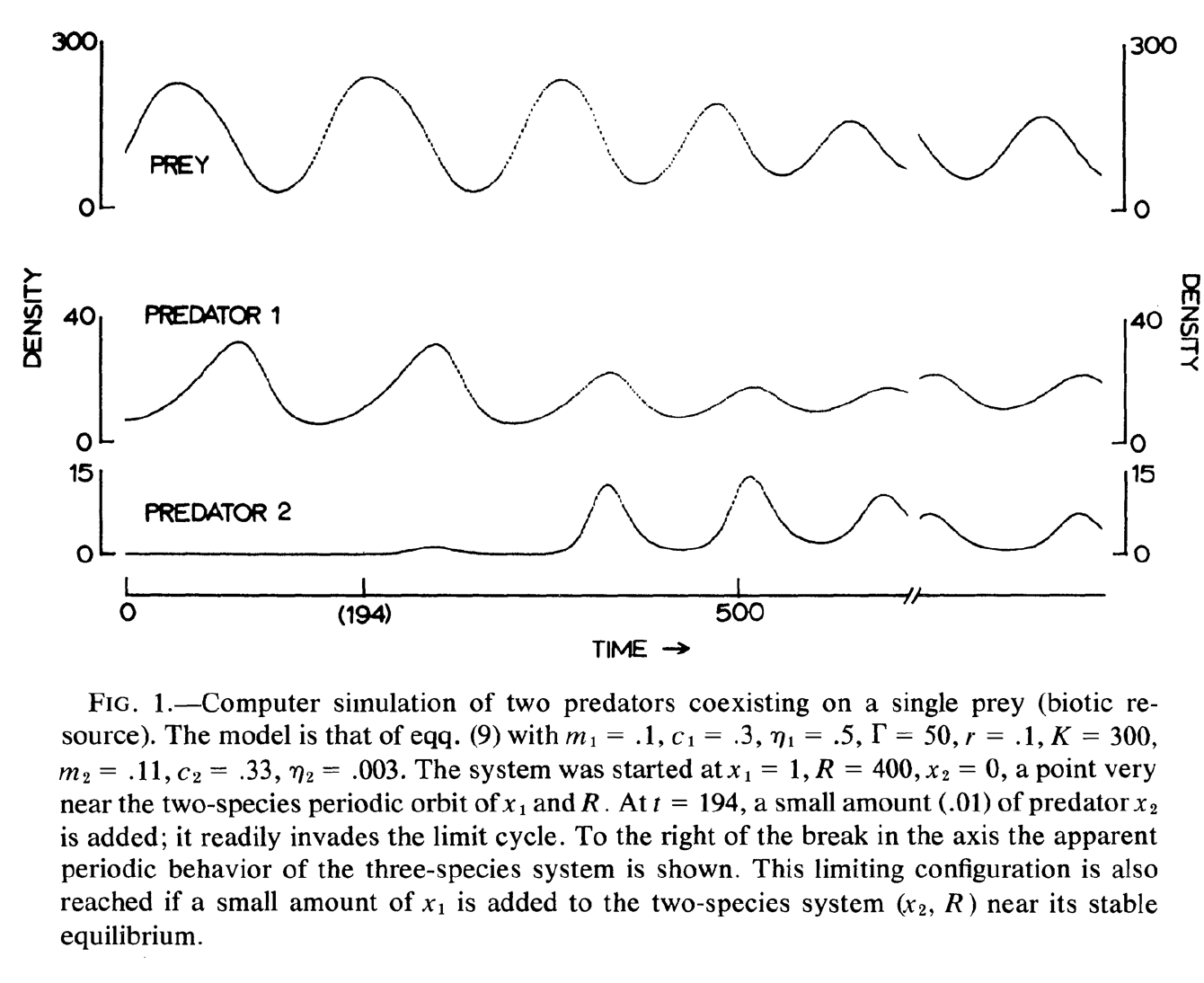} 
  \caption{Extrait de {\em Competitive Exclusion } de Armstron et MacGehee 1980. }\label{compRMA3} 
\end{figure}
Sur la figure \ref{compRMA3} on voit  les simulations proposées, avec les moyens de l'époque, par Armstrong et McGehee. 

Ainsi le PEC, qu'on pouvait considérer comme solidement établi {\em in vitro}, je veux dire en laboratoire dans des conditions précises et pour des microorganismes, se trouve être invalidé sur un modèle peut-être plus proche des situations naturelles. Cet article fait prendre conscience que  pour  définir mathématiquement la {\em coexistence} ce qu'il faut c'est interdire aux solutions de se rapprocher des faces de l'orthant.
 Un nouveau concept organisateur était né : celui  de {\em persistance} d'un système : Un système de $n$ espèces $x_i$ interagissant entre elles est dit {\em persistant} si quelque soit la condition initiale,  pour chacune des espèces $x_i(t)$, il existe un nombre strictement positif $a_i$ tel que, pour tout $t$, $x_i(t) \geq a_i$. Il existe diverses définitions techniques de la persistance, y compris pour des systèmes stochastiques (par exemple, \cite{SCH11}), mais ce qui compte du point de vue de l'écologie théorique c'est la prise de conscience de ce que :
 {\em "coexistence" n'est pas synonyme de "coexistence à l'équilibre"} ! 
  
  Ce travail de Armstron et MacGehee n'a évidemment pas échappé à Hsu, S. Hubbell et Waltman qui immédiatement (dès 1978) publient des articles comme \cite{HSU78}) où ils commencent à caractériser des domaines de paramètres pour lesquels il y a PEC ou pas.

 \begin{figure}[!tb]
   \centering
  \includegraphics[width=0.8\textwidth]{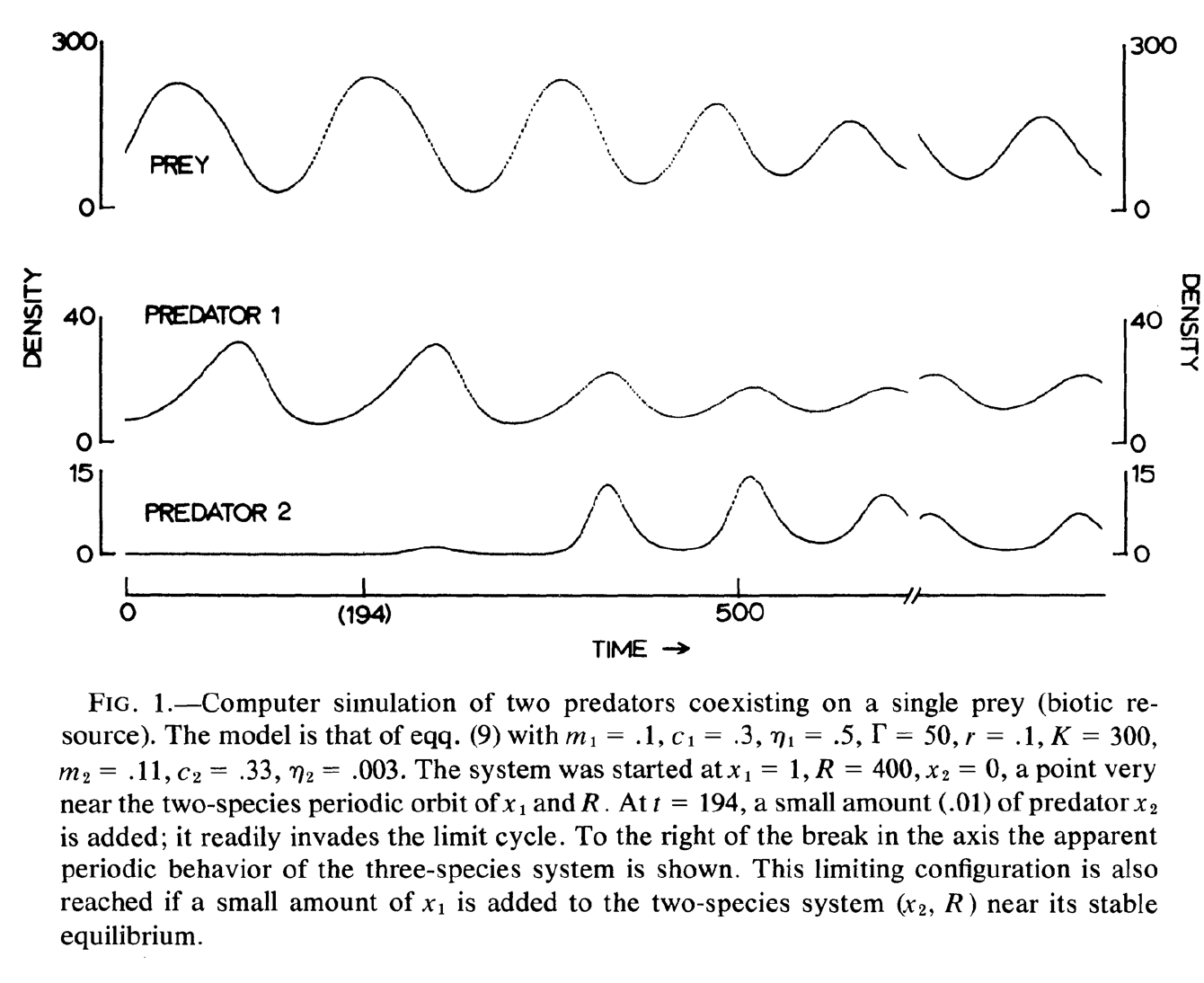} 
  \caption{Extrait de {\em Competitive Exclusion } de Armstron et MacGehee 1980. }
\end{figure}\label{simuArmstrong} 
 \subsection{Le temps des modèles}

Cette question de la persistance est attrayante pour les mathématiciens. En effet, plaçons nous dans le cas d'un modèle avec deux consommateurs et une ressource sur un site isolé. Le modèle est un système différentiel   à trois variables qui possède une propriété structurelle importante : Il laisse l'orthant  positif invariant $-$ les quantités de biomasse sont par essence positives$-$ ainsi que ses faces $-$ il n'y a pas de génération spontanée, si une espèce est absente elle ne peut apparaitre $-$ mais rien n'interdit que pour certaines solutions une variable (une quantité de biomasse) tende vers $0$, ou, plus subtilement, ait une limite inférieure nulle\footnote{c.a.d. qu'il existe une suite $ t_n $ tendant vers $+\infty$ telle que $x(t_n) \rightarrow 0$.} ce serait contraire à la persistance. Montrer qu'un système particulier de dynamique des populations est persistant est certainement un beau chalenge pour mathématicien ; le succès de cette question dans le monde  mathématique n'est pas étonnant.

Mais on peut se demander ce que serait devenu ce concept mathématique de persistance chez les biologistes en l'absence de simulations numériques de modèles aux paramètres plausibles. Armstrong et MacGehee l'avaient bien compris et proposaient déjà des simulations de modèles, comme on le voit sur la figure
extraite de l'article de The American Naturalist. A partir de 1980 les ordinateurs personnels se sont largement répandus dans le monde des écologues, d'abord pour effectuer des traitements statistiques et automatiser des expériences, puis, la diffusion de logiciels ''clef en mains'' de simulation aidant, pour simuler des ''modèles''. A partir de là on ne compte plus les thèses d'écologie qui comportent un ''modèle'' prétendant simuler un écosystème particulier ce qui impose une collaboration de plus en plus étroite entre mathématiciens et biologistes sur le thème de la dynamique des populations. On pourrait souvent critiquer la qualité de cette collaboration $-$ la course au contrats et l'obsession de l'évaluation ne favorisent pas l'émergence d'une saine interaction$-$ mais ce n'est pas ici mon sujet.

Il n'en demeure pas moins que cette collaboration a fait émerger des questions importantes comme celles :
\bit
\item de la (non) pertinence d'un environnement constant dans le temps ; les modèles deviennent non-stationnaires ;
\item  de la (non) pertinence du modèle différentiel déterministe pour traiter de questions d'extinction qui conduit au développement de modèles probabilistes ; 
\item de la (non) pertinence de l'hypothèse d'un milieu homogène qui conduit à considèrer des modèles spatialisés, soit en réseau d'iles interconnectées, soit de façon continue avec des modèles de type ''équations de diffusion réaction''.
\item de l'importance de la structuration physiologique des individus d'une population : structuration en âge, en taille, en stades etc...
\fit
 Il n'y a plus de limite à la sophistication des modèles jusqu'à l'intégration de toutes sortes d'échelles de temps et d'espace dans des modèles du type de ceux invoqués par le G.I.E.C. Mais,  encore une fois, là aussi, ce n'est pas mon sujet et, de toutes façons, ma compétence ne me permet pas d'en parler.

\textit{Dans le passage du PEC à la mise en avant de la {\em persistance} les mathématiques interviennent de deux façons. D'une part en  mathématisant efficacement des expériences de laboratoire, principalement en microbiologie ; on peut considérer que les expériences de Hansen et Hubbel avec leur arsenal théorique n'ont rien à envier aux sciences les plus "dures". D'autre part en proposant un modèle plausible, du même genre que ceux qui prouvent leur pertinence en microbiologie, et qui met à mal le PEC les mathématiques obligent les écologues à regarder les choses différemment. Paradoxalement, c'est en confirmant par une théorie solide le PEC dans les expériences de cultures en continu que les modèles mathématiques acquièrent auprès des écologues une autorité sans laquelle  le "contre exemple" d'Armsrong et MacGehee serait peut être passé inaperçu.}

\section{ 1989 : Mais le réel a toujours son mot à dire.\\Arditi-Ginzburg et la ratio-dépendance}\label{ratiodep}

\textit{L'explosion de la modélisation mathématique  s'explique aussi par le développement des sciences de l'ingénieur dans les sciences du vivant. Par exemple, si le chémostat peut être considéré comme un modèle simplifié d'un milieu naturel "ouvert" (comme un lac), c'est aussi le modèle du milieu particulièrement artificiel que constitue une station d'épuration des eaux usées. La bonne compréhension  de tels systèmes mobilise beaucoup de mathématiques, parfois sophistiquée, surtout quand il est question d'optimiser les procédés (\cite{HARM08,HARM17,HARM19,WAD16}).}

{\em Mais tous ces développements  mathématiques, pour utiles qu'il soient, ne sont pas à eux seuls garants de progrès de l'écologie théorique. 
Nous allons voir, pour terminer, comment des mathématiques qui étaient déjà  largement dans la boite à outils des mathématiciens de "l'âge d'or" (section \ref{agedor}) ont participé à un progrès significatif de l'évolution des concepts au tournant du XXIème siècle.}

\subsection{Les paradoxes du modèle de Gause-Rosenzweig-MacArthur}
Bien avant que la question de l'exclusion compétitive ait été remplacée par celle de la persistance,  on a considéré que modèle de Rosenzweig-MacArthur (\ref{RMA}) présentait des particularités  problématiques. En voici deux.
\subsubsection*{Le paradoxe de l'enrichissement.}
Reprenons le modèle (\ref{RMA}) :
 \beq \label{RMAter}
\begin{array}{rcl} 
\displaystyle  \frac{dx}{dt}& =&\displaystyle rx\left(1-\frac{x}{K} \right)- \left(\frac{\mu_{\max}x}{e+x}\right)y \\[8pt]
\displaystyle  \frac{dy}{dt}& =&\displaystyle c \left(\frac{\mu_{\max}x}{e+x} - m\right) y  \end{array}
\feq
La dynamique de la ressource $\frac{dx}{dt} =\displaystyle rx\left(1-\frac{x}{K}\right)$ est une logistique dont les solutions tendent vers l'équilibre $x_e = K$. 
\begin{figure}[!tb]
   \centering
  \includegraphics[width=1\textwidth]{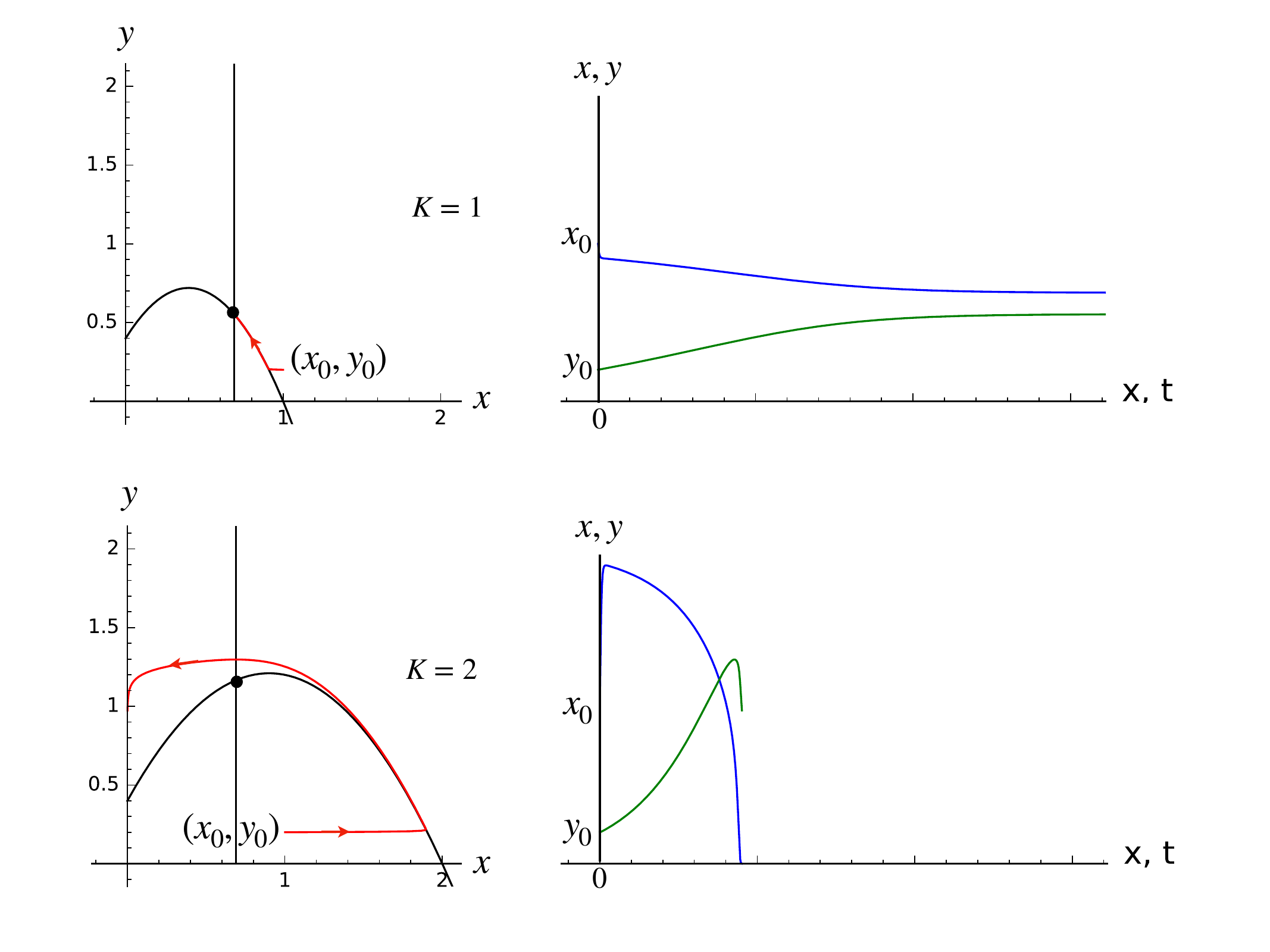} 
  \caption{Le modèle (\ref{RMAter})
  $\quad r = 4 \quad \mu_{\max} = 2 \quad e = 0.2 \quad c = 0.2\quad m = 1.55$. Haut : K = 1. Bas : K = 2}\label{enrich} 
\end{figure} 
Observons les simulations de la figure \ref{enrich} obtenues avec le modèle (\ref{RMAter}) pour les valeurs $$\quad r = 4 \quad \mu_{\max} = 2 \quad e = 0.2 \quad c = 0.2\quad m = 1.55$$ avec, $K = 1$ en haut, $K=2 $ en bas. A gauche l'espace des phases, à  droite les chroniques temporelles.
\ben
\item Pour $K= 1$, la solution issue de $(x_0,y_0) =(1,0.2)$ se dirige rapidement vers l'isocline $\frac{dx}{dt} = 0$ (la courbe en forme de parabole) puis longe cette dernière pour rejoindre un équilibre. Vu au cours du temps, la quantité de ressource $x(t)$, décroit, d'abord rapidement, puis lentement vers une valeur d'équilibre $\approx 0.65$ alors que la quantité de consommateur croit lentement vers la valeur $\approx 0.60$.
\item Pour $K=2$ on observe que l'isocline $\frac{dx}{dt} = 0$ a toujours une allure parabolique, mais beaucoup plus développée. Partant de la même condition initiale la solution rejoint rapidement l'isocline, la longe un certain temps, puis ''saute'' littéralement vers l'axe vertical, ce qui correspond à une disparition de la ressource, ce qu'on observe à droite : la ressource (en bleu) croît brusquement vers une valeur proche de $2$ puis décroit assez vite pour disparaître ; de son côté la quantité de consommateur croît vers une valeur $\approx 1.2$ beaucoup plus élevée que dans le cas précédent puis se met à décroitre rapidement quand la ressource a disparu\footnote{
Le la mathématicien.ne habitué.e à ce genre de modèle peut être choqué.e par mon \og puis ''saute'' littéralement vers l'axe vertical, ce qui correspond à une disparition de la ressource...\fg, lui, elle qui sait que la solution converge vers un cycle limite. Mathématiquement il n'y a pas disparition mais pratiquement oui car, dans cet exemple, le long du cycle, $\min x(t) \approx 10^{-6}$. C'est ce point de vue qui est intéressant dans la discussion du paradoxe.}
\fen
Il est paradoxal qu'une situation où la ressource est à priori deux fois plus abondante puisque, en l'absence de consommateur, elle croit jusqu'à la valeur 2 au lieu de 1,
conduise à l'extinction des deux espèce alors que pour $K = 1$ elles se maintiennent toutes les deux. Ce paradoxe a été mis en évidence par Rosenzweig lui même dans un article de 1971 {\em Paradox of Enrichment: Destabilization of Exploitation Ecosystems in Ecological Time} \cite{ROS71}\footnote{Science, Vol 171 pp. 385-387}. Il n'hésite pas à écrire dans le résumé de son article :
\dcom
Six reasonable models of trophic exploitation in a two-species ecosystem whose exploiters compete only by depleting each other's resource supply are presented. In each case, increasing the supply of limiting nutrients or energy tends to destroy the steady state. Thus man must be very careful in attempting to enrich an ecosystem in order to increase its food yield. There is a real chance that such activity may result in decimation of the food species that are wanted in greater abundance.
\fcom
On le voit, Rosenzweig n'hésite pas à tirer de modèles  spéculatifs des conclusions pratiques assez fortes. 
\subsubsection*{Réponse des  chaines trophiques à une augmentation de la production primaire.}
Que ce soit dans les modèle de Gause-Rozsenzweig-MacArthur    (\ref{RMA})  ou du chémostat (\ref{chem}) le prélèvement effectué par les consommateurs et par unité de temps est de la forme :
$$ - \mu(x)y$$
prélèvement transformé en une croissance de la biomasse du consommateur à travers un terme de rendement $c$ :
$$ +c\, \mu(x)y$$
Supposons que dans une chaîne trophique, à chaque niveau, le prélèvement soit de ce type. Pour une chaîne à quatre niveaux nous aurons :
 \beq \label{chaîne)}
\begin{array}{lcl} 
\displaystyle  \frac{dx}{dt}& =&\displaystyle U - \mu_1(x)y_1 \\[8pt]
\displaystyle  \frac{dy_1}{dt}& =&\displaystyle c_1\mu_1(x)y_1 -\mu_2(y_1) y_2\\[8pt]
\displaystyle  \frac{dy_2}{dt}& =&\displaystyle c_2\mu_2(y_1)y_2 -\mu_3(y_2) y_3\\[8pt]
\displaystyle  \frac{dy_3}{dt}& =&\displaystyle c_3\mu_3(y_2)y_3 -my_3
 \end{array}
\feq

La première espèce $x$ possède un taux de croissance $U$ constant (ce pourrait être un apport constant de nutriment par les affluents d'un lac). On peut calculer l'équilibre de ce système. On a :
\beq \label{chaîne0)}
\begin{array}{lclr} 
\displaystyle  0& =&\displaystyle U - \mu_1(x)y_1\quad \quad &(e_0) \\[8pt]
\displaystyle  0& =&\displaystyle c_1\mu_1(x)y_1 -\mu_2(y_1) y_2&(e_1)\\[8pt]
\displaystyle  0& =&\displaystyle c_2\mu_2(y_1)y_2 -\mu_3(y_2) y_3&(e_2)\\[8pt]
\displaystyle 0 & =&\displaystyle c_3\mu_3(y_2)y_3 -my_3&(e_3)
 \end{array}
\feq

\ben
\item En ajoutant  $(e_3)+c_3(e_2)+c_2c_3(e_1)+ c_1c_2c_3(e_0) $ on obtient :
$ 0 = c_1c_2c_3 U - my_3 $
soit: $$y_{3e} = c_1c_2c_3 U/m$$
\textbf{ L'équilibre $y_{3e}$  croit linéairement avec U.}
\item La dernière équation fixe $y_2$ tel que $c_3\mu_3(y_2) = m$ soit  à la valeur :
$$ y_{2e}  = \mu_3^{-1}(m/c_3)$$
\textbf{ L'équilibre $y_{2e}$  ne dépend pas de U.}
\item De l'équation $(e_2)$ on déduit $y_1$ en fonction de $y_2$ et $y_3$ :
$$y_{1e} = \mu_2^{-1}\left( \frac{ c_1U}{y_{2e} } \right)$$
\textbf{ L'équilibre $y_{1e}$  croît non linéairement avec U.}
\item Enfin de l'équation $(e_0)$ on déduit :
$$ x_e = \mu_1^{-1}\left( \frac{U}{y_{1e}}\right)$$
Si $\mu_1$ et $\mu_2$ sont des fonctions de Monod,\textbf{ l'équilibre} $x_e$  \textbf{ est décroissant}.
\fen 
\begin{figure}[!tb]
   \centering
  \includegraphics[width=1\textwidth]{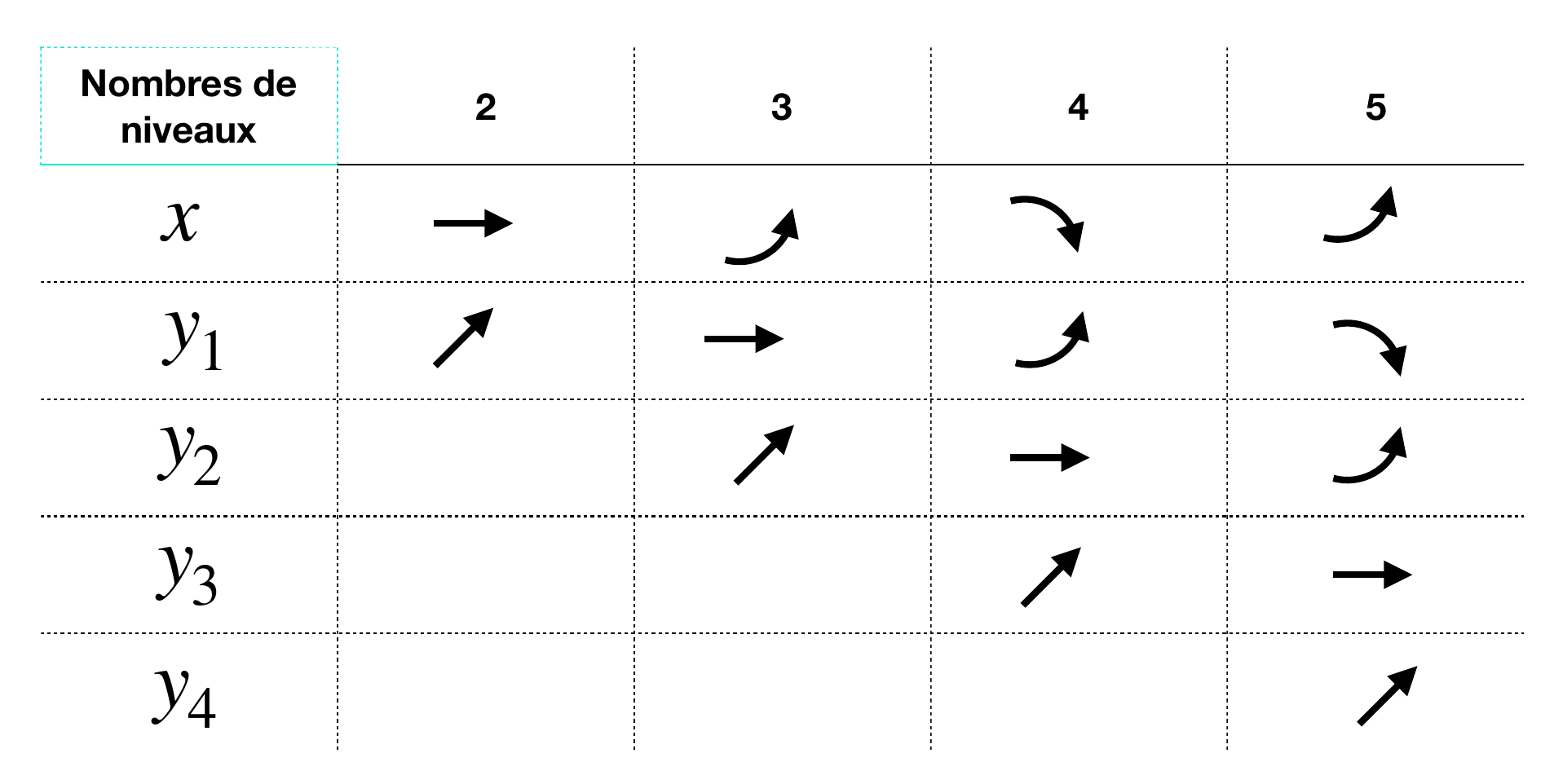} 
  \caption{Réponse d'une chaine trophique à une augmentation de la production primaire.  D'après Arditi-Ginzburg {\em How species interact} \cite{ARD12}.}\label{chaine} 
\end{figure}
Ces résultats, résumés dans le tableau de la figure \ref{chaine}  sont plutôt surprenants. Il semble que ce soit le prédateur supérieur qui impose une sorte de loi : constance, puis croissance, puis décroissance..., immuable sur les espèces qui le précèdent, indépendamment de leur nombre dans la chaîne !
\subsection{La ratio-dépendance}

Très insatisfaits de cette situation, Arditi et Ginzburg ont proposé, à la fin des années 1980 une nouvelle vision \cite{ARD89}\footnote{Coupling in predator-prey dynamics: ratio-dependence{ \em  Journal of theoretical biology}  139.3 (1989): 311-326.}. Voici comment ils présentent les choses dans leur livre 
de 2012 : {\em How Species Interact : \texttt{Altering the Standard View on Trophic Ecology}} \cite{ARD12}\footnote{OXFORD University press, 2012}
\dcom
The standard theory of predator-prey interactions taught in the common textbooks has changed little since the time of Lotka and Volterra in the 1920s. The main improvement was the introduction of Holling’s generalization of functional responses to nonlinear forms, together with the prey carrying capacity, which led to the paradox of enrichment and the limit cycles of the more complex Rosenzweig-MacArthur model.
The authors of this book attempted to suggest substantial changes in the 1970s, publishing independently first in Russian and in French. The implications of these suggestions went mostly unnoticed until our first common article on the ratio-dependent interaction (Arditi and Ginzburg 1989), which started a heated controversy. We actually wrote the article in 1987 but it took two years and three journal submissions to see it through publication. Today, this article is highly cited, and its citation rate has been increasing continuously for over 20 years.
\fcom
La polémique, dont on trouvera un compte-rendu détaillé dans {\em  From Lotka–Volterra to Arditi–Ginzburg: 90 years of evolving trophic functions} \cite{TUY20}\footnote{Tyutyunov, Y. V., \& Titova, L. I. (2020) {\em Biology Bulletin Reviews}, 10, 167-185.}  est maintenant apaisée, le livre a reçu un accueil très positif et le modèle de Arditi-Ginzburg commence à faire son apparition dans les livres de cours.\\

Arditi et Ginzburg remettent en cause le terme de prélèvement (de consommation) :
$$-\mu(x)y$$ 
qui apparait aussi bien dans le modèle du chémostat (avec $s$ à la place de $x$) que celui de Rosenzweig-MacArthur. Rappelons qu'à l'origine de ce terme il y a la loi d'action de masse :
$$ - r x y$$
où la fonction linéaire $x \mapsto r x$ a été remplacée par une fonction non linéaire (bornée) au motif (pertinent) que, contrairement à une réaction chimique, dans une ''réaction biologique'' de consommation il y a une limite à la vitesse d'absorption de la ressource par le consommateur (Holling). Mais même avec ce correctif le terme de consommation reste très étrange.

En effet, considérons, par exemple, un troupeau de moutons de 100 têtes dans un pré donné. Mettons que par unité de temps, selon le modèle ci-dessus, la quantité d'herbe prélevée soit :
$$ - \mu \times \mathrm{biomasse\; d'herbe} \times 100$$
alors, pour 10 000 moutons dans le même pré elle serait cent fois plus grande, pour $10^6$ moutons, à supposer que tous puissent tenir dans le pré, elle serait $10\;000$ fois plus grande ! Il y a quelque chose qui ne va pas. 

Arditi-Ginzburg proposent de dire que, ce qui compte pour le taux de prélèvement, ce n'est pas la quantité absolue de ressource disponible $x$, mais  {\em la quantité de ressource disponible par consommateur} : Le quotient $x/y$.\\\\ Ils proposent à la place du modèle (\ref{RMA}) le :\\
\begin{center}
 \fbox{
 \begin{minipage}[c]{11cm}
 \begin{center}
 \textbf{Modèle {\em ratio-dépendant} de Arditi-Ginzburg}
 \end{center}
\beq \label{AG}
\begin{array}{rcl} 
\displaystyle  \frac{dx}{dt}& =&\displaystyle f(x) - \mu \left (x/y\right) y \\[8pt]
\displaystyle  \frac{dy}{dt}& =&\displaystyle  \left(c\mu \left(x/y\right) - m \right)y
  \end{array}
\feq
où $x \mapsto f(x)$ est nulle en $0$, croissante puis décroissante et nulle pour $x =K$ et $x \mapsto(\mu(x)$ est nulle en $0$ , croissante et bornée.
\end{minipage}}
\end{center}
Pour ce modèle il n'y a plus de paradoxe de l'enrichissement parce que l'isocline du consommateur $\frac{dy}{dt} = 0$ n'est plus une droite verticale, mais une droite passant par l'origine comme il est expliqué sur la figure \ref{AGrat}.  
\begin{figure}[!tb]
   \centering
  \includegraphics[width=1\textwidth]{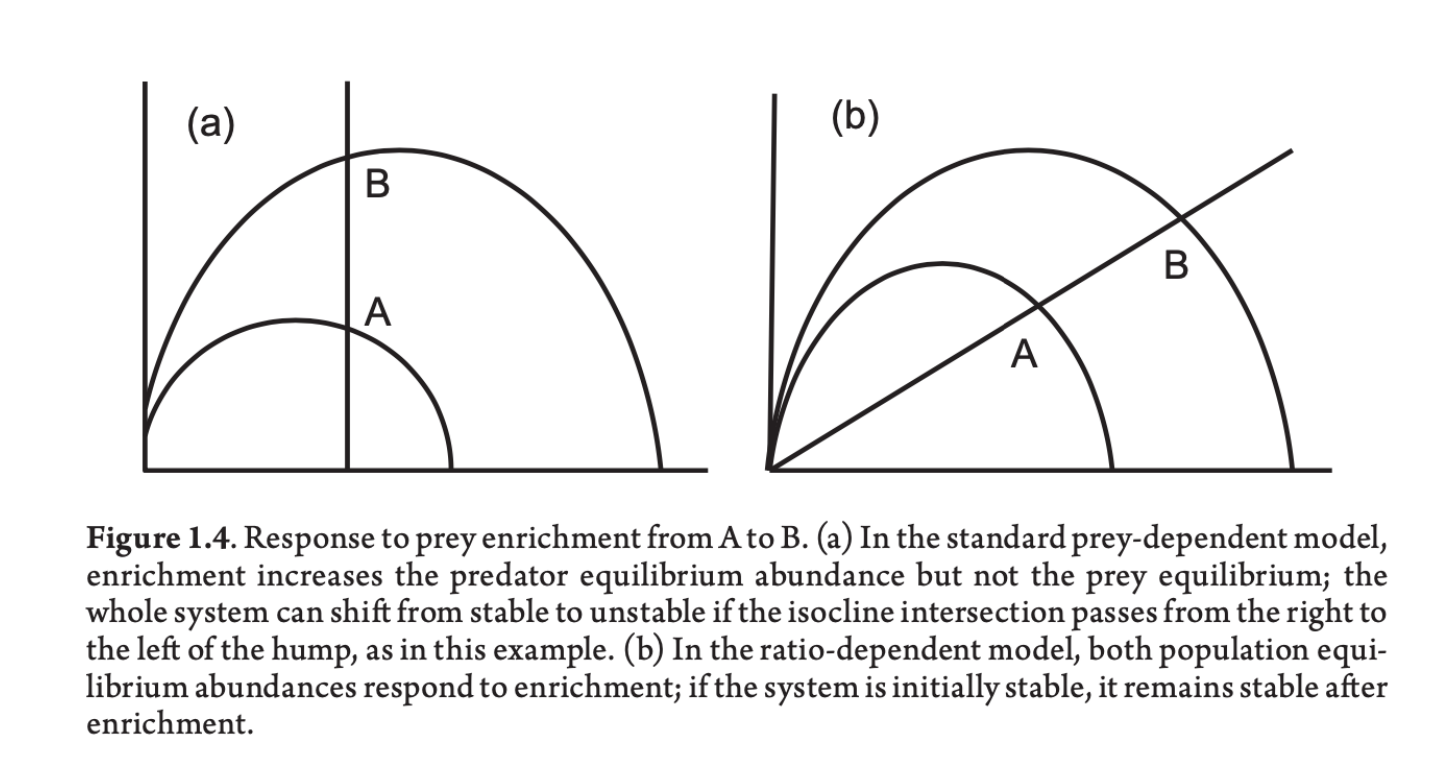} 
  \caption{Extrait de {\em How species interact \cite{ARD12}}. }\label{AGrat} 
\end{figure}
En ce qui concerne la réponse d'une chaine trophique à une augmentation de la productivité primaire les calculs sont les suivants.
 \beq \label{chaîneRatio)}
\begin{array}{lcl} 
\displaystyle  \frac{dx}{dt}& =&\displaystyle U - \mu_1(x/y_1)y_1 \\[8pt]
\displaystyle  \frac{dy_1}{dt}& =&\displaystyle c_1\mu_1(x/y_1)y_1 -\mu_2(y_1/y_2) y_2\\[8pt]
\displaystyle  \frac{dy_2}{dt}& =&\displaystyle c_2\mu_2(y_1/y_2)y_2 -\mu_3(y_2/y_3) y_3\\[8pt]
\displaystyle  \frac{dy_3}{dt}& =&\displaystyle c_3\mu_3(y_2/y_3)y_3 -my_3
 \end{array}
\feq
 A l'équilibre on a :
\beq \label{chaîneRatio0)}
\begin{array}{lclr} 
\displaystyle  0& =&\displaystyle U - \mu_1(x/y_1)y_1\quad \quad &(e_0) \\[8pt]
\displaystyle  0& =&\displaystyle c_1\mu_1(x/y_1)y_1 -\mu_2(y_1/y_2) y_2&(e_1)\\[8pt]
\displaystyle  0& =&\displaystyle c_2\mu_2(y_1/y_2)y_2 -\mu_3(y_2/y_3) y_3&(e_2)\\[8pt]
\displaystyle 0 & =&\displaystyle c_3\mu_3(y_2/y_3)y_3 -my_3&(e_3)
 \end{array}
\feq
\ben
\item En ajoutant, comme nous l'avons fait dans le cas ''ressource dépendant''   $(e_3)+c_3(e_2)+c_2c_3(e_1)+ c_1c_2c_3(e_0) $ on obtient :
$ 0 = c_1c_2c_3 U - my_3 $
soit: $$y_{3e} = c_1c_2c_3 U/m = $$
\item Maintenant que nous avons établit que $y_{3e}$ croit proportionnellement à $U$ de l'équation $(e_3)$ nous tirons :
$$\frac{y_2}{y_3} = \mu_3^{-1}\left(   \frac{m}{c_3} \right) \Longrightarrow y_{2e} = \mu_3^{-1}\left( \frac{m}{c_3}\right) y_{3e} = a_3 c_1c_2c_3 U/m $$
avec $a_3 = \mu_3^{-1}\left( \frac{m}{c_3}\right)$.
\item De $y_{2e} $ en fonction de $y_{3e}$ et de $(e_2)$ on tire :
$$c_2\mu_2(y_1/y_2) a_3y_{3e}-\frac{m}{c_3}{  y_{3e}} \Longrightarrow y_{1e} = \mu_{2}^{-1}\left(\frac{m}{a_3c_2c_3}\right)y_{2e}$$
donc $y_{1e} $ est une fonction linéaire de $y_{2e}$  et par suite linéaire de $U$
\fen 
Ainsi, à tous les niveaux de la chaine on a une réponse proportionnelle à la production primaire $U$.

Voyons enfin ce qu'il en est de la question qui nous intéresse ici : le PEC. En suivant la même logique que pour le modèle de compétition de Gause-Rosenzweig-McArthur un modèle possible de compétition ratio-dépendant est :
\beq \label{competeAG1}
\begin{array}{lcl} 
\displaystyle  \frac{dx}{dt}& =&f(x) - \mu_1(x/y_1) y_1 - \mu_2(x/y_2)y_2\\[8pt]
\displaystyle  \frac{dy_1}{dt}& =& (c_1\mu_1(x/y_1) - m_1)y_1 \\[8pt]
\displaystyle  \frac{dy_2}{dt}& =& (c_2\mu_2(x/y_2) - m_2)y_2 \\[8pt]
  \end{array}
\feq
Dans la recherche d'un équilibre, des deux dernières équations on tire :
$$ y_{ie} = \frac{x}{\mu_i^{-1} (m_i/c_i) } \quad \quad i = 1,2$$
qui reporté dans la première détermine $x_e$. Il n'y a plus de fatalité, comme dans le modèle ressource dépendant, à ce que pour annuler les deux dernières équations une des deux biomasse, au moins, soit égale à $0$. Toutefois ce modèle n'est pas vraiment satisfaisant. La logique qui dans le cas d'un seul consommateur nous a fait dire que le taux de prélèvement doit être une fonction de la ''quantité de ressource par unité de consommateur'' nous oblige a décider, quand deux consommateurs se partagent la même ressource, que le taux de prélèvement du $i-$ème consommateur doit être une fonction de la ressource divisée par {\em la somme des quantités de consommateurs} soit, de façon plus précise :
$$ - \mu_i\left( \frac{x}{a_{i1}y_1+a_{i2}y_2}\right)y_i$$
où les coefficients $a_{ij}$ expriment le fait que les deux espèces ont une relation différente à la ressource. L'analyse mathématique de ce genre de modèle est encore possible et il en ressort que sous des conditions particulières un équilibre de coexistence est possible, mais je ne m'étends pas plus sur ce point qui participe plus de la recherche contemporaine  que de l'histoire du PEC. 

Le modèle ratio dépendant a été difficilement accepté au début. On lui reprochait, entre autres, de ne par être défini pour $y = 0$ le terme $x/y$ n'ayant pas de sens. Ce point a été réglé avec la collaboration d'un mathématicien (\cite{JOS99}). Mais la critique la plus forte était  que la "ratio dépendance" ne reposait pas sur la description d'un  ''mécanisme'' de la ''capture'' d'une proie par un prédateur. Une longue polémique s'en est suivie, que je ne relate pas car nous sortons de la question du PEC. 
Elle a eu le mérite de faire comprendre qu'il n'est pas pertinent de reproduire à l'échelle macroscopique de l'évolution temporelle sur des durées longue des deux biomasses $x(t)$ et $y(t)$, le schéma à l'échelle microscopique de la rencontre de "l'individu consommateur'' avec "l'individu ressource''. La dynamique des populations était enfin sortie du paradigme de la loi d'action de masse.

	Comme ils le signalent dans leur livre, Arditi et Ginzburg n'ont pas été les premiers à montrer la pertinence de la ratio-dépendance. Les microbiologistes avaient perçu l'importance de la densité de la biomasse des bactéries et proposaient dès 1959 le modèle de Contois \cite{CON59}\footnote{Kinetics of bacterial growth: relationship between population density and specific growth rate of continuous cultures. {\em Microbiology}, 21(1), 40-50.}
	$\frac{1}{P} \frac{dP}{dt} = \frac{r_pS}{BP+S}$
où $P$ est la biomasse de bactéries, $S$ la concentration en substrat, à la place du classique modèle de Monod $\frac{1}{P} \frac{dP}{dt} = \frac{r_pS}{K+S}$. Comme $ \frac{r_pS}{BP+S} =  \frac{r_pS/P}{B+S/P}$ on voit que le modèle de Contois est {\em ratio-dépendant}. Une fois de plus la micro-biologie 
	indiquait la voie. Arditi et Ginzburg l'expliquent ainsi
	\dcom
	An interesting point, which should be instructive in the ecological context, is that the Contois model of bacterial growth (despite being rejected by a first journal) was rather easily accepted by microbiologists, as applicable to different conditions from those of the already-established Monod model. In contrast, our homologous ratio-dependent model was received with skepticism by a number of theoretical ecologists, who generally stay loyal to the standard predator-prey model (i.e., the Lotka-Volterra model with its Rosenzweig-MacArthur variant; see section 1.2). We think that this difference in attitudes is explained by the fact that theory and application are much more divorced in ecology than they are in microbiology. Microbiologists accept pragmatically the theoretical developments that result from experiments. However, because it is not easy to set up real-scale well-controlled experiments involving large organisms, theoretical ecology has developed as a more or less closed discipline, with theoretical predictions often taking the status of facts.	\fcom
	
	Notons que maintenant la dynamique des populations de micro organismes a largement intégré le champ de l'écologie théorique. Ainsi en témoigne l'article de synthèse de 2004 {\em Big questions, small worlds: microbial model systems in ecology} \cite{JES04} dont voici le résumé
	\dcom
	Although many biologists have embraced microbial model systems as tools to address genetic and physiological questions, the explicit use of microbial communities as model systems in ecology has traditionally been more restricted. Here, we highlight recent studies that use laboratory-based microbial model systems to address ecological questions. Such studies have signifi- cantly advanced our understanding of processes that have proven difficult to study in field systems, including the genetic and biochemical underpinnings of traits involved in ecological interactions, and the ecological differences driving evolutionary change. It is the simplicity of microbial model systems that makes them such powerful tools for the study of ecology. Such simplicity enables the high degrees of experimental control and replication that are necessary to address many questions that are inaccessible through field observation or experimentation.
	\fcom

\section{Conclusion : Le roman mathématique}

Lorsque la question de la dynamique des  populations d'espèces différentes en relation commence à se poser au tout debut du XXème siècle l'écologie naissante n'est pas du tout mathématisée ; les processus sont décrits dans la langue naturelle par des naturalistes qui observent des populations d'espèces évoluées (comme les oiseaux dans le cas de Grinnell) aux comportement collectifs non dénués d'intelligence. Il n'est pas étonnant qu'un concept comme celui de {\em compétition} qui était apparu  bien avant, en particulier dans le domaine de l'économie, pour décrire le comportement de populations humaines, ait ainsi été utilisé.

Il faut attendre 20 ans pour qu'apparaissent les premières tentatives de mathématisation de cette question du PEC où émergent naturellement les noms de Lotka, Gause et Volterra dont les travaux vont renforcer cette vision que la compétition entre les espèces est essentielle pour expliquer la dynamique des populations. Notons que dès cette époque les mathématiques étaient en mesure de rendre compte de  la possibilité de la {\em coexistence}, la brève remarque de Lotka dans \cite{LOT32}  le montre, mais cela n'a pas eu lieu,  {\em l'exlusion par la compétition} reste le préjugé dominant.

Après une longue éclipse de plus d'un quart de siècle $-$ de Kolmogorov  1936 à Rosenzweig-MacArthur  1963 $-\;$la question de la compétition fait un retour tonitruant avec 
  Hardin et son  célèbre {\em The competitive Exclusion Principle}. Nous avons vu que cet article n'est pas une contribution à l'écologie mais plutôt un essai sur l'idée de ''compétition'' en général,  notamment  chez Darwin et en théorie économique.  Nous sommes  assez loin de la science et l'on peut se demander pourquoi  cet article a eu un tel succès au point de devenir un "marqueur"  de la théorie écologique.  Je ne sais pas répondre à cette question. 
 
  En revanche il est une autre question sur laquelle je me propose d'émettre une hypothèse. Pourquoi, cette longue éclipse des mathématiques ?
   Il y a naturellement des raisons externes. La physique qui accumule les succès attire les esprits brillants, la guerre, puis le début de la guerre froide, autant de raisons qui peuvent expliquer en partie un éloignement des mathématiques de l'écologie. 
   Inversement, l'horreur de  la création de l'arme atomique, qui motivera la conversion à la biologie moléculaire du physicien  Leo Szilard, et le gigantisme de la techno-physique d'après guerre motiveront un retours vers des sciences plus ''molles". Mais je ne pense pas que ces causes externes suffisent à expliquer le long effacement des mathématiques de l'écologie théorique. Il y a une raison interne.
      En effet, nous avons vu qu'en 1936 la théorie des systèmes dynamiques, au sens où nous l'entendons maintenant, n'existe pas encore. Or elle seule, avec ses concepts qualitatifs $-$ stabilité structurelle, bifurcation, généricité, attracteur, chaos déterministe $-$ permet d'aller plus loin que la cinétique chimique de Lotka et Volterra. Ce n'est que vers 1970 que la théorie mathématique des systèmes dynamiques a commencé à constituer un corps de doctrine à la disposition des mathématiciens ordinaires. Ces dernier vont pouvoir, aidés des ordinateurs, nouer des relations avec l'écologie qui vont se traduire par une explosion d'articles de "modélisation".
  
  Inversement, on peut se demander quelle a été l'importance des problèmes posés par  la dynamique des populations, et plus généralement les questions de biologie, dans les motivations des mathématiciens de la constitution de la théorie des systèmes dynamique ? Dans leur article {\em Writing the History of Dynamical Systems and Chaos: {\em Longue Durée} and Revolution, Disciplines and Cultures}\footnote{Historia Mathematica 29 (2002), 273–339} les historiens des sciences D. Aubin et A. Dahan Dalmedico mentionnent bien la dynamique des populations comme source de motivation, mais, en dehors de Lotka et Volterra, ne donnent pas d'exemples d'interaction avant les années 1970-80 avec R. May. Il reste encore bien des questions et des domaines à explorer.

Mais, pour incomplète qu'elle soit,  ce que l'histoire du PEC que j'ai développée met bien en évidence c'est l'importance du {\em ''roman mathématique''}. Avant d'expliquer ce que j'entends par ''roman mathématique'' je demande au lecteur ou la lectrice de comparer l'usage qui est fait des mathématiques dans la théorie du chémostat exposée sous-section \ref{th-chemostat} et celui qu'en font Armstrong et MacGehe sous-section \ref{Armstrong-MacGehee}. Dans le premier cas nous pouvons affirmer que les mathématiques interviennent dans une théorie qui a toutes les caractéristiques d'un théorie physique.
\bito
\item Une expérience parfaitement reproductible à travers un dispositif expérimental précis.
\item Deux grandeurs mesurables avec précision : $s$ la concentration en substrat, $x$ la concentration en micro-organismes (la biomasse).
\item L'équation différentielle \eqref{chem} qui relie les quantités $s(t)$ et $x(t)$.
\fit
Dans le second cas les variables $x$, $y_1$ et $y_2$ du système d'équations \eqref{compRMA} ne représentent aucune quantités mesurables dans un dispositif expérimental ou observables dans un système naturel. Ce système n'est là que pour montrer que la ''coexistence'' ne se réduit pas à la ''coexistence à l'équilibre'', il peut y avoir coexistence sous forme dynamique. C'est ici qu'on peut parler de ''roman mathématique''.

Ce qu'est le "roman mathématique", Volterra nous l'a très bien expliqué dans ce passage de \cite{VOL28} (c'est moi qui souligne)
\dcom
Permit me to indicate how the question can be considered: Let us seek to \underline{express} in \underline{words} the way the phenomenon \underline{proceeds roughly}: afterwards let us \underline{translate} these words into \underline{mathematical language}. This leads to the formulation of \underline{differential equations}. If then we allow ourselves to be \underline{guided by the methods of analysis} we are led \underline{much farther } than the \underline{ language and ordinary reasoning} would be able to carry us and can formulate precise mathematical laws. 
\fcom
que je me permets de paraphraser :
\bito
\item On «décrit» avec des «mots» de la «langue ordinaire» comment le phénomène se «déroule grossièrement».
\item On «traduit» les mots dans le «langage mathématique» ce qui conduit à écrire des «équations différentielles».
\item Alors, si l'on se permet de se «laisser guider par les méthodes de l'analyse», on peut aller beaucoup «plus loin» et formuler des «lois mathématiques» précises auxquelles le «langage et le raisonnement  ordinaire» n'aurait pas permis d'accéder.
\fit
Mais que faut-il entendre par "traduire dans le langage mathématique" une réalité qui aurait été décrite préalablement avec des "mots" ? On peut le voir ainsi.\\\\
Dans le langage ordinaire, nous racontons une histoire, une sorte de conte :
\dcom
\textit{Il était une fois, dans une forêt de la côte pacifique de l'Amérique du Nord, deux espèces de mésanges à dos marrons, les unes à flancs clairs les autres à flancs foncés. Ces deux espèces avaient des habitudes très semblables, en particulier elles se nourrissaient à peu près des mêmes insectes. A un moment donné il s'est trouvé que les espèces à flancs clairs ont eu un succès reproductif plus grand que les autres, leur population est devenue plus importante, obligeant l'espèce à flancs foncé à émigrer ....}
\fcom
La '' traduction mathématique '' n'est pas une traduction où un dictionnaire nous dirait "mésange à flancs clairs '' se traduit par $x_1$ et "mésange à flancs foncés" par $x_2$ mais plutôt  l'élaboration d'une autre histoire, d'un autre conte, mais  mathématique cette fois :
\dcom
\textit{Il était une fois le système différentiel :
\beq 
\begin{array}{rcl} 
\displaystyle  \frac{dx_1}{dt}& =&\displaystyle b_1 x_1\left(1-( a_{11}x_1+a_{12} x_2) \right) \\[8pt]
\displaystyle  \frac{dx_2}{dt}& =&\displaystyle b_2 x_2\left(1-(a_{21}  x_2+a_{22}x_2 )\right)
  \end{array}
\feq
où nous convenons de voir dans les grandeurs $x_1(t) $ et $x_2(t)$ la taille de deux populations qui interragissent. Le coefficient $a_{11} $ exprime la force de la ''compétition intra-spécifique" à l'intérieur de l'espèce $x_1 $ et $a_{12}$ celle de la pression que l'espèce $2$ exerce sur l'espèce 1.}
\fcom
 C'est l'\textbf{interprétation initiale}\footnote{C'est ce qu'on appelle le ''modèle de compétition'' de Volterra.}.
Maintenant il existe une ''loi mathématique'', ou plutôt un théorème, qui dit que 
\dcom
\textit{si $a_{11}a_{22} > a_{12}a_{21}$  alors il existe un équilibre stable où $x_1$ et $x_2$ sont positifs  tous les deux, }
\fcom ce que nous interprétons
\dcom  {\em lorsque la compétition intra-specifique est plus forte que la compétition inter-spécifique, il y a un équilibre de coexistence.}\fcom 
 C'est l\textbf{'interprétation déduite.}\footnote{Il s'agit là aussi d'un théorème mathématique élémentaire dont la démonstration et l'interprétation se trouvent dans tous les ouvrages élémentaires d'écologie.} 
 
Ce qui fait la force d'une "loi mathématique", c'est que si l'on est d'accord pour trouver pertinente l'interprétation initiale, on est obligé d'accepter l'interprétation déduite. C'est le propre des mathématiques que, dès lors que des axiomes ont été acceptés, les théorèmes qui s'en déduisent sont certains. La force des mathématiques est de faire des déductions que nul ne pourra contester ce qui n'est pas le cas du raisonnement dans la langue naturelle.

Mais si les déductions mathématiques ne peuvent pas être contestées, il n'en est pas de même des interprétations.
Il faut des arguments pour étayer une interprétation. Lotka et Volterra les ont puisés dans une science de leur temps qu'il connaissaient bien : La chimie et sa loi d'action de masse.
 L'analogie entre des proies et des prédateurs d'une part et des molécules qui se rencontrent de l'autre, est bien sûr grossière. Lotka, Volterra et Gause le reconnaissent volontiers, mais c'est une première approximation qui permet d'avancer\footnote{Gause reproduit dans son laboratoire cette approximation  avec des paramécies qui phagocytent des bactéries transformant ainsi ce qui était un ''roman'' en un début de théorie physique.}. Les résultats sont encourageants et cette interprétation va tenir avant d'être remise en cause partiellement avec le modèle Gause-Rosenzweig-MacArthur puis plus radicalement avec le modèle Arditi-Ginzburg.
 
 
 
 Cette utilisation du "roman mathématique" ne va pas sans risque. Il est tentant de se réclamer de l'autorité de la "vérité mathématique" pour faire progresser ses idées dans, et hors du champs scientifique. Hardin est un bon exemple de ces dérives possibles, heureusement contrées par Hutchinson, adepte du "roman mathématique" mais conscient de ses limites,  que je cite à nouveau dans son {\em Paradox of the plancton} \cite{HUT61}.
 
 \dcom
 The principle of competitive exclusion has recently been under attack from a number of quarters. Since the principle can be deduced mathematically from a relatively simple series of postulates, which with the ordinary postulates of mathematics can be regarded as forming an axiom system, it follows that if the objections to the principle in any cases are valid, some or all the biological axioms introduced are in these cases incorrect.
 \fcom
 
  Ce mode d'utilisation des mathématiques que je viens d'appeler ''roman mathématique'' est très largement répandu en biologie et en sciences humaines sous l'appellation ''modélisation". On a peut-être remarqué que tout au long de cet article j'ai eu une certaine réticence à utiliser les expressions ''modèle'' et ''modélisation''. La raison en est que je trouve qu'il y a beaucoup de confusion autour d'elles. Il n'y a rien de commun entre le ''modèle standard'' en physique des particules et le modèle de prédation de Gause-Rozensweig-Mac Arthur, ni entre le ''modèle numérique'' d'un écoulement d'un fluide et le modèle ratio-dépendant d'Arditi-Ginzburg. Et même à l'intérieur de la dynamique des population entre le modèle du chémostat et le modèle de Gause-Rosenzweig-MasArthur. On peut s'intéresser aux raisons de cette confusion\footnote{Je l'ai fait dans  \cite{LOB19} et dans une version courte dans la revue Alliage \cite{LOB24}.}mais ce n'était pas l'objet de cet article. Je voulais simplement montrer que cet usage particulier des mathématiques que j'appelle ''roman mathématique'' a joué un rôle significatif dans l'évolution du concept de compétition en écologie théorique.

  {}
\appendix 
\section{ A propos de Witt} \label{witt}
Traduction par Deap L. de :
\url{https://de.wikipedia.org/wiki/Alexander_Adolfowitsch_Witt}

Né en 1902\\
Witt était le fils d’un employé de commerce et fréquentait le lycée à Moscou. À partir de 1920, il étudie à l’université Lomonossov tout en accomplissant son service militaire à l’école de photographie aérienne. À la fin de son service militaire, il est devenu chef du bureau statistique du personnel au siège de l’aviation. Ensuite, il a été de 1926 à 1929 avec Leonid Mandelstam à la Lomonosov, où il a obtenu son doctorat pendant cette période. Là, avec Alexandre Alexandrovitch Andronov, il est devenu l’un des principaux scientifiques dans le domaine des vibrations non linéaires. Il a également obtenu son habilitation auprès de Mandelstam et est devenu professeur à l’université Lomonossov. La monographie commune d’Andronov, de Chaikin et de Witt parut en 1937, lorsque Witt fut arrêté lors des purges staliniennes et condamné à cinq ans de prison le 4 juillet 1937. Il meurt peu après dans le camp de Kolyma (sa dernière lettre à son épouse date du 13 décembre 1937).

Mandelstam, qui proposa à Witt un séjour à l’étranger en 1929, loua particulièrement le talent mathématique de Witt. Lorsque, par exemple, dans son séminaire, une conférence sur l’équation de Schrödinger risquait d’échouer parce que le conférencier était malade, Mandelstam confia à Witt la tâche qu’il résout brillamment. [1] Witt a également été un pionnier dans le domaine des réactions d’oscillation chimique (comme la réaction de Belousov-Zhabotinsky), bien avant que celles-ci ne soient étudiées dans le cadre de la théorie du chaos, ainsi que dans le domaine de l’analyse mathématique des systèmes écologiques et de la dynamique des populations (un domaine déjà actif à cette époque par Vito Volterra et d’autres).

Il était marié depuis 1935 avec Olga Alexeïevna Witt et avait un fils Alexander Witt. Les parents de Witt ont été déportés au Kazakhstan en 1941, où ils sont morts peu après.

Il est officiellement réhabilité en 1957.

Schriften (publications)
A.A. Andronow, S.E. Chaikin, A.A. Witt Theorie der Schwingungen, Akademie Verlag 1965 (das Buch erschien zuerst russisch 1937 ohne Nennung von Witts Namen, der erst in der zweiten russischen Auflage 1959 hinzugefügt wurde)
mit G.F. Gause Behavior of mixed populations and the problem of natural selection, American Naturalist Bd. 69, 1935, S. 596–609
mit G.F. Gause, N.P. Smaragdova: Further studies of interaction between predators and prey, Journal of Animal Ecology, Bd. 5, 1936, S. 1–18

\end{document}